\documentstyle[aps,prl,twocolumn,epsfig]{revtex}
\begin{document}

\title{Pion Superfluidity and Meson Properties at Finite Isospin Density}
\author{Lianyi He, Meng Jin, and Pengfei Zhuang\\
        Physics Department, Tsinghua University, Beijing 100084, China}

\maketitle
\begin{abstract}
We investigate pion superfluidity and its effect on meson
properties and equation of state at finite temperature and isospin
and baryon densities in the frame of standard flavor SU(2) NJL
model. In mean field approximation to quarks and random phase
approximation to mesons, the critical isospin chemical potential
for pion superfluidity is exactly the pion mass in the vacuum, and
corresponding to the isospin symmetry spontaneous breaking, there
is in the pion superfluidity phase a Goldstone mode which is the
linear combination of the normal sigma and charged pion modes. We
calculate numerically the gap equations for the chiral and pion
condensates, the phase diagrams, the meson spectra, and the
equation of state, and compare them with that obtained in other
effective models. The competitions between pion superfluidity and
color superconductivity at finite baryon density and between pion
and kaon superfluidity at finite strangeness density in flavor
SU(3) NJL model are briefly discussed.
\end{abstract}

\noindent ${\bf PACS: 11.10.Wx, 12.38.-t, 25.75.Nq}$

\section {Introduction}
\label{s1}
It is generally believed that there exists a rich phase structure
of Quantum Chromodynamics (QCD) at finite temperature and baryon
density, for instance, the deconfinement process from hadron gas
to quark-gluon plasma, the transition from chiral symmetry
breaking to the symmetry restoration\cite{hwa}, and the color
superconductivity\cite{alford} at low temperature but high baryon
density. Recently, the study on the QCD phase structure is
extended to finite isospin density. The physical motivation to
study QCD at finite isospin density and the corresponding pion
superfluidity is related to the investigation of compact stars,
isospin asymmetric nuclear matter and heavy ion collisions at
intermediate energies. In early studies on dense nuclear matter
and compact stars, it has been suggested that charged pions and
even kaons are condensed at sufficiently high
density\cite{migdal,sawyer,camp,kaplan}.

While the perturbation theory of QCD can well describe the
properties of the new phases at high temperature and/or high
density, the study on the phase structure at moderate (baryon
or/and isospin) density depends on lattice QCD calculation and
effective models with QCD symmetries. While there is not yet
precise lattice result at finite baryon density due to the Fermion
sign problem\cite{karsch}, it is in principle no problem to do
lattice simulation at finite isospin density\cite{son}. Recently
it is found that\cite{kogut1,kogut2,kogut3} there is a phase
transition from normal phase to pion superfluidity at a critical
isospin chemical potential which is about the pion mass in the
vacuum, $\mu_I^c \simeq m_\pi$. The QCD phase structure at finite
isospin density is also investigated in many low energy effective
models, such as chiral perturbation
theory\cite{son,kogut4,split,loewe,michae}, Nambu--Jona-Lasinio
(NJL) model\cite{toub,bard,bard3f,frank,he}, random matrix
method\cite{klein1,klein2}, ladder QCD\cite{bard1}, and strong
coupling lattice QCD\cite{nishi}.

One of the models that enables us to see directly how the dynamic
mechanisms of chiral symmetry breaking and restoration operate is
the NJL model\cite{njl} applied to
quarks\cite{vogl,sandy,volkov,hatsuda}. The chiral phase
transition
line\cite{vogl,sandy,volkov,hatsuda,hufner,zhuang1,zhuang2} in the
temperature and baryon chemical potential ($T-\mu_B$) plane
calculated in the model is very close to the one obtained with
lattice QCD. Recently, this model is also used to investigate the
color superconductivity at moderate baryon
density\cite{schwarz,huang1,huang2,shov,liao,ebert,ruster,blasch}.
It is nature to extend the NJL model to finite isospin chemical
potential\cite{toub,bard}. In the frame of general NJL Lagrangian
with $U_A(1)$ breaking term\cite{frank}, it is analytically
proved\cite{he} at quark level that the critical isospin chemical
potential for pion superfluidity is exactly the pion mass in the
vacuum, $\mu_I^c = m_\pi$, and this relation is independent of the
model parameters, regularization scheme, $U_A(1)$ breaking
coupling, and color degrees of freedom $N_c$.

Unlike the effective models at hadron level where one can study
the effect of phase transitions on meson and diquark properties in
mean field approximation, in the NJL model applied to quarks, only
the order parameters of the phase transitions, namely the
condensates, can be described at mean field level, to study mesons
and diquarks should go beyond the mean field approximation. It is
well-known that\cite{sandy} in the mean field approximation to
quarks together with random phase approximation (RPA) to mesons in
the NJL model, one can obtain the hadronic mass spectra and the
static properties of mesons remarkably well, especially the
Goldstone mode corresponding to the chiral symmetry spontaneous
breaking. However, in the pion superfluidity or the color
superconductivity phase, the structure of the bubble summation in
RPA is much more complicated than that in the chiral breaking
phase. The correlation among the mesons and diquarks becomes very
important and leads to the Goldstone modes corresponding to the
isospin and color symmetry spontaneous breaking. In this paper, we
study analytically and numerically the pion superfluidity, its
effect on the meson properties, and its possible competition with
the color superconductivity and kaon condensation in the frame of
standard NJL model at finite temperature and baryon, isospin and
strangeness chemical potentials.

The paper is organized as follows. In Section \ref{s2}, we
establish the mean field theory of the NJL model at finite
temperature and baryon and isospin densities. In Section \ref{s3},
we study the pion superfluidity at zero temperature, zero baryon
chemical potential, but finite isospin chemical potential, and
calculate the chiral and pion condensates and thermodynamic
functions. In Section \ref{s4}, we study the temperature behavior
of the pion superfluidity at zero baryon chemical potential and
show the phase diagrams in the temperature and isospin chemical
potential plane. In Section \ref{s5}, we investigate the meson
modes in both normal phase and superfluidity phase in RPA. In
Section \ref{s6}, the bosonized version of the $SU(2)$ NJL model
is discussed at finite isospin chemical potential with pion
condensation. In Section \ref{s7}, we study the baryon chemical
potential effect and the competition between pion and diquark
condensations. In Section \ref{s8}, we briefly discuss the flavor
$SU(3)$ NJL model at finite isospin and strangeness chemical
potentials including both pion and kaon condensations. We
summarize in Section \ref{s9}.

\section {Mean Field Theory of NJL Model at Finite $\mu_I$}
\label{s2}
The standard approach for dealing with the thermodynamics of a
variable particles is via the grand canonical ensemble. The key
quantity for a system with baryon number conservation and isospin
number conservation is the partition function defined by
\begin{equation}
\label{mf1} Z(T,\mu_I,\mu_B,V)= \text{Tr} e^{-\beta(H-\mu_B B-\mu_I
I_3)}\ .
\end{equation}
Here $V$ is the volume of the system, $\beta$ the inverse
temperature $\beta = 1/T$, $\mu_B$ and $\mu_I$ are the baryon and
isospin chemical potentials, and $B$ and $I_3$ the conserved
baryon number and isospin number operators. If we take only quark
field $\psi$ as elementary field of the system, the operators can
be expressed as
\begin{eqnarray}
\label{mf2} B &=& \frac{1}{N_c}\int
d^{3}{\bf x}\bar{\psi}\gamma_{0}\psi\ ,\nonumber\\
I_{3} &=& \frac{1}{2}\int d^3{\bf x}\bar{\psi}\gamma_{0}\tau^{3}\psi
,
\end{eqnarray}
The factors $1/N_c$ and $1/2$ reflect the fact that $N_c$ quarks
make a baryon and quark's isospin quantum number is $1/2$. In the
imaginary time formulism of finite temperature field theory, the
partition function can be represented as
\begin{equation}
\label{mf3}
Z(T,\mu_I,\mu_B,V)=\int[d\bar{\psi}][d\psi]e^{\int_{0}^{\beta}d\tau\int
d^{3}{\bf x}\left({\cal
L}+\bar{\psi}\hat{\mu}\gamma_{0}\psi\right)}\ ,
\end{equation}
where ${\cal L}$ is the Lagrangian density of the system, and
$\hat{\mu}$ the quark chemical potential matrix in flavor space
$\hat{\mu}= \text{diag}(\mu_u,\mu_d)$ with the $u$ and $d$ quark
chemical potentials,
\begin{eqnarray}
\label{mf4}
\mu_u &=& \frac{\mu_B}{N_c}+\frac{\mu_I}{2}\ ,\nonumber\\
\mu_d &=& \frac{\mu_B}{N_c}-\frac{\mu_I}{2}\ .
\end{eqnarray}

The flavor $SU(2)$ NJL Lagrangian density is defined
as\cite{njl,vogl,sandy,volkov,hatsuda}
\begin{equation}
\label{mf5} {\cal L} =
\bar{\psi}\left(i\gamma^{\mu}\partial_{\mu}-m_0\right)\psi
+G\left[\left(\bar{\psi}\psi\right)^2+\left(\bar{\psi}i\gamma_5\mbox{\boldmath{$\tau$}}\psi\right)^2
\right]\ ,
\end{equation}
with scalar and pseudoscalar interactions corresponding to
$\sigma$ and ${\bf \pi}$ excitations, or equivalently,
\begin{eqnarray}
\label{mf6} {\cal L} &=&
\bar{\psi}\left(i\gamma^{\mu}\partial_{\mu}-m_0\right)\psi\\
&+&
G\Big[\left(\bar{\psi}\psi\right)^2+\left(\bar{\psi}i\gamma_5\tau_3\psi\right)^2+2\left(\bar\psi
i\gamma_5\tau_+\psi\right)\left(\bar\psi i\gamma_5\tau_-\psi
\right)\Big]\ ,\nonumber
\end{eqnarray}
with $\pi_+$ and $\pi_-$ excitations instead of $\pi_1$ and
$\pi_2$, and $\tau_{\pm} = \left(\tau_1 \pm
i\tau_2\right)/\sqrt{2}$.

The above Lagrangian density has the symmetry $U_B(1)\bigotimes
SU_I(2)\bigotimes SU_A(2)$ corresponding to baryon number gauge
symmetry, isospin symmetry and chiral symmetry, respectively.
However, in the presence of a nonzero isospin chemical potential,
the isospin symmetry $SU_I(2)$ breaks down to $U_I(1)$ global
symmetry which is related to Bose-Einstein condensation of charged
pions, and the chiral symmetry $SU_A(2)$ breaks down to $U_{IA}(1)$
global symmetry which is associated with the chiral condensation of
sigma meson. Therefore, in the case with $\mu_I\neq 0$, chiral
symmetry restoration at finite temperature and chemical potentials
means only degenerate of $\sigma$ and $\pi_0$ meson masses. At zero
baryon chemical potential, the ``Fermi surfaces" of the $u (d)$ and
anti-$d (u)$ quarks coincide. Therefore, at sufficiently high
$\mu_I>0$ the condensate of $u$ and anti-$d$ quarks is favored, and
at sufficiently high $\mu_I<0$ the condensate of $d$ and anti-$u$
quarks is favored. Since at low isospin chemical potential (and low
temperature and low baryon chemical potential) there is no
deconfinement, the condensates can be identified as Bose-Einstein
condensates of charged pions, $\pi^+$ and $\pi^-$. Introducing the
chiral condensate,
\begin{eqnarray}
\label{mf7} \langle\bar{\psi}\psi\rangle &=& \sigma = \sigma_u
+\sigma_d\
,\nonumber\\
\sigma_u &=&\langle\bar u u\rangle\ ,\nonumber\\
\sigma_d &=&\langle\bar d d\rangle\ ,
\end{eqnarray}
and two pion condensates,
\begin{eqnarray}
\label{mf8} \langle\bar{\psi}i\gamma_5\tau_+\psi\rangle &=&
\sqrt{2}\langle\bar{u}i\gamma_5d\rangle=\pi^+={\pi\over \sqrt
2}e^{i\theta}\ ,\nonumber\\
\langle\bar{\psi}i\gamma_5\tau_-\psi\rangle &=&
\sqrt{2}\langle\bar{d}i\gamma_5u\rangle=\pi^-={\pi\over \sqrt
2}e^{-i\theta}\ ,
\end{eqnarray}
a nonzero condensate $\sigma$ means spontaneous chiral symmetry
breaking, and a nonzero condensate $\pi$ means spontaneous isospin
symmetry breaking. The phase factor $\theta$ related to the
condensates $\pi_+$ and $\pi_-$ indicates the direction of the
$U_I(1)$ symmetry breaking. If the system is in global thermal
equilibrium, $\theta$ is a constant and it does not change any
physical result. For convenience we choose $\theta=0$ in the
following. As is well-known and will be seen later, there are
corresponding Goldstone modes in the region of spontaneous chiral
breaking and the region of spontaneous isospin breaking.

Defining the meson fluctuations
$\delta_\sigma,\delta_{\pi_0},\delta_{\pi_+},\delta_{\pi_-}$ as
\begin{eqnarray}
\label{mf9}
\bar\psi\psi &=& \sigma +\delta_\sigma \ ,\nonumber\\
\bar{\psi}i\gamma_5\tau_3\psi &=& \delta_{\pi_0}\ ,\nonumber\\
\bar{\psi}i\gamma_5\tau_+\psi &=& \pi^++\delta_{\pi_+}\
,\nonumber\\
\bar{\psi}i\gamma_5\tau_-\psi &=& \pi^-+\delta_{\pi_-}\ ,
\end{eqnarray}
and keeping only the linear terms in $\delta$, the partition
function in mean field approximation is simplified as
\begin{equation}
\label{mf10}
Z(T,\mu_I,\mu_B,V)=\int[d\bar{\psi}][d\psi]e^{\int_{0}^{\beta}d\tau\int
d^{3}{\bf x}{\cal L}_{mf}}\ ,
\end{equation}
with the mean field Lagrangian density
\begin{eqnarray}
\label{mf11} {\cal L}_{mf}
&=&\bar{\psi}\left[i\gamma^\mu\partial_\mu-\left(m_0-2G\sigma\right)+\mu\gamma_0
+2G\pi i\gamma_5\tau_1\right]\psi\nonumber\\
&-&G\left(\sigma^2 + \pi^2\right)\ ,
\end{eqnarray}
from which the inverse quark propagator matrix in the flavor space
as a function of quark momentum can be derived directly,
\begin{equation}
\label{mf12}
{\cal S}_{mf}^{-1}(k)=\left(\begin{array}{cc} \gamma^\mu k_\mu+\mu_u\gamma_0-M_q & 2iG\pi\gamma_5\\
2iG\pi\gamma_5 & \gamma^\mu
k_\mu+\mu_d\gamma_0-M_q\end{array}\right)\ ,
\end{equation}
with the effective quark mass
\begin{equation}
\label{mf13} M_q=m_0-2G\sigma\ .
\end{equation}
Using the method of massive energy projectors\cite{huang1}, we
obtain explicitly the mean field quark propagator
\begin{equation}
\label{mf14}
{\cal S}_{mf}(k)= \left(\begin{array}{cc} {\cal S}_{uu}(k)&{\cal S}_{ud}(k)\\
{\cal S}_{du}(k)&{\cal S}_{dd}(k)\end{array}\right)\ ,
\end{equation}
with the four matrix elements
\begin{eqnarray}
\label{mf15} {\cal S}_{uu}(k) &=& {k_0+E_k+\mu_d\over
(k_0-E^-_-)(k_0+E^-_+)}\Lambda_+\gamma_0\nonumber\\
&+& {k_0-E_k+\mu_d\over (k_0-E^+_-)(k_0+E^+_+)}\Lambda_-\gamma_0\ ,\nonumber\\
{\cal S}_{dd}(k) &=& {k_0-E_k+\mu_u\over
(k_0-E^-_-)(k_0+E^-_+)}\Lambda_-\gamma_0\nonumber\\
&+& {k_0+E_k+\mu_u\over (k_0-E^+_-)(k_0+E^+_+)}\Lambda_+\gamma_0\ ,\nonumber\\
{\cal S}_{ud}(k) &=& {2iG\pi\over
(k_0-E^-_-)(k_0+E^-_+)}\Lambda_+\gamma_5\nonumber\\
&+& {2iG\pi\over (k_0-E^+_-)(k_0+E^+_+)}\Lambda_-\gamma_5\ ,\nonumber\\
{\cal S}_{du}(k) &=& {2iG\pi\over
(k_0-E^-_-)(k_0+E^-_+)}\Lambda_-\gamma_5\nonumber\\
&+& {2iG\pi\over (k_0-E^+_-)(k_0+E^+_+)}\Lambda_+\gamma_5\ ,
\end{eqnarray}
where $E^\pm_\mp$ are effective quark energies
\begin{eqnarray}
\label{mf16}
E^\pm_\mp &=& E_k^\pm \mp \frac{\mu_B}{3}\ ,\nonumber\\
E_k^\pm &=& \sqrt{\left(E_k\pm {\mu_I/2}\right)^2+4G^2\pi^2}\
,\nonumber\\
E_k &=& \sqrt{|{\bf k}|^2+M_q^2}\ ,
\end{eqnarray}
and $\Lambda_\pm$ are the energy projectors
\begin{equation}
\label{mf17} \Lambda_{\pm}({\bf k}) = {1\over
2}\left(1\pm{\gamma_0\left({\bf \gamma}\cdot{\bf
k}+M_q\right)\over E_k}\right)\ .
\end{equation}

\subsection {Gap Equations}
The quark propagator (\ref{mf14}) is the background of
calculations for quarks in mean field approximation and also for
mesons in RPA. From the definitions of the chiral and pion
condensates (\ref{mf7}) and (\ref{mf8}), it is easy to express
them in terms of the matrix elements of the quark propagator,
\begin{eqnarray}
\label{mf18}
\sigma_u &=& - N_c\int {d^4k\over (2\pi)^4} \text{Tr}_D\left[i {\cal S}_{uu}(k)\right]\ ,\nonumber\\
\sigma_d &=& - N_c\int {d^4k\over (2\pi)^4} \text{Tr}_D\left[i {\cal S}_{dd}(k)\right]\ ,\nonumber\\
\pi &=& N_c\int {d^4 k\over (2\pi)^4} \text{Tr}_D \left[\left({\cal
S}_{ud}(k)+{\cal S}_{du}(k)\right)\gamma_5\right]\ ,
\end{eqnarray}
where the trace $\text{Tr}_D$ is taken in Dirac space and the four
momentum integration is defined as $\int {d^4k\over (2\pi)^4}
=iT\sum_n\int {d^3\bf k\over (2\pi)^3}$ at finite temperature.
Obviously, in the study without considering color superconductivity,
the color degrees of freedom in the NJL model is trivial, and the
trace in color space simply contributes a factor $N_c$. Performing
the trace in spin space (the trace including the energy projectors
are presented in Appendix \ref{a1}) and the Matsubara frequency
summation we have
\begin{eqnarray}
\label{mf19} \sigma_u &=& \int{d^3{\bf k}\over
(2\pi)^3}{N_cM_q\over
E_k}\Big[ f(E^-_-)+f(-E^-_+)-f(E^+_-)\nonumber\\
&-&f(-E^+_+)+{E_k-\mu_I/2\over
E_k^-}\left(f(E^-_-)-f(-E^-_+)\right)\nonumber\\
&+&{E_k+\mu_I/2\over E_k^+}\left(f(E^+_-)-f(-E^+_+)\right)
\Big]\ ,\nonumber\\
\sigma_d &=& \int{d^3{\bf k}\over (2\pi)^3}{N_cM_q\over E_k}\Big[
-f(E^-_-)-f(-E^-_+)+f(E^+_-)\nonumber\\
&+&f(-E^+_+)+{E_k-\mu_I/2\over
E_k^-}\left(f(E^-_-)-f(-E^-_+)\right)\nonumber\\
&+&{E_k+\mu_I/2\over E_k^+}\left(f(E^+_-)-f(-E^+_+)\right)
\Big]\ ,\nonumber\\
\pi &=& -4N_c G\pi\int{d^3{\bf k}\over (2\pi)^3}\Big[{1\over
E_k^-}\left(f(E^-_-)-f(-E^-_+)\right)\nonumber\\
&+&{1\over E_k^+}\left(f(E^+_-)-f(-E^+_+)\right)\Big]\ ,
\end{eqnarray}
with the Fermi-Dirac distribution function
\begin{equation}
\label{mf20} f(x) = {1\over e^{\beta x}+1}\ .
\end{equation}
This group of gap equations has a symmetry that it is invariant
under the transformations $\sigma_u\rightarrow\sigma_d,
\sigma_d\rightarrow\sigma_u, \mu_I\rightarrow-\mu_I$. Therefore,
we can only concentrate on the case $\mu_I>0$ and the results for
the case $\mu_I<0$ can be obtained easily. It is also necessary to
note that for $\mu_B=0$ or $\mu_I=0$, there is always
$\sigma_u=\sigma_d$,  since in this case the chemical potential
difference between $\bar u $ and $u$ is the same as the difference
between $\bar d$ and $d$.

In any case we can combine the first two gap equations into one
determining the chiral condensate $\sigma$,
\begin{eqnarray}
\label{mf21} \sigma &=& \int{d^3{\bf k}\over
(2\pi)^3}{2N_cM_q\over E_k} \Big[{E_k-\mu_I/2\over
E_k^-}\left(f(E^-_-)-f(-E^-_+)\right)\nonumber\\
&+&{E_k+\mu_I/2\over E_k^+}\left(f(E^+_-)-f(-E^+_+)\right)\Big]\ .
\end{eqnarray}

We now consider the limit of $\mu_I = 0$. In this case the gap
equations are reduced to
\begin{eqnarray}
\label{mf22} && \sigma\left[1+\int{d^3{\bf k}\over
(2\pi)^3}{8N_cG\over \sqrt{E_k^2+4G^2\pi^2}}
\left(f(E^-_-)-f(-E^-_+)\right)\right]\nonumber\\
&&=\int{d^3{\bf k}\over (2\pi)^3}{4m_0N_c\over
\sqrt{E_k^2+4G^2\pi^2}}
\left(f(E^+_-)-f(-E^+_+)\right)\ ,\nonumber\\
&& \pi\left[1+\int{d^3{\bf k}\over (2\pi)^3}{8N_cG\over
\sqrt{E_k^2+4G^2\pi^2}}
\left(f(E^-_-)-f(-E^-_+)\right)\right]\nonumber\\
&&=0\ .
\end{eqnarray}
Only in chiral limit, $m_0 = 0$, there is possibly nonzero $\pi$,
and the gap equation
\begin{eqnarray}
\label{mf23} && 1+\int{d^3{\bf k}\over (2\pi)^3}{8N_cG\over
\sqrt{k^2+4G^2(\sigma^2+\pi^2)}}
\left(f(E^-_-)-f(-E^-_+)\right)\nonumber\\
&&= 0
\end{eqnarray}
determines the combination of the two condensates $\sigma^2 +
\pi^2$. In real world with nonzero quark mass, $m_0 \neq 0$, the
pion condensate is forced to vanish, $\pi =0$ at $\mu_I = 0$, and
$\sigma$ satisfies the well-known NJL gap equation
\begin{eqnarray}
\label{mf24} &&\sigma-\int{d^3{\bf k}\over (2\pi)^3}{4N_cM_q\over
E_k} \left[f(E_k-{\mu_B\over 3})-f(-E_k-{\mu_B\over
3})\right]\nonumber\\
&&= 0\ .
\end{eqnarray}

\subsection {Thermodynamics }
The thermodynamic potential in mean field approximation
\begin{equation}
\label{mf25} \Omega=-{T\over V}\ln Z=-{T\over V}\text{Tr}\ln{{\cal
S}^{-1}_{mf}}+G\left(\sigma^2+\pi^2\right)
\end{equation}
can be evaluated with the effective quark energies
\begin{eqnarray}
\label{mf26} &&\Omega(T,\mu_I,\mu_B|\sigma_u,\sigma_d,\pi) =
+G\left(\sigma^2+\pi^2\right)\\
&&\ \ \ \ \ \ \ \ -N_c\int {d^3 {\bf k}\over (2\pi)^3}\Big[E^-_-
-E^-_+ +E^+_-
-E^+_+\nonumber\\
&&\ \ \ \ \ \ \ \ + 2T\Big(\ln\left(1+e^{- E^-_-/T}\right)
+\ln\left(1+e^{ E^-_+/T}\right)\nonumber\\
&&\ \ \ \ \ \ \ \ +\ln\left(1+e^{- E^+_-/T}\right)+\ln\left(1+e^{
E^+_+/T}\right)\Big)\Big]\ .\nonumber
\end{eqnarray}
The gap equations (\ref{mf19}) for the condensates are equivalent
to the extremum condition of the thermodynamic potential,
\begin{equation}
\label{mf27} \frac{\partial\Omega}{\partial\sigma_u}=0\ ,\ \ \
\frac{\partial\Omega}{\partial\sigma_d}=0\ ,\ \ \
\frac{\partial\Omega}{\partial\pi}=0\ .
\end{equation}
Note that when there exist multi roots of the gap equations, only
the solution which satisfies the minimum condition
\begin{equation}
\label{mf28} \frac{\partial^2\Omega}{\partial\sigma_u^2}\geq 0\ ,\
\ \ \frac{\partial^2\Omega}{\partial\sigma_d^2}\geq 0\ ,\ \ \
\frac{\partial^2\Omega}{\partial\pi^2}\geq 0
\end{equation}
is physical.

Once $\Omega$ is known, the thermodynamic functions that measure
the bulk properties of matter can be obtained. For an infinite
system, these are the pressure $p$, the entropy density $s$, the
charge number densities $n_B$ and $n_I$, the flavor number
densities $n_u$ and $n_d$, the energy density $\epsilon$, and the
specific heat $c$, that are defined as
\begin{eqnarray}
\label{mf29}
p &=& -\Omega\ ,\nonumber\\
s &=& -\frac{\partial\Omega}{\partial T}\ ,\nonumber\\
n_B &=& -\frac{\partial \Omega}{\partial\mu_B}\ ,\ \ \ \ \ n_I =
-\frac{\partial \Omega}{\partial\mu_I}\ ,\nonumber\\
n_u &=& -\frac{\partial \Omega}{\partial\mu_u}\ ,\ \ \ \ \ n_d =
-\frac{\partial \Omega}{\partial\mu_d}\ ,\nonumber\\
\epsilon &=& -p+Ts+\mu_In_I+\mu_Bn_B\ ,\nonumber\\
c &=& \frac{\partial\epsilon}{\partial T}\ .
\end{eqnarray}

The flavor number densities $n_u=\langle\bar u\gamma_0 u\rangle$
and $n_d = \langle\bar d\gamma_0 d\rangle$ can also be obtained
directly from the matrix elements of the quark propagator,
\begin{eqnarray}
\label{mf30} n_u &=& -N_c\int{d^4 k\over (2\pi)^4} \text{Tr}_D
\left[i{\cal
S}_{uu}(k)\gamma_0\right]\nonumber\\
&=& N_c \int{d^3{\bf k}\over
(2\pi)^3}\Big[f(E^-_-)+f(-E^-_+)+f(E^+_-)\nonumber\\
&+&f(-E^+_+)+{E_k-\mu_I/2\over
E_k^-}\left(f(E^-_-)-f(-E^-_+)\right)\nonumber\\
&-&{E_k+\mu_I/2\over E_k^+}
\left(f(E^+_-)-f(-E^+_+)\right)-2\Big]\ ,\nonumber\\
n_d &=& -N_c\int{d^4 k\over (2\pi)^4} \text{Tr}_D \left[i{\cal
S}_{dd}(k)\gamma_0\right]\nonumber\\
&=& N_c \int{d^3{\bf k}\over
(2\pi)^3}\Big[f(E^-_-)+f(-E^-_+)+f(E^+_-)\nonumber\\
&+&f(-E^+_+)-{E_k-\mu_I/2\over
E_k^-}\left(f(E^-_-)-f(-E^-_+)\right)\nonumber\\
&+&{E_k+\mu_I/2\over E_k^+} \left(f(E^+_-)-f(-E^+_+)\right)-2\Big]\
.
\end{eqnarray}

Each pure number density is the difference between the
corresponding quark number density and antiquark number density,
\begin{eqnarray}
\label{mf31} n_u &=&
n_u^+ - n_u^-\ ,\nonumber\\
n_d &=& n_d^+ - n_d^-\ .
\end{eqnarray}
With the help of the positive and negative energy projectors
$\Lambda_+$ and $\Lambda_-$,  we separate $n^\pm_{u,d}$ from
$n_{u,d}$,
\begin{eqnarray}
\label{mf32} n_u ^+ &=& N_c \int{d^3{\bf k}\over
(2\pi)^3}\Big[f(E^-_-)+f(-E^-_+)\nonumber\\
&+&{E_k-\mu_I/2\over
E_k^-}\left(f(E^-_-)-f(-E^-_+)\right)\Big]\ ,\nonumber\\
n_u ^- &=& -N_c \int{d^3{\bf k}\over
(2\pi)^3}\Big[f(E^+_-)+f(-E^+_+)\nonumber\\
&-&{E_k+\mu_I/2\over
E_k^+}\left(f(E^+_-)-f(-E^+_+)\right)-2\Big]\ ,\nonumber\\
n_d^+ &=& N_c \int{d^3{\bf k}\over
(2\pi)^3}\Big[f(E^+_-)+f(-E^+_+)\nonumber\\
&+&{E_k+\mu_I/2\over
E_k^+}\left(f(E^+_-)-f(-E^+_+)\right)\Big]\ ,\nonumber\\
n_d^- &=& -N_c \int{d^3{\bf k}\over
(2\pi)^3}\Big[f(E^-_-)+f(-E^-_+)\nonumber\\
&-&{E_k-\mu_I/2\over E_k^-}\left(f(E^-_-)-f(-E^-_+)\right)-2\Big]\ .
\end{eqnarray}
It is easy to obtain the relations between the flavor number
densities and the charge number densities,
\begin{eqnarray}
\label{mf33}
n_I &=& \frac{1}{2}(n_u-n_d)\ ,\nonumber\\
n_B &=& \frac{1}{3}(n_u+n_d)\ .
\end{eqnarray}

\section {Pure Isospin Effect in Mean Field Approximation}
\label{s3}
To solve the gap equations and calculate the thermodynamic
functions numerically, we should first fix the model parameters.
Because of the contact interaction between quarks that is
introduced in the model, there is of course no confinement. A
further consequence of this feature is that the model is
non-renormalizable, and it is necessary to introduce a regulator
$\Lambda$ that serves as a length scale in the problem, and which
can be thought of as indicating the onset of asymptotic
freedom\cite{bern}. In the following we take a hard three-momentum
cutoff $\Lambda$, which and the other two model parameters, the
effective coupling constant $G$ and the current quark mass $m_0$,
can be fixed by fitting the pion mass $m_\pi =0.134$ GeV, the pion
decay constant $f_\pi = 0.093$ GeV, and the quark condensate
density $\sigma_u = \sigma_d = (-0.25$ GeV$)^3$ in the vacuum.
They are\cite{zhuang1} $m_0 = 0$, $\Lambda = 0.65$ GeV and
$G=5.01$ GeV$^{-2}$ in the chiral limit, and $m_0 = 0.005$ GeV,
$\Lambda = 0.653$ GeV, and $G=4.93$ GeV$^{-2}$ in the real word.

In this section we concentrate on the pure isospin effect with
$T=\mu_B=0$. In this case the difference between the $u$ quark and
$d$ quark condensates disappears and we need to calculate the
total chiral condensate $\sigma=2\sigma_u=2\sigma_d$ only.

\subsection {Pion and Chiral Condensates}
At zero temperature and baryon chemical potential the gap
equations for the chiral and pion condensates are reduced to
\begin{eqnarray}
\label{iso1} && \sigma + 2N_c M_q\int{d^3{\bf k}\over
(2\pi)^3}{1\over E_k}\left({E_k-\mu_I/2\over
E_k^-}+{E_k+\mu_I/2\over E_k^+}\right) = 0\ ,\nonumber\\
&& \pi\left[1 - 4N_c G\int{d^3{\bf k}\over (2\pi)^3}\left({1\over
E_k^-}+{1\over E_k^+}\right)\right] = 0\ .
\end{eqnarray}
It is clear that the order parameter $\pi = 0$ is always a
solution of the second gap equation, it corresponds to the phase
with isospin symmetry at low isospin chemical potential. The other
order parameter $\sigma$ describing chiral properties in this
phase is $\mu_I$ independent,
\begin{equation}
\label{iso2} \sigma +4N_c M_q \int{d^3{\bf k}\over (2\pi)^3}
{1\over E_k} = 0\ .
\end{equation}
This means that the chiral condensate keeps its vacuum value in
the isospin symmetric phase with $\pi = 0$.

At the critical isospin chemical potential $\mu_I^c$ where the
isospin symmetry starts to break spontaneously and the pion
condensate appears, the solution $\pi =0$ should satisfies the
equation
\begin{equation}
\label{iso3} 1-8N_c G\int{d^3{\bf k}\over (2\pi)^3}{1\over
E_k}{E_k^2\over E_k^2-\left(\mu_I^c\right)^2/ 4} = 0\ .
\end{equation}
From the comparison with the well-known pole
equation\cite{vogl,sandy,volkov,hatsuda,hufner,zhuang1}
determining the pion mass in the vacuum,
\begin{equation}
\label{iso4} 1 - 8N_c G\int{d^3{\bf k}\over (2\pi)^3}{1\over
E_k}\frac{E_k^2}{E_k^2-m_\pi^2/4}= 0\ ,
\end{equation}
where the quark mass $M_q$ hidden in the quark energy $E_k$ is the
same as that in (\ref{iso3}) and controlled by the gap equation
(\ref{iso2}), we find explicitly that the critical isospin
chemical potential $\mu_I^c$ for the pion condensation phase
transition at $T=\mu_B=0$ is exactly equal to the pion mass in the
vacuum,
\begin{equation}
\label{iso5} \mu_I^c = m_\pi\ .
\end{equation}

For $\mu_I > \mu_I^c$ the isospin dependence of the two order
parameters is governed by the first equation of (\ref{iso1}) and
\begin{equation}
\label{iso6} 1 - 4N_c G\int{d^3{\bf k}\over (2\pi)^3}\left({1\over
E_k^-}+{1\over E_k^+}\right) = 0\ .
\end{equation}

\begin{figure}
\centering \includegraphics[width=2.5in]{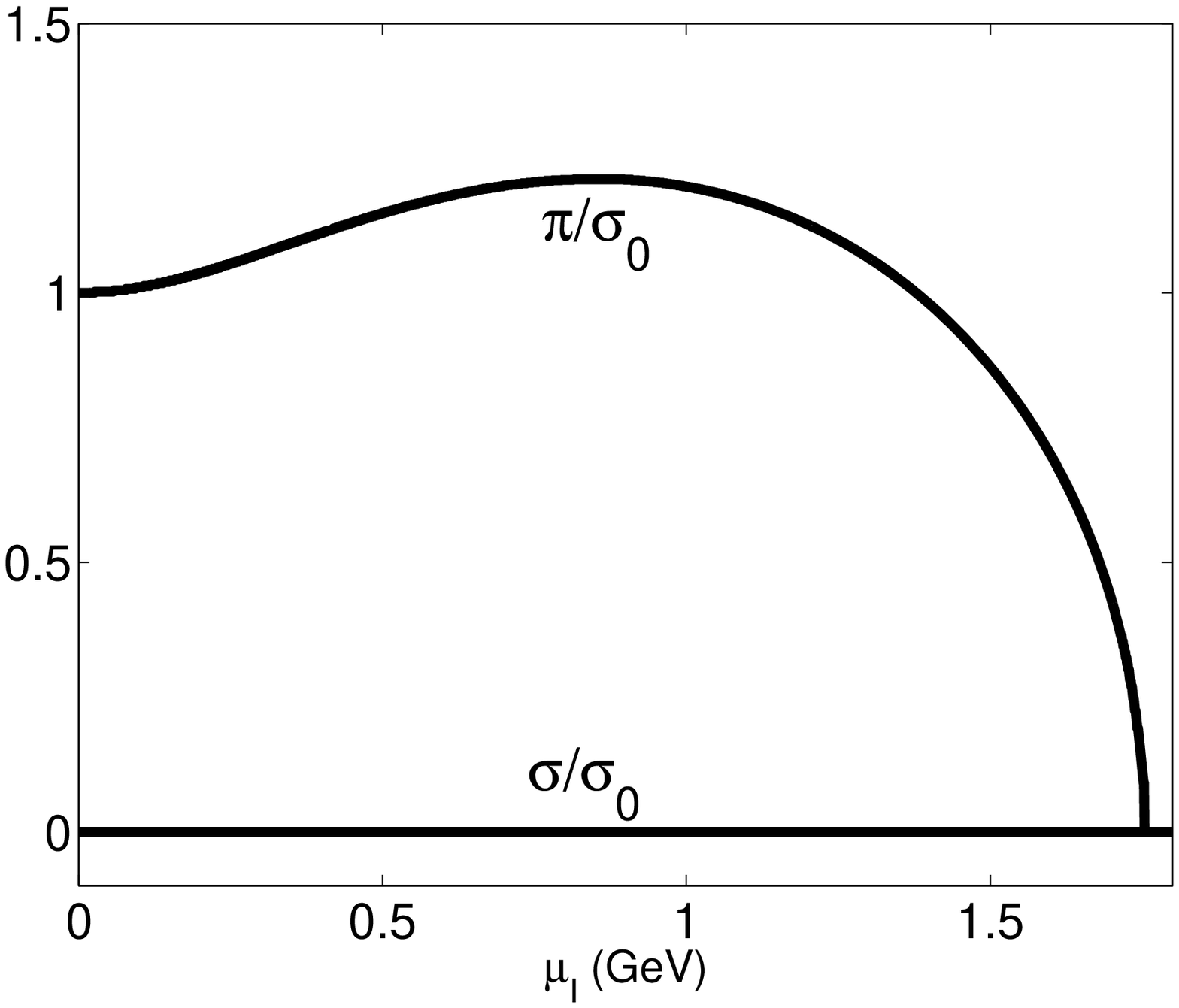}%
\hspace{0.5in}%
\includegraphics[width=2.5in]{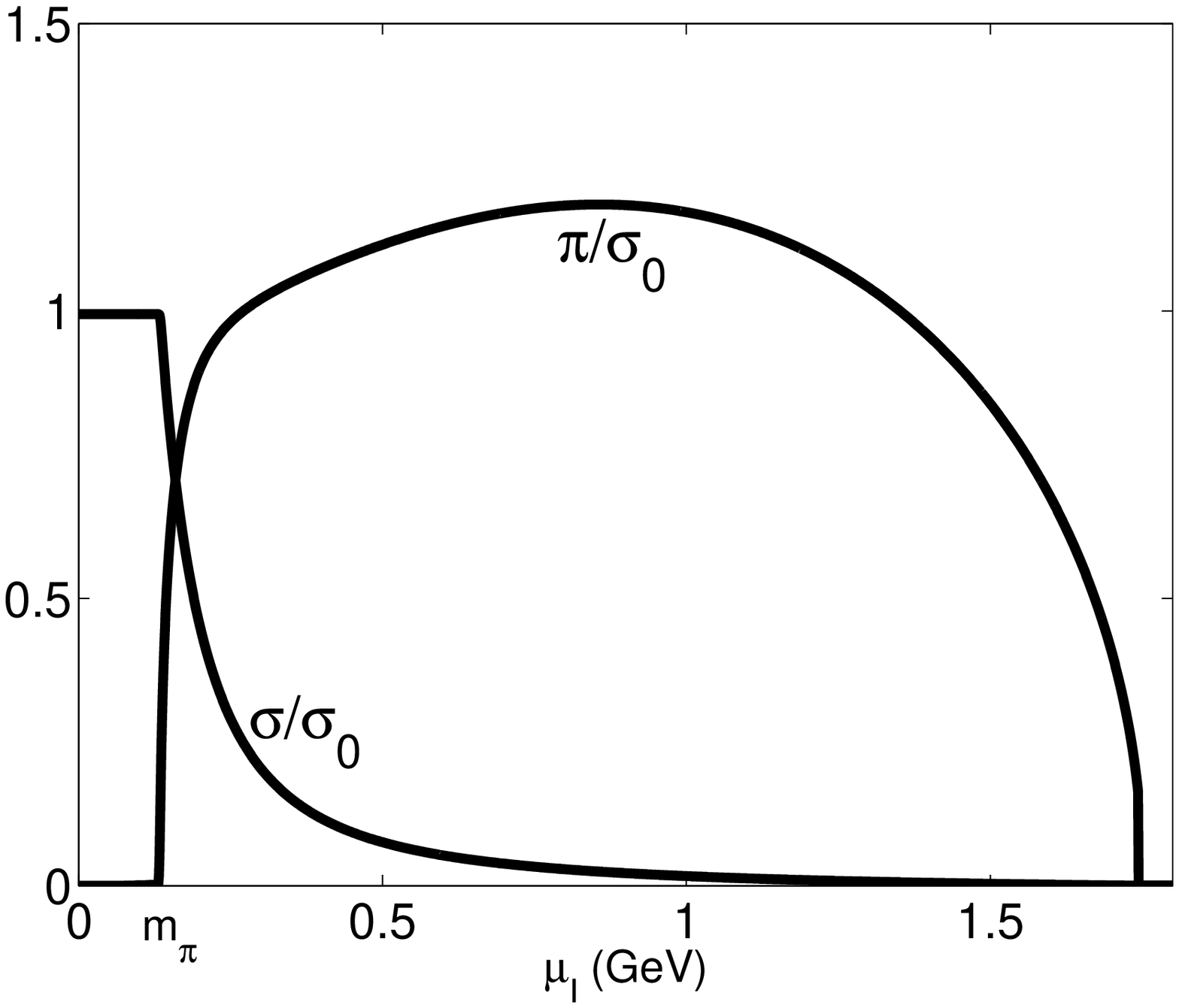} \caption{The
chiral and pion condensates $\sigma$ and $\pi$, scaled by the
chiral condensate $\sigma_0$, as function of isospin chemical
potential $\mu_I$ in the chiral limit (upper panel) and real world
(lower panel) at $T=\mu_B=0$. } \label{fig1}
\end{figure}

The condensates $\sigma$ and $\pi$ scaled by the chiral condensate
$\sigma_0$ in the vacuum are shown in Fig.(\ref{fig1}) as
functions of $\mu_I$. In chiral limit, pions are Goldstone
particles of chiral symmetry spontaneous breaking, $m_\pi =0$.
Therefore, at $\mu_I =0$ the pion condensate jumps up from the
vacuum value $\pi =0$ to a finite value, and the chiral condensate
jumps down from the vacuum value $\sigma_0 = 2(-0.25$ GeV$)^3$ to
zero and then keeps zero in the region $\mu_I
>0$ due to the constraint of the gap equations
(\ref{iso1}). These sudden changes at the critical point
$\mu_I^c=0$ means that in chiral limit, no matter how small the
isospin chemical potential is, the chiral symmetry is restored and
pion Bose-Einstein condensate occurs. In the real world with
nonzero current quark mass, the critical point is at $\mu_I^c
=m_\pi=0.134$ GeV and the both chiral and pion superfluidity phase
transitions are of second order. When the isospin chemical
potential is small, the vacuum state is not disturbed. Only when
it exceeds the pion mass, the pion condensate goes up with $\mu_I$
and at the same time the chiral condensate drops down with
$\mu_I$. The two condensates coexist in a wide region. Different
from the result obtained with chiral perturbation theory\cite{son}
where the pion condensate at finite $\mu_I$ is always less than
the chiral condensate in the vacuum, $\pi/\sigma_0 < 1$, $\pi$ in
our calculation can be larger than $\sigma_0$ at some intermediate
isospin chemical potential. The pion condensate tends to zero at
about $\mu_I =1.75 $ GeV. While the value of the endpoint of pion
condensation in our calculation is closely related to the finite
momentum cutoff $\Lambda$, the disappearance of pion condensation
at sufficiently high isospin chemical potential can be understood
in the frame of asymptotic freedom of QCD.

What is the isospin effect on the chiral condensate if we do not
consider isospin spontaneous breaking? In this case, the only
condensate, the chiral condensate $\sigma$ is controlled by the
gap equation
\begin{eqnarray}
\label{iso7} \sigma + 4N_c M_q\int{d^3{\bf k}\over
(2\pi)^3}{1\over E_k}\Theta(E_k-|\mu_I|/2) = 0,
\end{eqnarray}
the isospin chemical potential dependence of $\sigma$ is similar
to its well-known baryon chemical potential dependence. In chiral
limit, the phase transition point is at
\begin{eqnarray}
\label{iso8} \mu_I=2\sqrt{\Lambda^2-\frac{\pi^2}{2GN_c}}\ .
\end{eqnarray}

\subsection {Dispersion Relations }
\begin{figure}
\centering \includegraphics[width=2.5in]{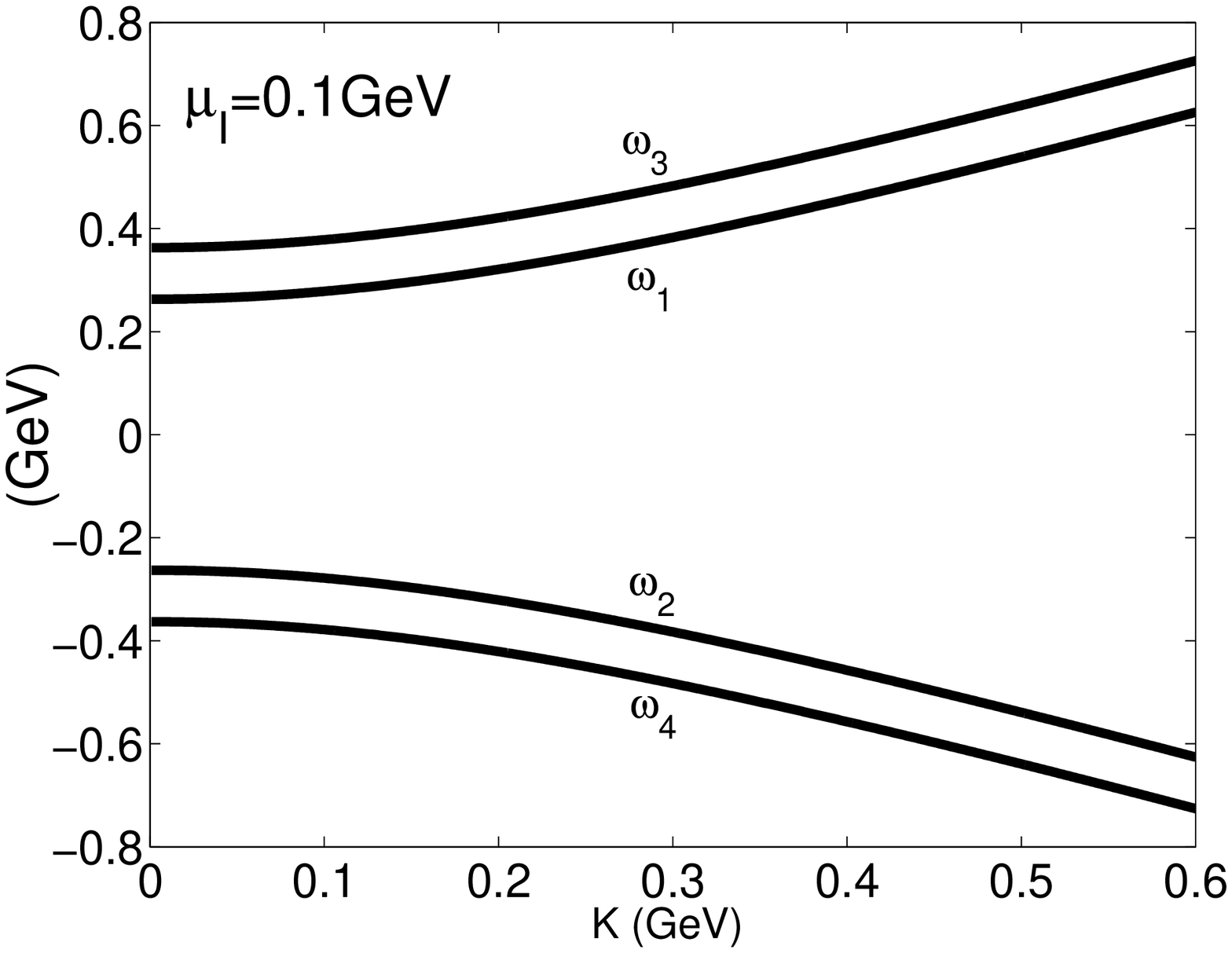}%
\hspace{0.5in}%
\includegraphics[width=2.5in]{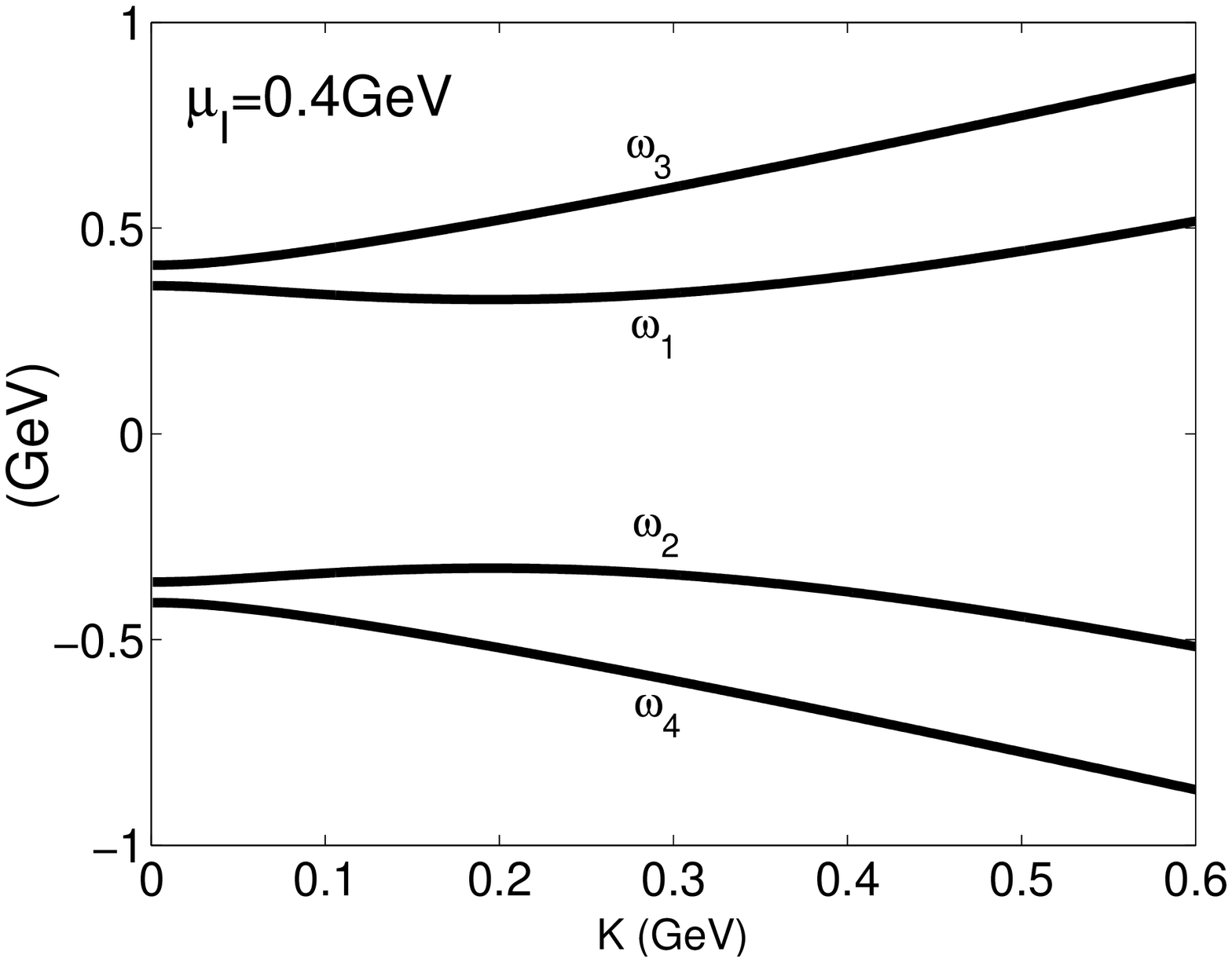}
\caption{The dispersion relations of the quasi-particles
$\omega_i, \ i=1,2,3,4$ in the normal phase with $\mu_I=0.1$ GeV
(upper panel) and pion superfluidity phase with $\mu_I = 0.4$ GeV
(lower panel) at $T=\mu_B=0$.} \label{fig2}
\end{figure}

The dispersion relations of quasi-particles are given by the poles
of the quark propagator (\ref{mf14}),
\begin{eqnarray}
\label{iso9} && \omega_1(|{\bf k}|) = E^-_-(|{\bf k}|)\ ,\ \ \
\omega_2(|{\bf k}|) = -E^-_+(|{\bf k}|)\ ,\nonumber\\
&& \omega_3(|{\bf k}|) = E^+_-(|{\bf k}|)\ ,\ \ \ \omega_4(|{\bf
k}|) = -E^+_+(|{\bf k}|)\ .
\end{eqnarray}
In the case with $\mu_B =0$, it is easy to see the symmetry of
$\omega_1 = -\omega_2$ and $\omega_3 = -\omega_4$. The dispersion
relations in this case are shown in Fig.(\ref{fig2}) at $T=0$. In
both the isospin symmetric phase with $\mu_I=0.1$ GeV $<m_\pi$ and
the symmetry breaking phase with $\mu_I = 0.4$ GeV $>m_\pi$, any
quasi-particle excitation needs energy. In the normal phase with
$\pi=0$, the chiral condensate $\sigma$ opens the gap for the
quasi-particle excitations, and in the pion superfluidity phase
with $\mu_I>m_\pi$, the gap is mainly due to the pion condensate
$\pi\ne 0$. In the normal phase, the minima of the quasi-particle
energies are located at $|{\bf k}|=0$, but in the superfluidity
phase, the location of the minima of $\omega_1$ and $\omega_2$ are
shifted to a finite momentum when $\mu_I$ is large enough.

\begin{figure}
\centering \includegraphics[width=2.2in]{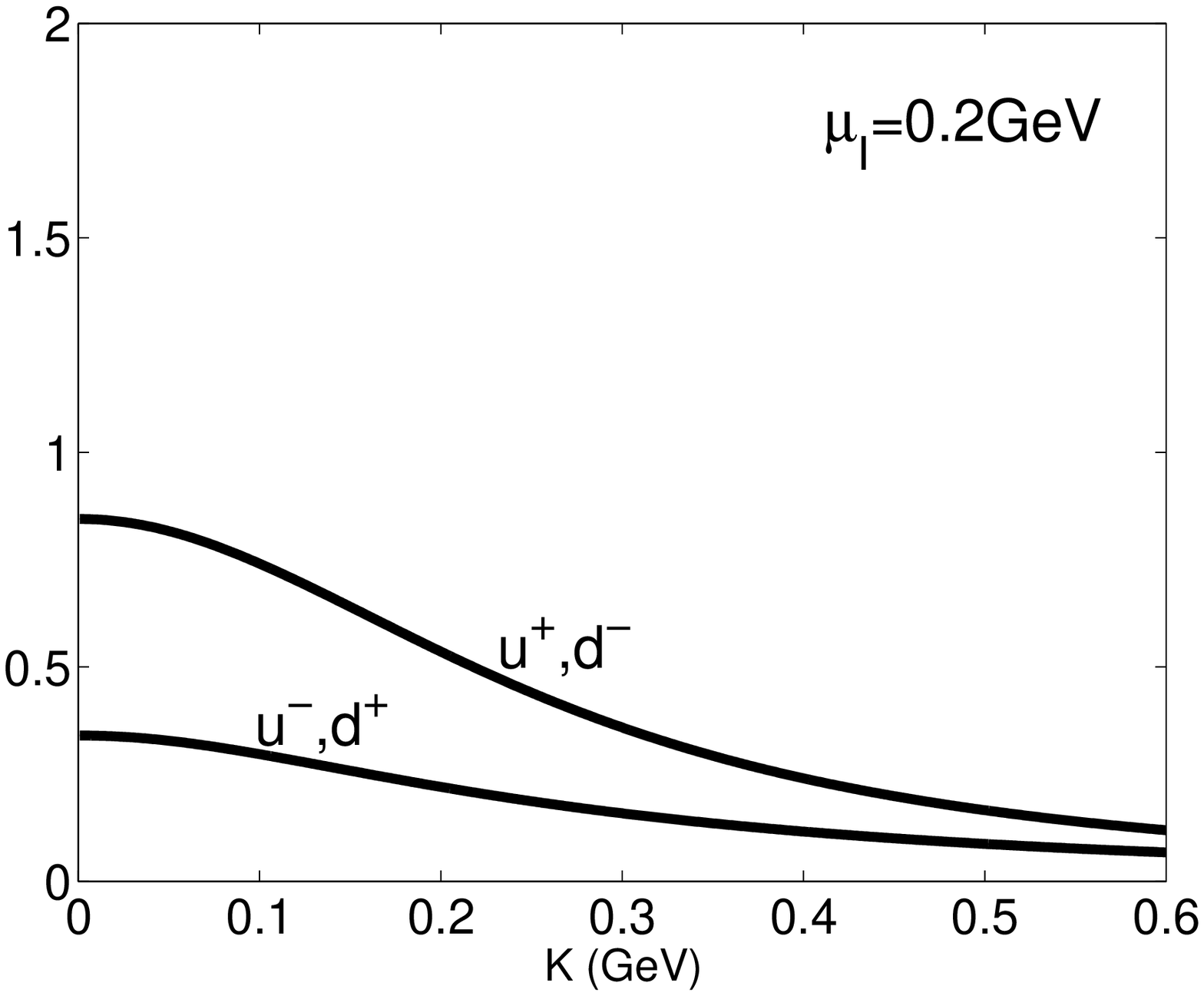}%
\hspace{0.5in}%
\includegraphics[width=2.2in]{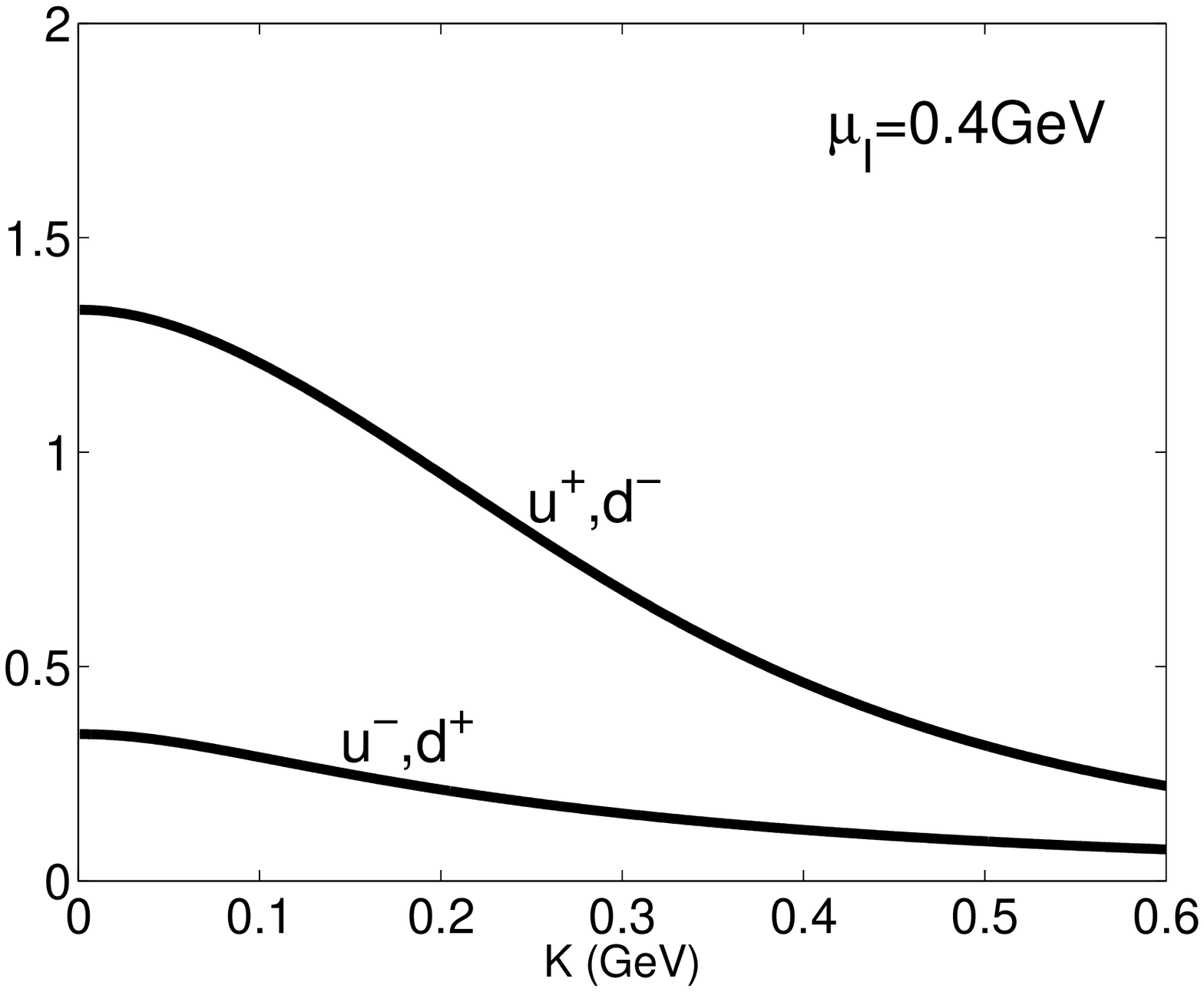}
\caption{The quark occupation numbers $n_{u,d}^\pm(|{\bf k}|)$ as
functions of momentum in pion superfluidity phase with $\mu_I=0.2$
GeV (upper panel) and $\mu_I = 0.4$ GeV (lower panel) at
$T=\mu_B=0$.} \label{fig3}
\end{figure}

The quark occupation numbers $n_u^+(|{\bf k}|)$, $n_u^-(|{\bf
k}|)$, $n_d^+(|{\bf k}|)$ and $n_d^-(|{\bf k}|)$ in momentum space
are just the integrated functions in (\ref{mf32}). At $T=\mu_B=0$
they are reduced to
\begin{eqnarray}
\label{iso10} n_u ^+ (|{\bf k}|)&=& n_d ^- (|{\bf k}|) =
1-{E_k-\mu_I/2\over
E_k^-}\ ,\nonumber\\
n_d ^+ (|{\bf k}|)&=& n_u ^- (|{\bf k}|) = 1-{E_k+\mu_I/2\over
E_k^-}\ .
\end{eqnarray}
In the normal phase we have $n_u^+(|{\bf k}|) = n_u^-(|{\bf k}|) =
n_d^+(|{\bf k}|) = n_d^-(|{\bf k}|) = 0$, since the ground state
in this case is just the vacuum state and there are no quark and
anti-quark excitations. In the superfluidity phase, there are
excited quarks and anti-quarks, the occupation numbers are shown
in Fig.(\ref{fig3}) as functions of momentum. Note that the
maximal occupation number is $2$ because of the spin degenerate.

\subsection {Bulk Properties}
\begin{figure}
\centering
\includegraphics[width=2.5in]{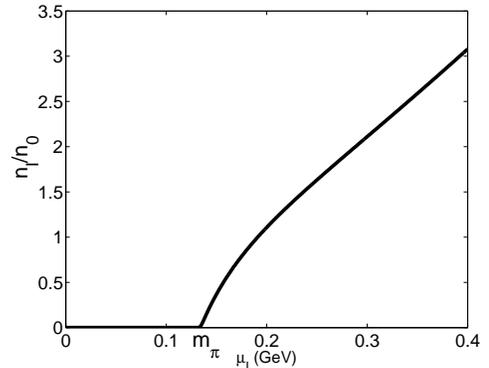}
\caption{The isospin density $n_I$, scaled by the normal nuclear
density $n_0$, as a function of $\mu_I$ at $T=\mu_B=0$.}
\label{fig4}
\end{figure}

\begin{figure}
\centering \includegraphics[width=2.5in]{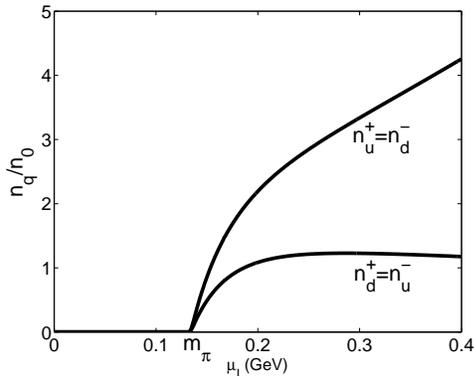} \caption{The
flavor densities $n_{u,d}^\pm$, scaled by the normal nuclear
density $n_0$, as functions of $\mu_I$ at $T=\mu_B=0$.}
\label{fig5}
\end{figure}

We first calculate the isospin density. At $T=\mu_B=0$ it is
\begin{equation}
\label{iso11} n_I = N_c \int{d^3{\bf
k}\over(2\pi)^3}\left[{E_k+\mu_I/2\over E_k^+}-{E_k-\mu_I/2\over
E_k^-}\right],
\end{equation}
the numerical result, scaled by the normal nuclear density
$n_0=0.17/fm^3$, is shown in Fig.(\ref{fig4}). Again in the normal
phase with $\mu_I <m_\pi$ the unchanged ground state leads to a
zero isospin density. Only when $\mu_I$ exceeds the vacuum pion
mass, $n_I$ increases monotonously with $\mu_I$. This nonzero net
isospin density is due to the Bose-Einstein condensation of
charged pions which results in different flavor densities. To see
this clearly, we plot the flavor densities, scaled by the normal
nuclear density $n_0$, as functions of isospin chemical potential
in Fig.(\ref{fig5}). In the superfluidity phase the relation
$n_u^+ = n_d^- > n_d^+ = n_u^-$ leads to the net isospin density,
and then the number of $\pi_+$ in the system should be larger than
the number of $\pi_-$.

Physically, the thermodynamic potential $\Omega(T,\mu_I,
\mu_B,|\sigma_u(T,\mu_I,\mu_B),\sigma_d(T,\mu_I,\mu_B),\pi(T,\mu_I,\mu_B))$
corresponds to the pressure except for a sign, and only the
pressure relative to the physical vacuum $\Omega_{vac}^{phys}$ can
be measured. The physical vacuum is defined to be
\begin{equation}
\label{iso12}
\Omega_{vac}^{phys}=\Omega(0,0,0|\sigma_u(0,0,0),\sigma_d(0,0,0),\pi(0,0,0))\
,
\end{equation}
which for the mean field approximation corresponds to
\begin{equation}
\label{iso13}
\left(\Omega_{vac}^{phys}\right)_{mf}=\frac{(M_q(0,0,0)-m_0)^2}{4G}-4N_c\int\frac{d^3{\bf
k}}{(2\pi)^3}E_k\ .
\end{equation}

We introduce the rescaled thermodynamic potential
\begin{equation}
\label{iso14}
\bar{\Omega}(T,\mu_I,\mu_B|\sigma_u,\sigma_d,\pi)=\Omega(T,\mu_I,\mu_B|\sigma_u,\sigma_d,\pi)
-\Omega_{vac}^{phys}\ ,
\end{equation}
the measurable pressure $p$ and the energy density $\epsilon$ are
related to it. Of course, the redefinition of $\Omega$ will not
change the number densities and the entropy density since they are
the derivatives of $\Omega$ with respect to $\mu$ and $T$.

\begin{figure}
\centering \includegraphics[width=2.5in]{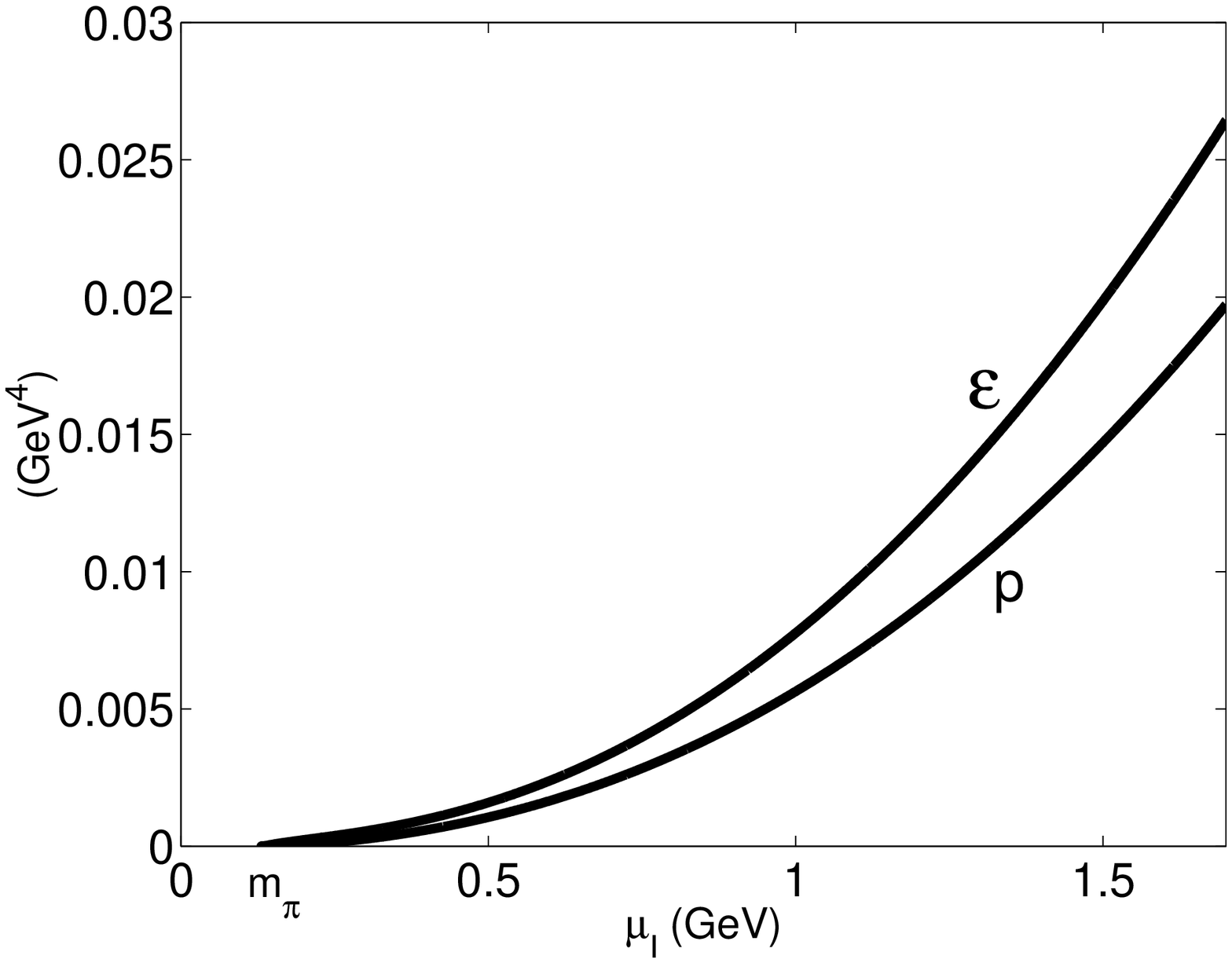}%
\hspace{0.5in}%
\includegraphics[width=2.5in]{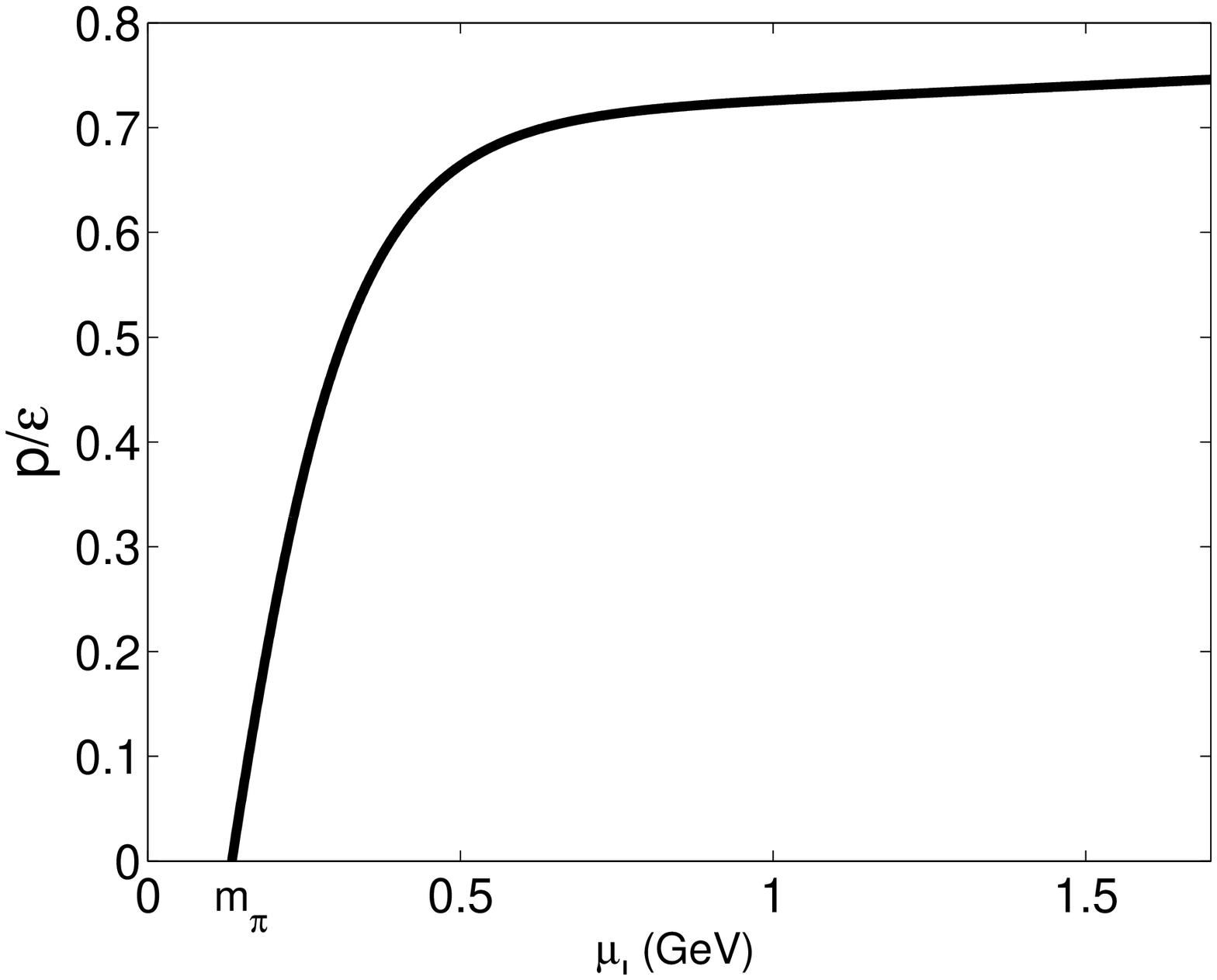}
\caption{The pressure $p$ and energy density $\epsilon$ (upper
panel) and the ratio of $p$ to $\epsilon$ (lower panel) as
functions of $\mu_I$ at $T=\mu_B=0$.} \label{fig6}
\end{figure}

The pressure and energy density are shown in Fig.(\ref{fig6}) at
$T=\mu_B=0$. From the redefinition (\ref{iso14}) they are zero in
the vacuum and keep zero in the normal phase, since the isospin
effect is not strong enough to disturb the vacuum state. In the
pion superfluidity phase, the ratio of $p$ to $\epsilon$ which
describes the equation of state of the system goes up with
increasing $\mu_I$ first and then gets saturated, $p/\epsilon \sim
0.7$, at $\mu_I\sim 0.5$ GeV. This behavior indicates that the
system is far from an ideal gas which satisfies the equation of
state $p={1\over 3}\epsilon$.

\begin{figure}
\centering \includegraphics[width=2.5in]{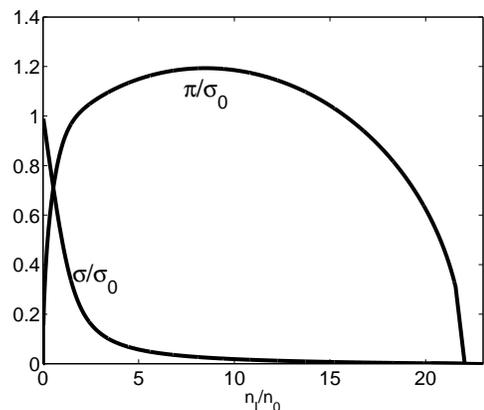} \caption{The
chiral and pion condensates $\sigma$ and $\pi$, scaled by the
chiral condensate $\sigma_0$ in the vacuum, as functions of
isospin density $n_I$ at $T=\mu_B=0$.} \label{fig7}
\end{figure}

\begin{figure}
\centering \includegraphics[width=2.5in]{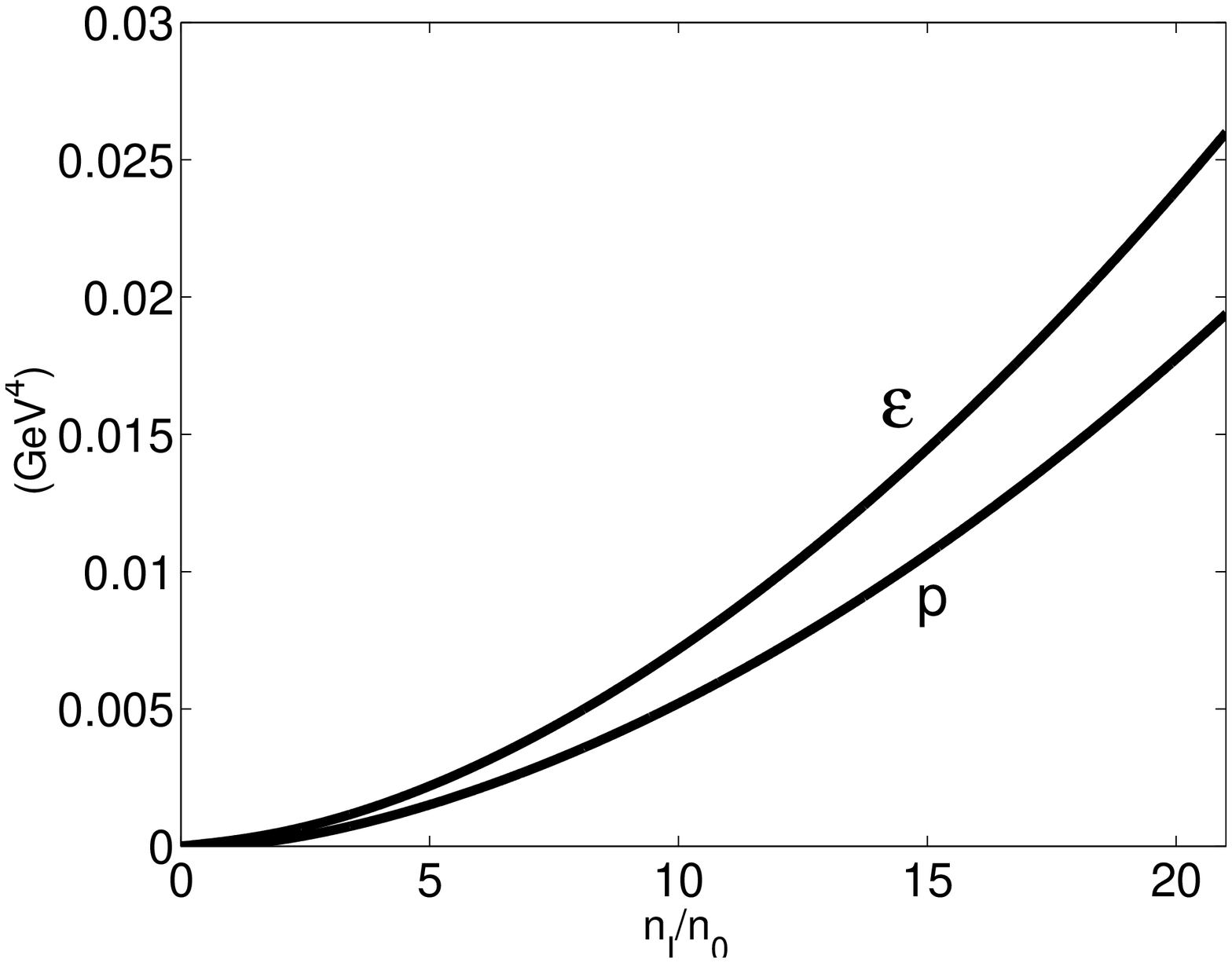}%
\hspace{0.5in}%
\includegraphics[width=2.5in]{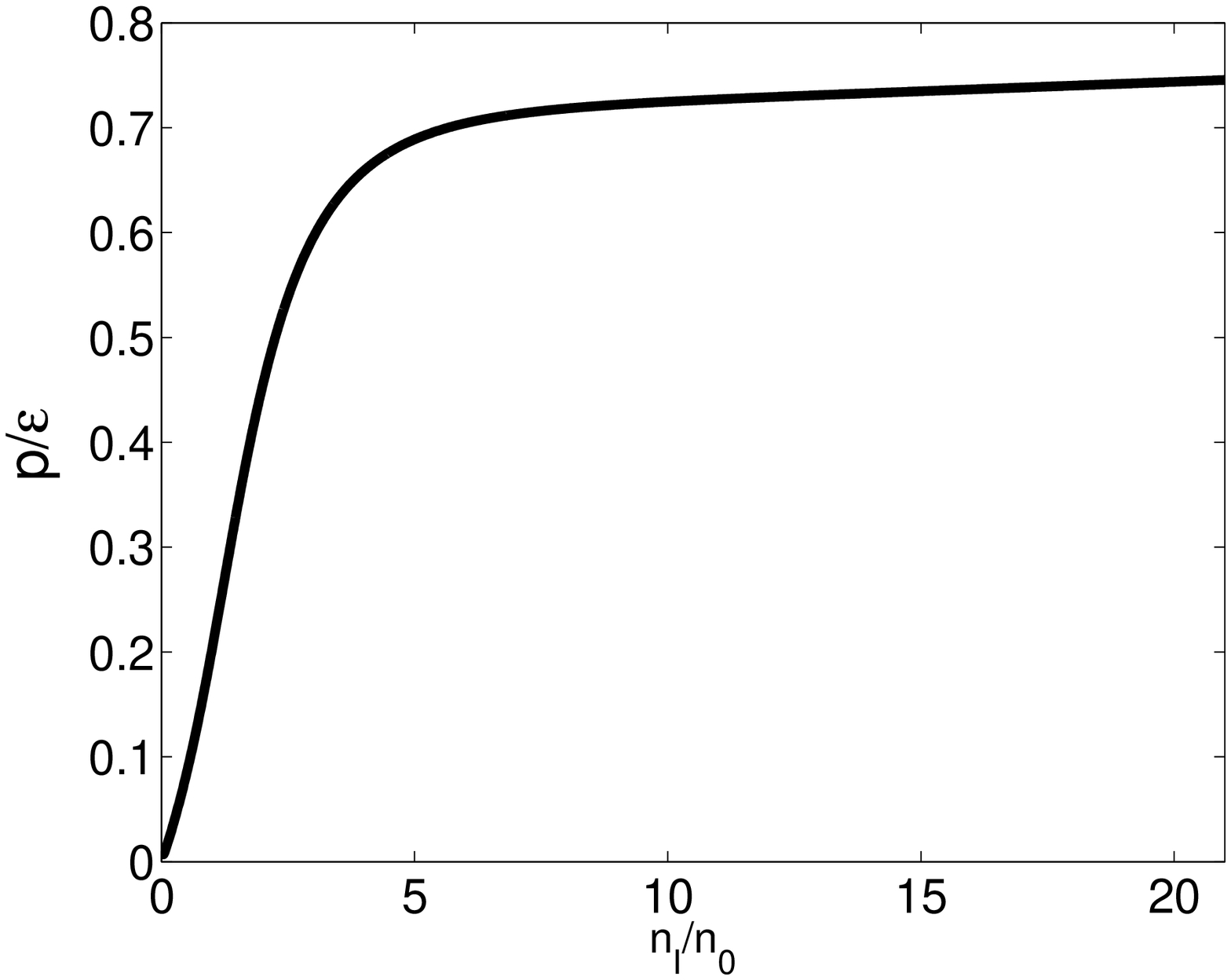}
\caption{The pressure $p$ and energy density $\epsilon$ (upper
panel)and the ratio of $p$ to $\epsilon$ (lower panel) as
functions of isospin density $n_I$ at $T=\mu_B=0$.} \label{fig8}
\end{figure}

To conclude this subsection, we remark that the isospin number
density $n_I$ is perhaps a more physical variable than the isospin
chemical potential $\mu_I$, since $n_I$ can be directly measured.
The chiral and pion condensates $\sigma$ and $\pi$, scaled by the
chiral condensate $\sigma_0$ in the vacuum, the pressure $p$ and
energy density $\epsilon$, and the ratio of $p$ to $\epsilon$ are
shown again in Figs.(\ref{fig7}) and (\ref{fig8}). Different from
their $\mu_I$-dependence, Figs.(\ref{fig1}) and (\ref{fig6}),
where the isospin symmetric and symmetry breaking phases are shown
simultaneously and are separated from each other at $\mu_I=m_\pi$,
the isospin symmetric phase is now contracted at a point $n_I=0$.
The isospin density corresponding to the saturated equation of
state is about $5$ times the normal nuclear density.

\section {Temperature Effect in Mean Field Approximation}
\label{s4}
We now study the temperature behavior of the chiral and pion
condensates and the thermodynamic functions, and discuss the phase
diagram in the $T-\mu_I$ plane in mean field approximation at
$\mu_B=0$.
\subsection {Chiral and Pion Condensates}
\begin{figure}
\centering \includegraphics[width=2.5in]{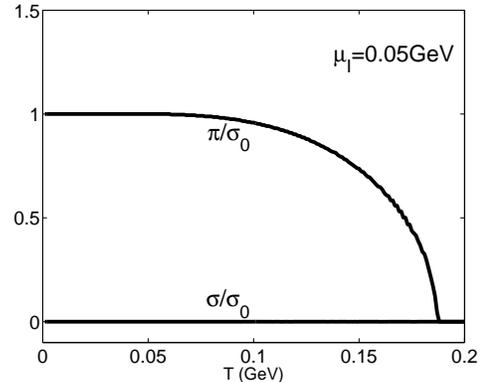} \caption{The
chiral and pion condensates $\sigma$ and $\pi$, scaled by the
chiral condensate $\sigma_0$ in the vacuum, as functions of
temperature $T$ at $\mu_B=0$ and $\mu_I=0.05$ GeV in the chiral
limit. }\label{fig9}
\end{figure}

\begin{figure}
\centering \includegraphics[width=2.5in]{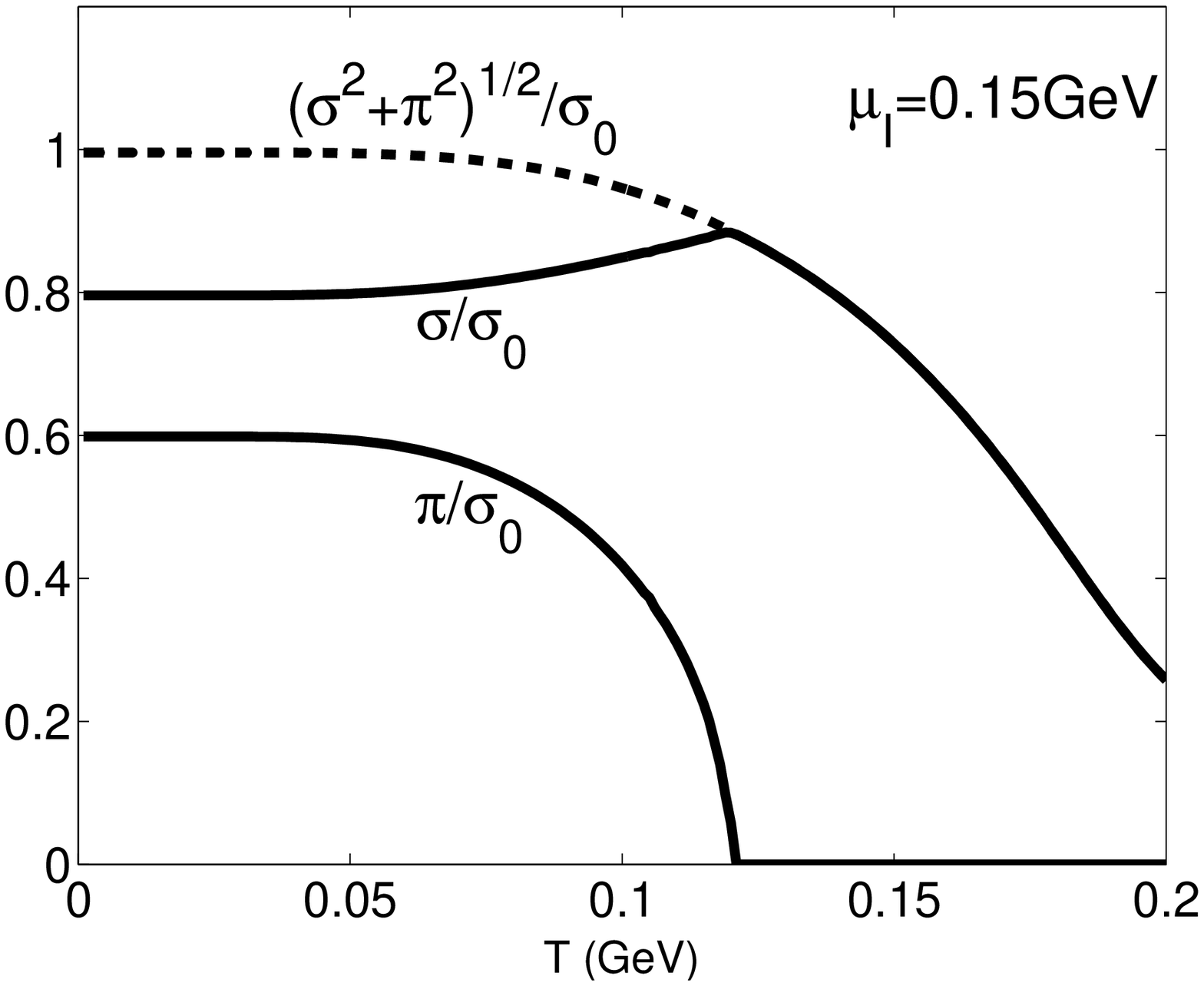}%
\hspace{0.5in}%
\includegraphics[width=2.5in]{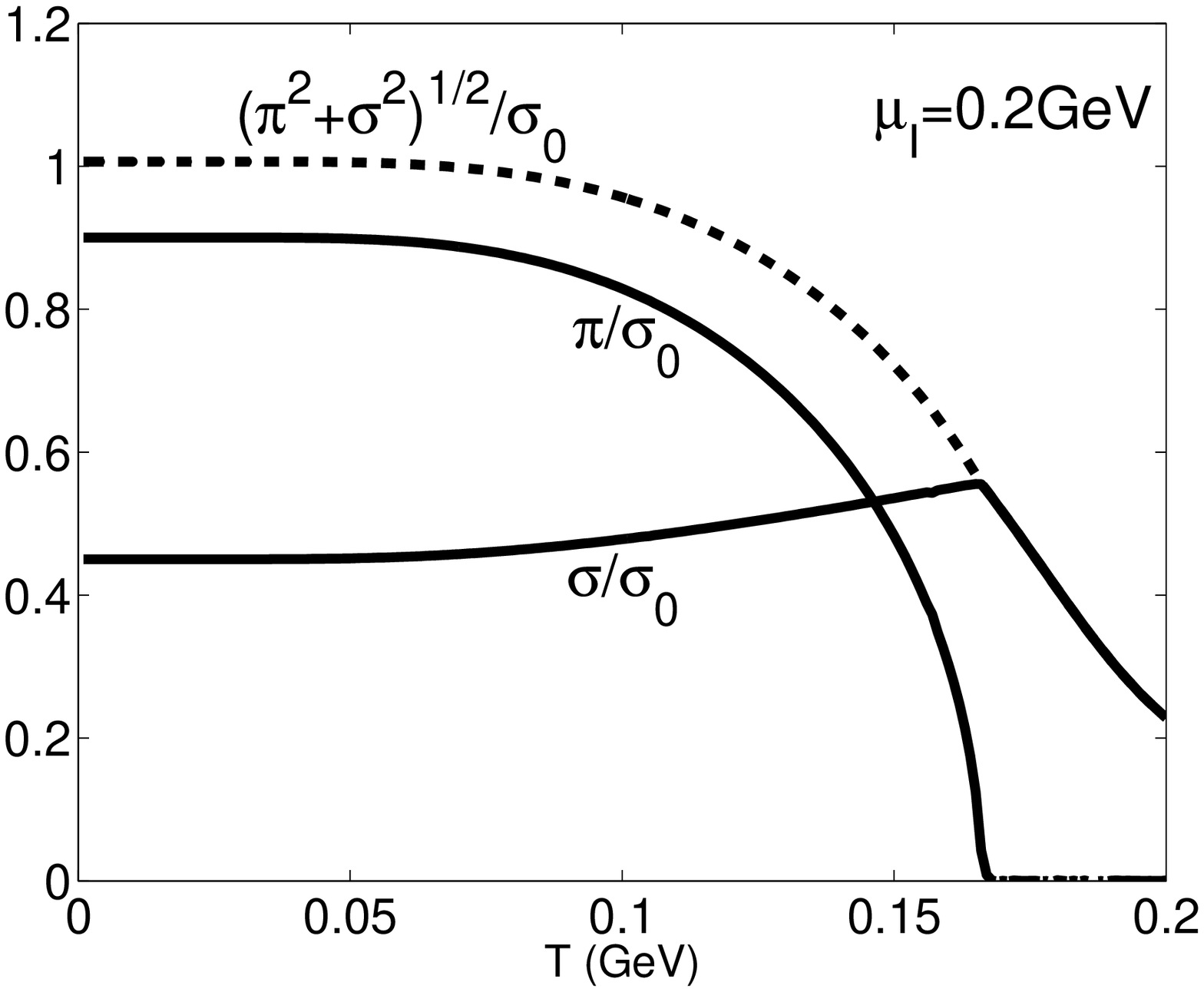}
\caption{The chiral and pion condensates $\sigma$ and $\pi$,
scaled by the chiral condensate $\sigma_0$ in the vacuum, as
functions of temperature $T$ at $\mu_B=0$ and $\mu_I=0.15$ GeV
(upper panel) and $\mu_I=0.2$ GeV (lower panel) in the real world.
The dashed lines are for the total condensate
$\left(\sigma^2+\pi^2\right)^{1/2}/\sigma_0$.}\label{fig10}
\end{figure}

The temperature dependence of the two condensates $\sigma$ and
$\pi$, again scaled by the chiral condensate $\sigma_0$ in the
vacuum, at a fixed isospin chemical potential $\mu_I=0.05$ GeV is
shown in Fig.(\ref{fig9}) in the chiral limit. In the chiral limit
any small isospin chemical potential will force the broken chiral
symmetry to be restored and the isospin symmetry to be broken at
$T=0$. Therefore, the chiral condensate keeps zero, and the pion
condensate is continuously melted in the hot medium. Finally at a
critical temperature $T_c\sim 0.18$ GeV the pion condensate
vanishes and the isospin symmetry is restored. At small isospin
chemical potential the value of the critical temperature for
isospin restoration is almost the same for chiral restoration in
the study without considering pion
condensation\cite{sandy,zhuang1}. In Fig.(\ref{fig10}) we show the
pion and chiral condensates as functions of $T$ at $\mu_I=0.15$
GeV and $0.2$ GeV. For $\mu_I> m_\pi$, the pion condensate is
already nonzero at the beginning, then drops down due to the
temperature effect, and finally disappears at a critical
temperature $T_c$. Like the standard BCS theory, the critical
temperature and the pion condensate at $T=0$ satisfies the linear
relation,
\begin{equation}
\label{t1} T_C(\mu_I)\sim 0.57\pi(T=0,\mu_I)\ .
\end{equation}

The relative strength of the two condensates at the beginning
depends on the isospin chemical potential, $\sigma(T=0) >
\pi(T=0)$ at small $\mu_I$ and $\pi(T=0) > \sigma(T=0)$ at large
$\mu_I$. Qualitatively different from the study on chiral symmetry
restoration without considering pion condensation where the chiral
condensate decreases with increasing temperature monotonously,
here $\sigma$ goes up continuously in the coexistence region of
the two condensates, and drops down only in the region where the
isospin symmetry is restored. However, we found that when $\mu_I$
is close to the critical value $\mu_I^c =m_\pi$, say $m_\pi <
\mu_I < 0.2$ GeV, the total condensate $\sqrt{\sigma^2+\pi^2}$
(the dashed line in Fig.(\ref{fig10}), behaviors in a similar way
like the $\sigma$ in the normal study without considering pion
condensation.

\subsection {Number Densities}
\begin{figure}
\centering \includegraphics[width=2.5in]{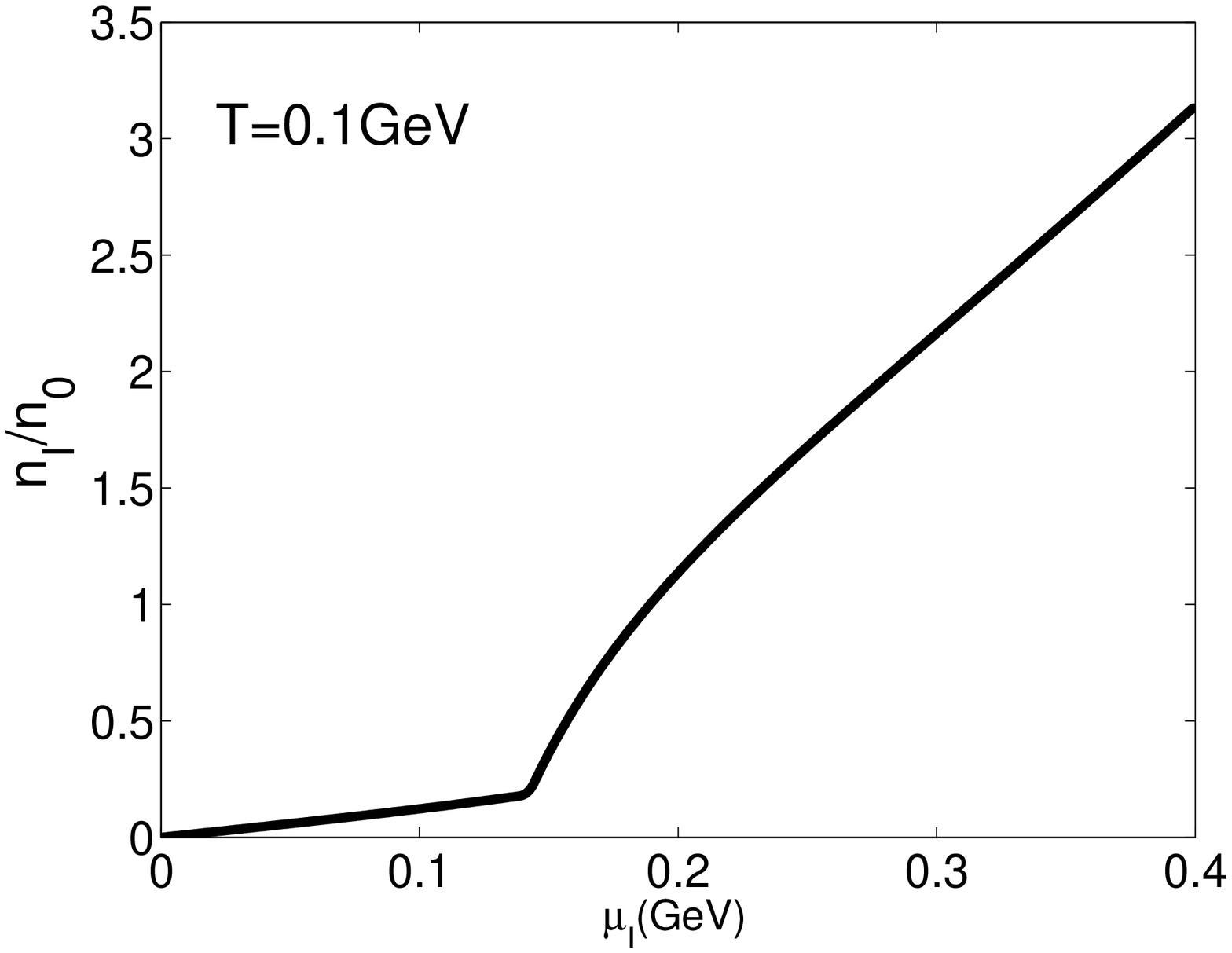}%
\hspace{0.5in}%
\includegraphics[width=2.5in]{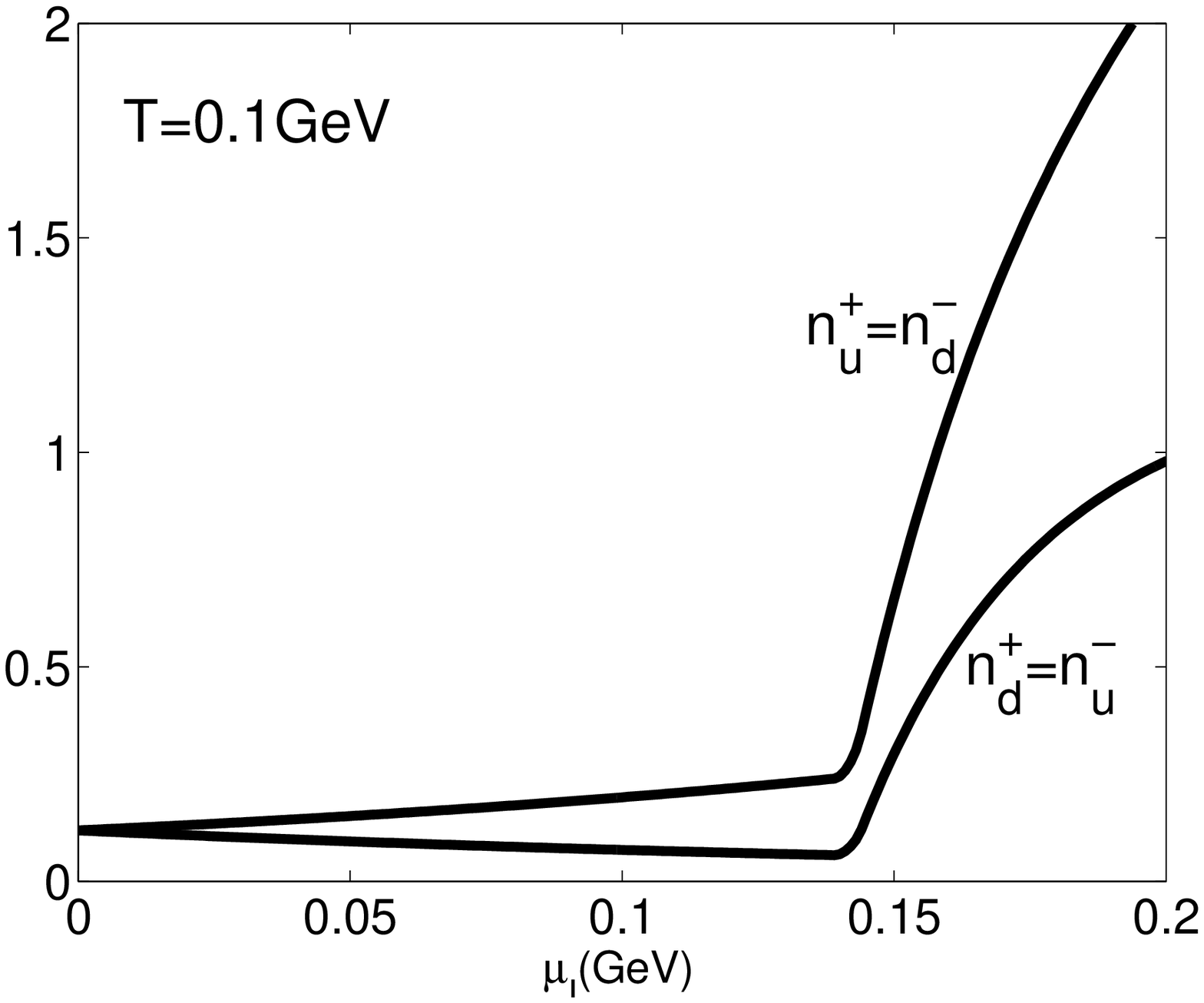}
\caption{The isospin density $n_I$ (upper panel) and flavor
densities $n_{u,d}^\pm$ (lower panel), scaled by the normal
nuclear density $n_0$, as functions of isospin chemical potential
$\mu_I$ at $\mu_B =0$ and $T=0.1$ GeV.} \label{fig11}
\end{figure}

The isospin density $n_I$ and flavor densities $n_{u,d}^\pm$,
scaled by the normal nuclear density $n_0$, are shown in
Fig.(\ref{fig11}) as functions of isospin chemical potential
$\mu_I$ at $\mu_B=0$ but $T=0.1$ GeV. At finite temperature, there
is thermal excitation of particles. Since $T=0.1$ GeV is not very
high, the critical isospin chemical potential $\mu_I^c$ for pion
condensation is still approximately the vacuum pion mass $m_\pi$.
In the region of $\mu_I<\mu_I^c$, the small but finite isospin
density is purely due to the thermal excitation, while in the
region of $\mu_I>\mu_I^c$ the rapidly increasing density is
dominated by the Bose-Einstein condensation of charged pions. The
thermal excitation depends strongly on the isospin chemical
potential carried by the particle. At $\mu_I<\mu_I^c$, the
densities of quarks with positive isospin chemical potential
increase with $\mu_I$, while the densities of quarks with negative
isospin chemical potential decrease with $\mu_I$.

\subsection { Phase Diagram in $T-\mu_I$ Plane}
\begin{figure}
\centering \includegraphics[width=2.5in]{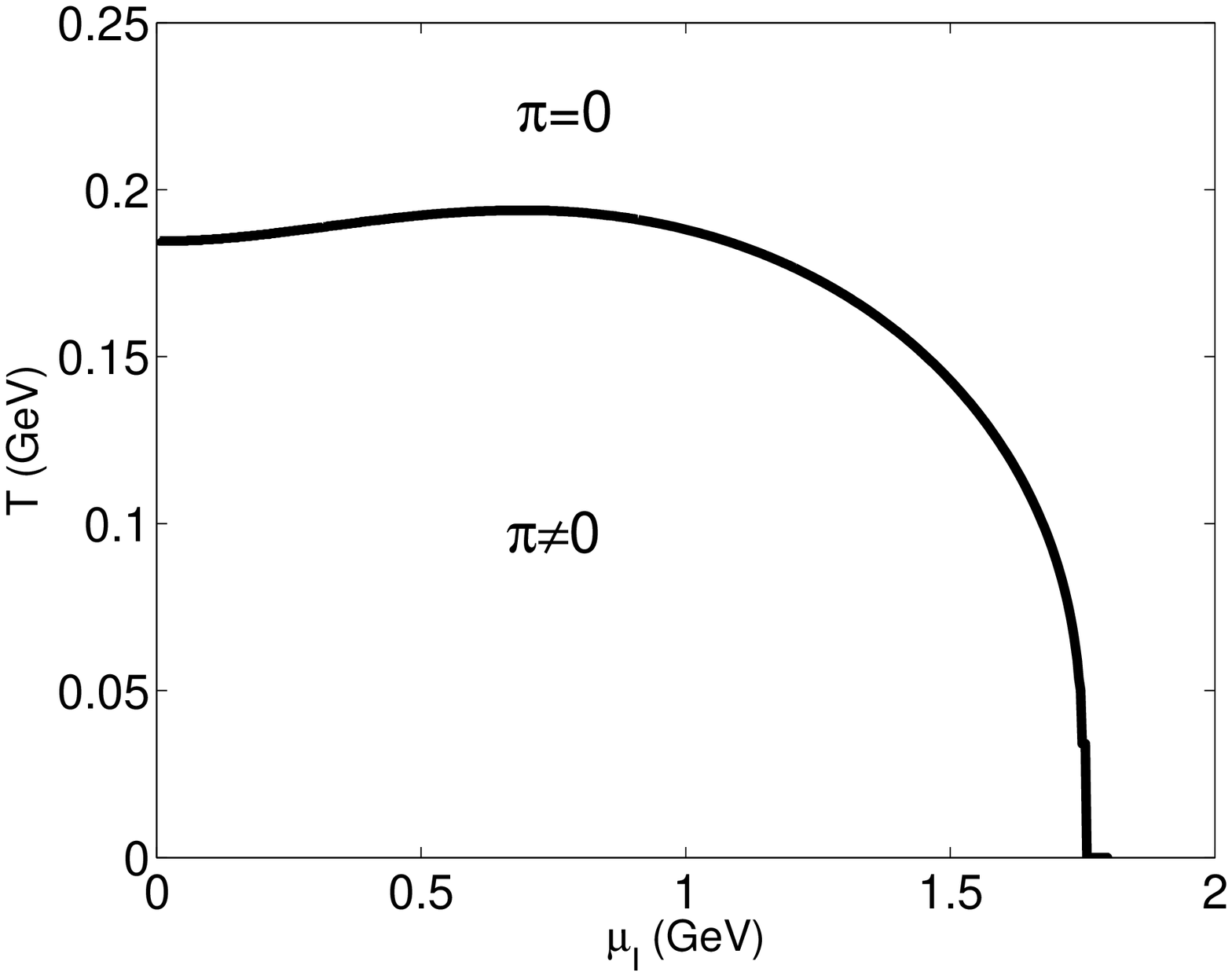}%
\hspace{0.5in}%
\includegraphics[width=2.5in]{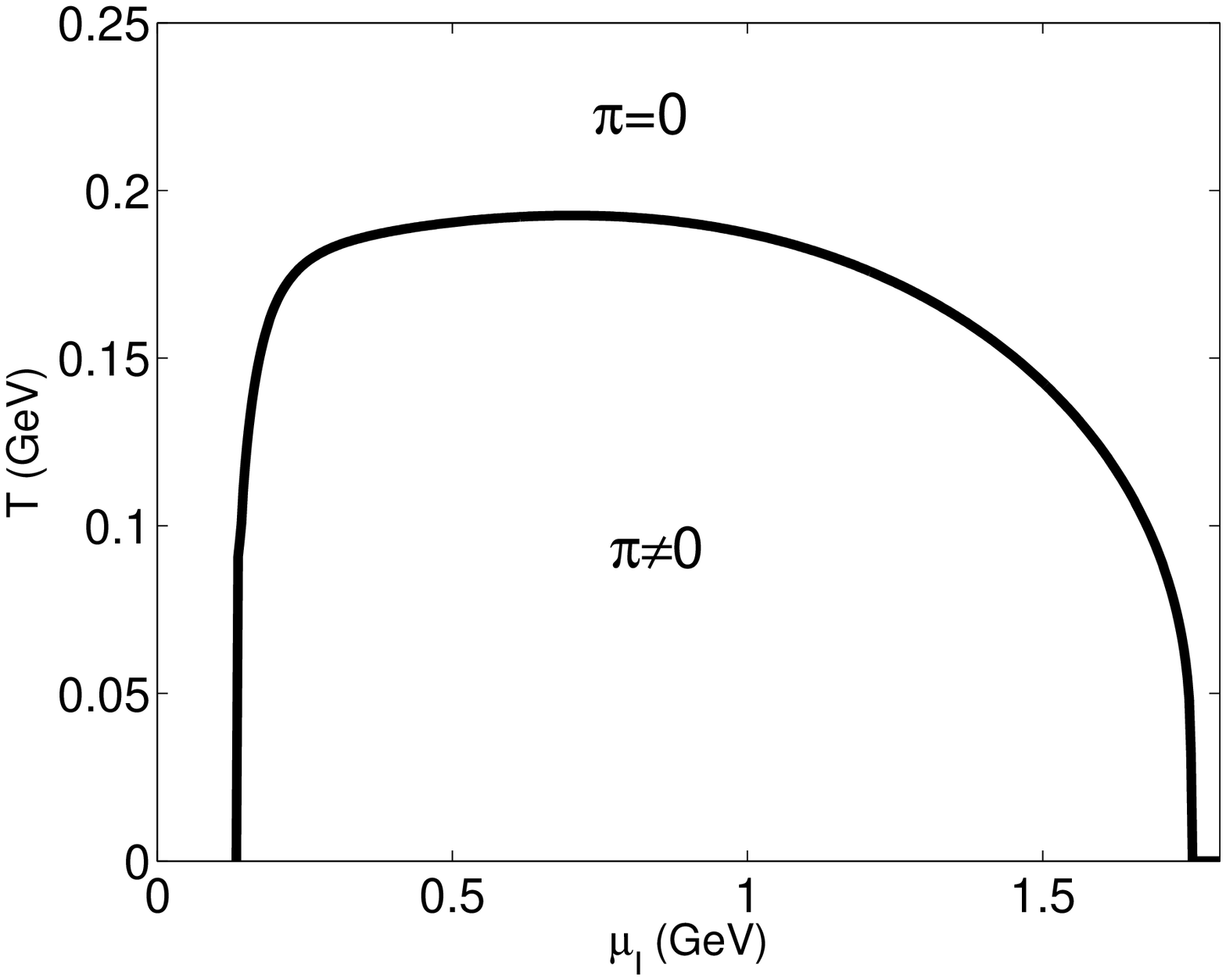}
\caption{The phase diagram in the $T-\mu_I$ plane in the chiral
limit (upper panel) and real world (lower panel) at $\mu_B=0$.}
\label{fig12}\end{figure}

In the chiral limit when $m_0=0$, the solution of the last gap
equation of (\ref{mf19}) for $\pi$ with $\sigma_u =\sigma_d =0$
separates the region of isospin symmetry breaking with $\pi\ne 0$
from the region of symmetry restoration with $\pi=0$. The phase
transition line demarcating these two regions is given in the
upper panel of Fig.(\ref{fig12}) in the $T-\mu_I$ plane for
$\mu_B=0$. Since an infinite small isospin effect can restore the
chiral symmetry, there is $\sigma =0$ in the whole plane in the
chiral limit. In the real world with nonzero current quark mass,
the system is in isospin symmetric phase not only at high
temperature and/or high isospin chemical potential but also at low
isospin chemical potential with $\mu_I \le m_\pi$. Since the
chiral symmetry phase transition is not well defined in the case
with nonzero current quark mass, we did not show the chiral phase
transition line. In principle, there is $\sigma\ne 0$ in the whole
$T-\mu_I$ plane.

\section {Mesons at Finite Isospin Density}
\label{s5}
We now investigate meson properties at finite isospin chemical
potential. In the NJL model, the meson modes are regarded as
quantum fluctuations above the mean field. The meson modes can be
calculated in the frame of RPA\cite{sandy}. When the mean field
quark propagator is diagonal, for instance the case with only
chiral condensation, the summation of bubbles in RPA selects its
specific channel by choosing at each stage the same proper
polarization function, a meson mode which is determined by the
pole of the corresponding meson propagator is related to its own
polarization function $\Pi_{MM}(k)$\cite{hufner,zhuang1} only,
\begin{equation}
\label{meson1}
1-2G\Pi_{MM}(k)=0\ .
\end{equation}
However, for the quark propagator with off-diagonal elements, like
the cases of $\eta$ and $\eta'$ meson spectrum\cite{sandy}, pion
superfluidity considered here, and color superconductivity in
Section \ref{s7}, we must consider all possible channels in the
bubble summation in RPA. Using matrix notation, the effective
quark propagator (\ref{mf14}) leads to the meson modes determined
by
\begin{equation}
\label{meson2}
\det\left(1-2G\Pi(k)\right)=0\ ,
\end{equation}
with the polarization function matrix
\begin{eqnarray}
\label{meson3} &&1-2G\Pi= \\
&&\left(\begin{array}{cccc}
1-2G\Pi_{\sigma\sigma}&-2G\Pi_{\sigma\pi_+}
&-2G\Pi_{\sigma\pi_-}&-2G\Pi_{\sigma\pi_0}\\
-2G\Pi_{\pi_+\sigma}&1-2G\Pi_{\pi_+\pi_+}&-2G\Pi_{\pi_+\pi_-}&-2G\Pi_{\pi_+\pi_0}\\
-2G\Pi_{\pi_-\sigma}&-2G\Pi_{\pi_-\pi_+}&1-2G\Pi_{\pi_-\pi_-}&-2G\Pi_{\pi_-\pi_0}\\
-2G\Pi_{\pi_0\sigma}&-2G\Pi_{\pi_0\pi_+}&-2G\Pi_{\pi_0\pi_-}&1-2G\Pi_{\pi_0\pi_0}\\
\end{array}\right).\nonumber
\end{eqnarray}
Here the polarization functions defined as
\begin{equation}
\label{meson4} \Pi_{MM^\prime}(k) = i\int{d^4p\over (2\pi)^4}
\text{Tr}\left(\Gamma_M^* {\cal S}_{mf}(p+k)\Gamma_{M^\prime} {\cal
S}_{mf}(p)\right)\ ,
\end{equation}
with the vertexes
\begin{eqnarray}
\label{meson5} \Gamma_M &=& \left\{\begin{array}{ll}
1 & M=\sigma\\
i\tau_+\gamma_5 & M=\pi_+ \\
i\tau_-\gamma_5 & M=\pi_- \\
i\tau_3\gamma_5 & M=\pi_0
\end{array}\right.\ ,\nonumber\\
\Gamma_M^* &=& \left\{\begin{array}{ll}
1 & M=\sigma\\
i\tau_-\gamma_5 & M=\pi_+ \\
i\tau_+\gamma_5 & M=\pi_- \\
i\tau_3\gamma_5 & M=\pi_0 \\
\end{array}\right.\ ,
\end{eqnarray}
are shown in Appendix \ref{a2}. Here the trace $\text{Tr}$ is taken
in color, flavor and Dirac spaces.

As will be seen in the following, in the normal phase with
vanished pion condensate, the polarization matrix is diagonal at
finite isospin chemical potential and one can then directly use
the method developed in \cite{hufner,zhuang1} to calculate the
masses of $\sigma,\pi_+,\pi_-$ and $\pi_0$. In the superfluidity
phase with nonzero $\pi$ and off diagonal matrix elements in the
mean field quark propagator, the collective excitations of the
system will in principle not be the original meson modes in the
vacuum but their mixture. Especially it is expected that there
will be a Goldstone mode in the superfluidity phase corresponding
to the spontaneously isospin symmetry breaking.

In the following we will study the meson properties in the normal
phase($\pi=0$) and superfluidity phase($\pi\neq0$) separately.

\subsection {Meson Properties in Normal Phase}
In the vacuum, due to isospin symmetry, one can regard
$\pi_1,\pi_2$ or $\pi_+,\pi_-$ or any other linear combination of
$\pi_1,\pi_2$ as eigen collective modes. In the normal phase with
vanished pion condensate, the polarization function matrix is
diagonal with respect to the basis $(\sigma, \pi_+, \pi_-, \pi_3)$
but not diagonal with respect to the basis $(\sigma, \pi_1, \pi_2,
\pi_3)$. This means that in the normal phase, the eigen collective
modes are $\sigma,\pi_+,\pi_-,\pi_0$. Thus the dispersion relation
for each mode is determined through the equation
\begin{equation}
\label{meson6}
1-2G\Pi_{MM}(k) = 0\ .
\end{equation}
The meson mass is defined as the root $k_0$ at ${\bf k=0}$.

Doing the trace in flavor and color spaces, we obtain the
polarization functions in terms of the matrix elements of the
quark propagator,
\begin{eqnarray}
\label{meson7} \Pi_{\sigma\sigma} (k) &=& i N_c\int {d^4p\over
(2\pi)^4} \text{Tr}_D \Big({\cal S}_{uu}(p+k){\cal
S}_{uu}(p)\\
&+&{\cal
S}_{dd}(p+k){\cal S}_{dd}(p)\Big)\ ,\nonumber\\
\Pi_{\pi_0\pi_0}(k) &=& -iN_c\int {d^4p\over (2\pi)^4} \text{Tr}_D
\Big(\gamma_5{\cal S}_{uu}(p+k)\gamma_5{\cal
S}_{uu}(p)\nonumber\\
&+&\gamma_5{\cal
S}_{dd}(p+k)\gamma_5{\cal S}_{dd}(p)\Big)\ ,\nonumber\\
\Pi_{\pi_+\pi_+}(k) &=& -2iN_c\int {d^4p\over (2\pi)^4} \text{Tr}_D
\Big(\gamma_5{\cal S}_{uu}(p+k)\gamma_5{\cal S}_{dd}(p)\Big)\ ,\nonumber\\
\Pi_{\pi_-\pi_-}(k) &=& -2iN_c\int {d^4p\over (2\pi)^4} \text{Tr}_D
\Big(\gamma_5{\cal S}_{dd}(p+k)\gamma_5{\cal S}_{uu}(p)\Big)\
,\nonumber
\end{eqnarray}
now the trace $\text{Tr}_D$ is taken only in Dirac space. At zero
temperature and zero baryon density, the polarization functions at
$\bf k=0$ are simplified as
\begin{eqnarray}
\label{meson8}
&&\Pi_{\sigma\sigma}(k_0)=4N_c\int\frac{d^3{\bf
p}}{(2\pi^3)}\frac{1}{E_p}\frac{E_p^2-M_q^2}{E_p^2-k_0^2/4}\ ,\nonumber\\
&&\Pi_{\pi_0\pi_0}(k_0)=4N_c\int\frac{d^3{\bf
p}}{(2\pi^3)}\frac{1}{E_p}\frac{E_p^2}{E_p^2-k_0^2/4}\ ,\nonumber\\
&&\Pi_{\pi_+\pi_+}(k_0)=4N_c\int\frac{d^3{\bf
p}}{(2\pi^3)}\frac{1}{E_p}\frac{E_p^2}{E_p^2-(k_0+\mu_I)^2/4}\ ,\nonumber\\
&&\Pi_{\pi_-\pi_-}(k_0)=4N_c\int\frac{d^3{\bf
p}}{(2\pi^3)}\frac{1}{E_p}\frac{E_p^2}{E_p^2-(k_0-\mu_I)^2/4}\ .
\end{eqnarray}
From the comparison with their expressions in the vacuum with
$\mu_I=0$, we obtain the $\mu_I$ dependence of the meson masses,
\begin{eqnarray}
\label{meson9}
M_\sigma(\mu_I) &=& m_\sigma\ ,\nonumber\\
M_{\pi_0}(\mu_I)&=& m_\pi\ ,\nonumber\\
M_{\pi_+}(\mu_I)&=& m_\pi-\mu_I\ ,\nonumber\\
M_{\pi_-}(\mu_I)&=& m_\pi+\mu_I\ .
\end{eqnarray}
The mesons which carry zero isospin charge keep their vacuum
masses, while the mass of the meson which carries positive
(negative) isospin charge decreases (increases) with $\mu_I$
linearly. These relations hold before $\mu_I=m_\pi$ where the pion
condensation starts.

After performing Matsubara frequency summation the meson
polarization functions at ${\bf k}=0$ are explicitly expressed as
\begin{eqnarray}
\label{meson10}
&&\Pi_{\sigma\sigma}(k_0)=-2N_c\int\frac{d^3{\bf
p}}{(2\pi^3)}\frac{1}{E_p}\frac{E_p^2-M_q^2}{E_p^2-k_0^2/4}\nonumber\\
&&\ \ \ \times\Big(f(E^-_-)+f(E^+_-)-f(-E^-_+)
-f(-E^+_+)\Big)\ ,\nonumber\\
&&\Pi_{\pi_0\pi_0}(k_0)=-2N_c\int\frac{d^3{\bf
p}}{(2\pi^3)}\frac{1}{E_p}\frac{E_p^2}{E_p^2-k_0^2/4}\nonumber\\
&&\ \ \ \times\Big(f(E^-_-)+f(E^+_-)-f(-E^-_+) -f(-E^+_+)
\Big)\ ,\nonumber\\
&&\Pi_{\pi_+\pi_+}(k_0)=4N_c\int\frac{d^3{\bf
p}}{(2\pi^3)}\nonumber\\
&&\ \ \ \times\Bigg[\frac{2E_p-\mu_I+k_0}{k_0^2-4(E_p-\mu_I/2)^2}(f(E^-_-)-f(-E^-_+))\nonumber\\
&&\ \ \ +\frac{2E_p+\mu_I-k_0}{k_0^2-4(E_p+\mu_I/2)^2}(f(E^+_-)-f(-E^+_+))\Bigg]\ ,\nonumber\\
&&\Pi_{\pi_-\pi_-}(k_0)=4N_c\int\frac{d^3{\bf
p}}{(2\pi^3)}\nonumber\\
&&\ \ \ \times\Bigg[\frac{2E_p-\mu_I-k_0}{k_0^2-4(E_p-\mu_I/2)^2}(f(E^-_-)-f(-E^-_+))\nonumber\\
&&\ \ \
+\frac{2E_p+\mu_I+k_0}{k_0^2-4(E_p+\mu_I/2)^2}(f(E^+_-)-f(-E^+_+))\Bigg]\
.
\end{eqnarray}
From the comparison of $\Pi_{\pi_+\pi_+}$ at $\mu_B=0$ and $k_0=0$
with the last gap equation of (\ref{mf19}) for the pion condensate
$\pi$ , the phase transition line from $\pi=0$ to $\pi\neq 0$ in
the $T-\mu_I$ plane at $\mu_B=0$ can be calculated by the equation
\begin{eqnarray}
\label{meson11}
1-2G\Pi_{\pi_+\pi_+}(k_0=0)=0\ .
\end{eqnarray}
Thus the $\pi_+$ mass is always zero on the phase boundary.

\subsection {Meson Properties in Superfluidity Phase}
The polarization functions $\Pi_{MM'}(k_0)$ at ${\bf k}=0$ are
calculated in Appendix \ref{a2} as explicit functions of
temperature $T$, baryon chemical potential $\mu_B$, and isospin
chemical potential $\mu_I$ in general case with nonzero chiral and
pion condensates. While $\sigma, \pi_+$ and $\pi_-$ are no longer
collective excitation modes due to the pion condensation, there is
still no mixing between $\pi_0$ and other mesons,
\begin{equation}
\label{meson12}
\Pi_{\pi_0\sigma}(k) = \Pi_{\pi_0\pi_+}(k) =
\Pi_{\pi_0\pi_-}(k) = 0\ ,
\end{equation}
its mass in the superfluidity phase is determined by
\begin{equation}
\label{meson13}
1-2G\Pi_{\pi_0\pi_0}(k_0=M_{\pi_0},{\bf k}=0) = 0\
.
\end{equation}
Taking the polarization function $\Pi_{\pi_0\pi_0}$ evaluated in
Appendix \ref{a2} and making comparison with the gap equation for
$\pi$ at $T=\mu_B=0$, we get analytically
\begin{equation}
\label{meson14}
M_{\pi_0}(\mu_I)=\mu_I\ ,\ \ \ \ \ \mu_I>m_\pi\ .
\end{equation}

The other meson masses are now determined by the equation
\begin{eqnarray}
\label{meson15} &&\det\left(\begin{array}{ccc}
1-2G\Pi_{\sigma\sigma}&-2G\Pi_{\sigma\pi_+}
&-2G\Pi_{\sigma\pi_-}\\
-2G\Pi_{\pi_+\sigma}&1-2G\Pi_{\pi_+\pi_+}&-2G\Pi_{\pi_+\pi_-}\\
-2G\Pi_{\pi_-\sigma}&-2G\Pi_{\pi_-\pi_+}&1-2G\Pi_{\pi_-\pi_-}\\
\end{array}\right)_{k_0=M}\nonumber\\
&&=0\
\end{eqnarray}
at ${\bf k}=0$, namely,
\begin{eqnarray}
\label{meson16}
&&\Big[\left((1-2G\Pi_{\pi_+\pi_+}(k_0))(1-2G\Pi_{\pi_-\pi_-}(k_0))
-4G^2\Pi_{\pi_+\pi_-}^2(k_0)
\right)\nonumber\\
&&\times(1-2G\Pi_{\sigma\sigma}(k_0))-16G^3\Pi_{\sigma\pi_+}(k_0)\Pi_{\sigma\pi_-}(k_0)\Pi_{\pi_+\pi_-}(k_0)\nonumber\\
&&-4G^2\Pi_{\sigma\pi_-}^2(k_0)(1-2G\Pi_{\pi_+\pi_+}(k_0))-4G^2\Pi_{\sigma\pi_+}^2(k_0)\nonumber\\
&&\times(1-2G\Pi_{\pi_-\pi_-}(k_0))\Big]_{k_0=M}=0\ .
\end{eqnarray}

\begin{figure}[ht]
\vspace*{+0cm} \centerline{\epsfxsize=2.5in \epsffile{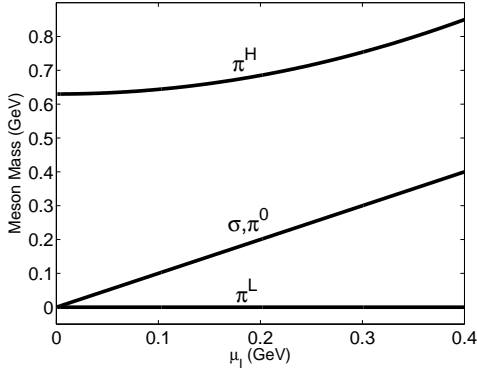}}
\caption{\it The meson mass spectra in the chiral limit as
functions of $\mu_I$ at $T=\mu_B=0$. } \label{fig13}
\end{figure}

We first discuss the masses in the chiral limit. In the chiral
limit, the chiral condensate $\sigma$ as well as the effective
quark mass $M_q$ are forced to be zero by any small isospin
chemical potential. From the explicit form of the polarization
functions evaluated in Appendix \ref{a2}, the mixing between
$\sigma$ and $\pi_\pm$ disappears, $\Pi_{\sigma\pi_\pm}=0$, then
$\sigma$ is still an eigen collective mode. Since the polarization
functions for $\sigma$ and $\pi_0$ are exactly the same in this
case, $\Pi_{\sigma\sigma} =\Pi_{\pi_0\pi_0}$, the mesons $\pi^0$
and $\sigma$ have the same mass in the superfluidity phase,
\begin{equation}
\label{meson17}
M_\sigma(T,\mu_I,\mu_B)=M_{\pi_0}(T,\mu_I,\mu_B)\ ,
\end{equation}
for any $T, \mu_I$ and $\mu_B$, which reflects the restoration of
chiral symmetry. In the case of $T=\mu_B=0$ we have
\begin{equation}
\label{meson18}
M_\sigma(\mu_I) =M_{\pi_0}(\mu_I) =\mu_I\ .
\end{equation}
The other two eigen modes are determined by
\begin{eqnarray}
\label{meson19}
&&\Big[(1-2G\Pi_{\pi_+\pi_+}(k_0))(1-2G\Pi_{\pi_-\pi_-}(k_0))\nonumber\\
&&\ \ -4G^2\Pi_{\pi_+\pi_-}^2(k_0) \Big]_{k_0=M}=0\ .
\end{eqnarray}
From the polarization functions shown in Appendix \ref{a2}, it is
easy to prove that $M=0$ is a solution of (\ref{meson19})
corresponding to the spontaneously isospin symmetry breaking. We
call the massless meson mode as $\pi_L$ and the other heavy mode
as $\pi_H$. The meson mass spectra in the chiral limit are shown
in Fig.(\ref{fig13}) as functions of $\mu_I$ at $T=\mu_B=0$.

\begin{figure}[ht]
\vspace*{+0cm} \centerline{\epsfxsize=2.5in \epsffile{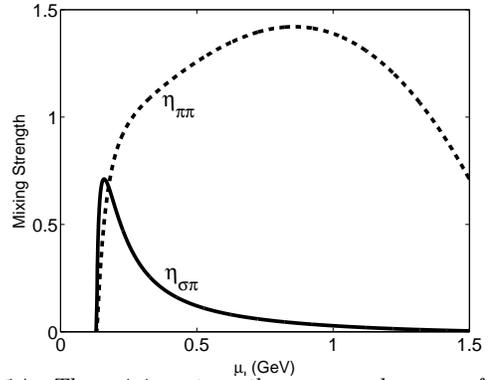}}
\caption{\it The mixing strengthes $\eta_{\sigma\pi}$ and
$\eta_{\pi\pi}$ as functions of isospin chemical potential $\mu_I$
at $T=\mu_B=0$.} \label{fig14}
\end{figure}

In the real world with nonzero current quark mass, there is not
only the mixing between the charged pions, but also a mixing
between $\sigma$ and the charged pions. From the explicit
expression of the polarization functions shown in Appendix
\ref{a2}, we have
\begin{eqnarray}
\label{meson20}
&& \Pi_{\sigma\pi_\pm} =
\Pi_{\pi_\pm\sigma}\propto
\sqrt{2}M_q|2G\pi|\ ,\nonumber\\
&& \Pi_{\pi_+\pi_-}=\Pi_{\pi_-\pi_+}\propto 4G^2\pi^2.
\end{eqnarray}
To quantitatively see the mixing degree among $\sigma, \pi_+$ and
$\pi_-$, we introduce two dimensionless parameters,
\begin{eqnarray}
\label{meson21}
&&
\eta_{\sigma\pi}=\frac{\sqrt{2}M_q|2G\pi|}{4G^2\sigma_0^2}\
,\nonumber\\
&& \eta_{\pi\pi}=\frac{4G^2\pi^2}{4G^2\sigma_0^2}\ ,
\end{eqnarray}
describing the strengthes of $\sigma-\pi$ mixing and $\pi-\pi$
mixing. Their isospin dependence is shown in Fig.(\ref{fig14}) at
$T=\mu_B=0$. The $\sigma-\pi$ mixing is much weaker than the
$\pi-\pi$ mixing, since the former is proportional to both chiral
condensate and pion condensate, but the latter to the square of
pion condensate, and in the superfluidity phase the chiral
symmetry is almost restored. The $\sigma-\pi$ mixing is important
only around the phase transition point $\mu_I^c=m_\pi$, but the
$\pi-\pi$ mixing is strong in a wide region after the phase
transition.

\begin{figure}
\centering \includegraphics[width=2.5in]{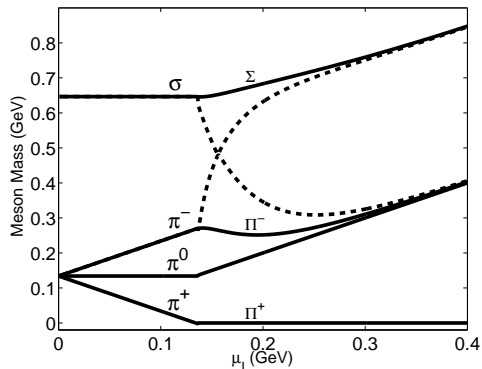} \caption{The
meson masses as functions of isospin chemical potential $\mu_I$ at
$T=\mu_B=0$ in the real world. The solid lines are the full
calculation and the dashed lines from the approximation neglecting
$\sigma-\pi$ mixing. } \label{fig15}
\end{figure}

The mass spectra calculated through solving the Eq.(\ref{meson16})
numerically are shown in Fig.(\ref{fig15}), together with
$M_{\pi_0}$, Eq.(\ref{meson14}) in the superfluidity phase and the
masses (\ref{meson9}) in the normal phase, as functions of $\mu_I$
at $T=\mu_B=0$. Again we obtain the Goldstone mode reflecting
correctly the spontaneous isospin symmetry breaking in the
superfluidity phase. In the normal phase $\sigma, \pi_+, \pi_-$
and $\pi_0$ themselves are eigen modes of the collective
excitation of the system, but in the superfluidity phase, except
for $\pi_0$ which is still an eigen mode, $\sigma, \pi_+$ and
$\pi_-$ are no longer eigen modes. Since the masses determined by
(\ref{meson16}) are continuously connected with the masses of
$\sigma, \pi_+$ and $\pi_-$ at the phase transition point $\mu_I =
m_\pi$, we call them $\Sigma, \Pi_+$ and $\Pi_-$, shown in
Fig.(\ref{fig15}). When $\mu_I$ is high enough, the pion
condensation will disappear due to the asymptotic freedom of QCD,
this turning point in the NJL model is about $\tilde\mu_I^c=1.75$
GeV, see Fig.(\ref{fig1}). Above this turning point the mass
spectra are again controlled by the polarization functions
(\ref{meson8}) and the $\mu_I$ dependence is similar to
(\ref{meson9}).

The mass spectra in the superfluidity phase calculated in the NJL
model here are very similar to the result predicted in chiral
perturbation theory\cite{son,kogut4},
\begin{eqnarray}
\label{meson22}
&& M_{\Pi_+}=0\ ,\nonumber\\
&& M_{\Pi_-}=\mu_I\sqrt{1+3m_\pi^4/\mu_I^4}\
,\nonumber\\
&& M_{\pi_0}=\mu_I\ .
\end{eqnarray}
However, if we neglect the $\sigma-\pi$ mixing in the
superfluidity phase, $\sigma$ is still the eigen mode and the mass
is calculated from the pole equation
\begin{equation}
\label{meson23}
1-2G\Pi_{\sigma\sigma}(k_0=M_\sigma,{\bf k}=0)=0\
,
\end{equation}
and the other two eigen modes which are linear combinations of
$\pi_+$ and $\pi_-$ are determined by the equation
(\ref{meson19}). One of the modes is still the Goldstone mode and
the other is a heavy one. The numerical results for $\sigma$ and
the heavy one are also shown in Fig.(\ref{fig15}) as dashed lines.
In this case, the broken $U_A(1)$ symmetry will be restored at
sufficiently high isospin density, indicating by the same $\sigma$
and $\pi_0$ mass. It is clear that the difference between the full
calculation and the approximation neglecting $\sigma-\pi$ mixing
is mainly in the narrow region above the phase transition point
$\mu_I=m_\pi$.

We now analytically prove the Goldstone mode in general case in
the NJL model. To this end we consider the polarization function
matrix
\begin{eqnarray}
\label{meson24}
&& 1-2G\Pi=\\
&& \left(\begin{array}{cccc} 1-2G\Pi_{00}&-2G\Pi_{01}
&-2G\Pi_{02}&-2G\Pi_{03}\\
-2G\Pi_{10}&1-2G\Pi_{11}&-2G\Pi_{12}&-2G\Pi_{13}\\
-2G\Pi_{20}&-2G\Pi_{21}&1-2G\Pi_{22}&-2G\Pi_{23}\\
-2G\Pi_{30}&-2G\Pi_{31}&-2G\Pi_{32}&1-2G\Pi_{33}\\
\end{array}\right)\ ,\nonumber
\end{eqnarray}
with respect to the basis $(\sigma, \pi_1, \pi_2, \pi_3)$.

From Appendix \ref{a2}, $\pi_3$ does not mix with other modes,
\begin{equation}
\label{meson25}
\Pi_{3n}(k) =\Pi_{n3}(k) =0\ ,\ \ \ n=0,1,2\ ,
\end{equation}
its mass is determined by its own polarization function at ${\bf
k}=0$,
\begin{equation}
\label{meson26}
1-2G\Pi_{33}(k_0=M_{\pi_0})=0\ .
\end{equation}

The masses of the other modes are solved from the equation
\begin{eqnarray}
\label{meson27}
&&\det\left(\begin{array}{ccc}
1-2G\Pi_{00}&-2G\Pi_{01}
&-2G\Pi_{02}\\
-2G\Pi_{10}&1-2G\Pi_{11}&-2G\Pi_{12}\\
-2G\Pi_{20}&-2G\Pi_{21}&1-2G\Pi_{22}\\
\end{array}\right)_{k_0=M}\nonumber\\
&&=0
\end{eqnarray}
at ${\bf k}=0$. The existence of the Goldstone mode means that
$M=0$ is a solution of the above equation. This is really true. In
fact, setting $k_0={\bf k}=0$ we have from the Appendix \ref{a2}
\begin{equation}
\label{meson28}
\Pi_{02}(0,{\bf 0})=\Pi_{20}(0,{\bf
0})=\Pi_{12}(0,{\bf 0})=\Pi_{21}(0,{\bf 0})=0\ ,
\end{equation}
the Goldstone mode can be checked by the equation
\begin{equation}
\label{meson29}
1-2G\Pi_{22}(k_0=0,{\bf k}=0)=0\ .
\end{equation}
Using the analytical form of $\Pi_{22}$ evaluated in Appendix
\ref{a2}, we find that the above equation is exactly the same as
the gap equation for the pion condensate. Therefore, in
superfluidity phase with $\pi\neq0$, there is always a Goldstone
mode. Note that this massless mode holds in the whole
superfluidity phase at finite temperature and isospin and baryon
chemical potentials.

The Goldstone mode will extremely change the thermodynamics of the
system. In mean field approximation, only quarks contribute to the
thermodynamic potential, see (\ref{mf26}). Since quarks are heavy
at low temperature, the thermodynamic functions scaled by the
temperature, for example, $p/T^4, \epsilon/T^4, s/T^3$ and $c/T^3$
approach to zero in the limit of $T\rightarrow 0$\cite{zhuang1}.
However, when the meson fluctuations are included, the massless
meson mode will lead to power laws of thermodynamic functions at
low temperature, these scaled functions will approach to nonzero
constants.

\section {Pion Superfluidity in Linear Sigma Model }
\label{s6}
In this section we study the linear sigma model at finite isospin
chemical potential and compare it with the NJL model. Since the
linear sigma model can be regarded as a bosonized version of the
NJL model, we want to know whether the former can reproduce the
results we have worked out with the latter with quarks as
elementary degrees of freedom. It is well-known that one can use
the bosonization technique to integrate out the quark degrees of
freedom and obtain the effective Lagrangian for the meson bound
states in the NJL model\cite{sandy,eguchi}. In the real world the
Lagrangian is just the one of the linear sigma model with an
explicit chiral breaking term,
\begin{eqnarray}
\label{sigma1} {\cal
L}&=&\frac{1}{2}[(\partial_\mu\sigma)^2+(\partial_\mu\vec{
\pi})^2]+\frac{1}{2}\left(2g_{\pi}^2
f_\pi^2-m_\pi^2\right)(\sigma^2+\vec{\pi}^2)\nonumber\\
&-&\frac{g_\pi^2}{2}(\sigma^2+\vec{\pi}^2)^4-f_\pi m_\pi^2\sigma\
,
\end{eqnarray}
where $g_\pi = g_{\pi q\bar q}$ is the pion-quark-antiquark
coupling constant, $f_\pi$ the pion decay constant, and $m_\pi$
the pion mass in the vacuum, $g_\pi$ and $f_\pi$ can be calculated
in the NJL model in RPA.

\subsection {Zero Temperature}

Introducing the isospin chemical potential $\mu_I$ corresponding
to the third component of isospin charge
\begin{equation}
\label{sigma2} I_3 =\int d^3{\bf x}
(\pi_1\partial_t\pi_2-\pi_2\partial_t\pi_1) \ ,
\end{equation}
and the chiral and pion condensates
\begin{eqnarray}
\label{sigma3}
&& \langle\sigma\rangle=\xi\ ,\nonumber\\
&& \langle\pi^+\rangle=\langle\pi^-\rangle=\frac{\rho}{\sqrt{2}}\
,
\end{eqnarray}
we obtain the effective Lagrangian at tree level,
\begin{eqnarray}
\label{sigma4}
 {\cal L}_{eff} &=&-U(\xi, \rho)+
\frac{1}{2}\Big[(\partial_\mu\sigma)^2 +(\partial_\mu {\bf
\pi}_3)^2+\left(\partial_t\pi_1-\mu_I\pi_2\right)^2\nonumber\\
&+&\left(\partial_t\pi_2+\mu_I\pi_1\right)^2- \left({\bf
\nabla}\pi_1\right)^2-\left({\bf
\nabla}\pi_2\right)^2\Big]\nonumber\\
&-&g_{\pi}^2\Big[\left(3\xi^2+\rho^2-f_\pi^2+{m_\pi^2\over
2g_\pi^2}\right)
\sigma^2\nonumber\\
&+&\left(\xi^2+3\rho^2-f_\pi^2+{m_\pi^2\over
2g_\pi^2}\right)\pi_1^2\nonumber\\
&+&\left(\xi^2+ \rho^2-f_\pi^{2}+{m_\pi^2\over
2g_\pi^2}\right)(\pi_2^2+\pi_3^2)+4\xi\rho\sigma\pi_1\Big]
\nonumber\\
&-&2g_{\pi}^2\left(\xi\sigma+\rho\pi_1\right)\left(\sigma^2+{\bf
\pi}^2\right) -\frac{g_{\pi}^2}{2}\left(\sigma^2+{\bf
\pi}^2\right)^2\ ,
\end{eqnarray}
with the classical potential
\begin{eqnarray}
\label{sigma5} \label{potential2} U(\xi,\rho) &=&
-\frac{1}{2}\left(2g_{\pi}^2
f_\pi^2-m_\pi^2\right)\left(\xi^2+\rho^2\right)
+\frac{g_\pi^2}{2}\left(\xi^2+\rho^2\right)^2\nonumber\\
&-&f_\pi m_\pi^2\xi-{1\over 2}\mu_I^2\rho^2\ ,
\end{eqnarray}
where the two condensates are determined by minimizing the
potential,
\begin{equation}
\label{sigma6} \frac{\partial U}{\partial \xi}=0,\ \ \ \
\frac{\partial U}{\partial \rho}=0\ .
\end{equation}
We have chosen pion condensate to be real as in the NJL model. In
this section we use $\xi$ and $\rho$ standing for the chiral and
pion condensates in order to avoid confusion with the definitions
in the NJL model.

In the vacuum the constraints (\ref{sigma6}) on the potential give
$\xi=f_\pi$ and $\rho=0$, and the sigma mass can be read out from
the quadratic term in the sigma field in the effective Lagrangian
(\ref{sigma4}), $m_\sigma^2=4g_\pi^2f_\pi^2 +m_\pi^2$.

At finite isospin density the two condensates in the ground state
satisfy $\xi=0, \rho=f_\pi\sqrt{1+2\mu_I^2/m_\sigma^2}$ in the
chiral limit. This is consistent with the NJL model: Any small
isospin density can force the chiral symmetry to be restored in
the chiral limit. In the real world, we obtain $\xi=f_\pi, \rho=0$
for $\mu_I < m_\pi$ which is the same as in the vacuum,
and\cite{camp}
\begin{eqnarray}
\label{sigma7}
&& \xi = f_\pi {m_\pi^2\over \mu_I^2}\ ,\nonumber\\
&& \rho=f_\pi\sqrt{1-\left({m_\pi^2\over
\mu_I^2}\right)^2+2\frac{\mu_I^2-m_\pi^2}{m_\sigma^2-m_\pi^2}}\ ,
\end{eqnarray}
for $\mu_I > m_\pi$. This result is qualitatively consistent with
the NJL model and the chiral perturbation theory. Especially the
critical behaviors of the two condensates around the phase
transition point $\mu_I^c=m_\pi$ in the three effective models are
almost the same.

The thermodynamic functions can be evaluated from the potential
$U$. The isospin density $n_I$, pressure $p$ and energy density
$\epsilon$ are analytically written as
\begin{eqnarray}
\label{sigma8}
n_I&=&-\frac{\partial U}{\partial
\mu_I}=\mu_I\rho^2\nonumber\\
&=&\mu_If_\pi^2\left(1-\frac{m_\pi^4}{\mu_I^4}+
2\frac{\mu_I^2-m_\pi^2}{m_\sigma^2-m_\pi^2}\right)\ ,\nonumber\\
p&=&-U(\xi,\rho)+U(f_\pi,0)\nonumber\\
&=&\frac{1}{2}f_\pi^2(\mu_I^2-m_\pi^2)^2
\left(\frac{1}{\mu_I^2}+\frac{1}{m_\sigma^2-m_\pi^2}\right)\ ,\nonumber\\
\epsilon&=&-p+\mu_I
n_I\nonumber\\
&=&\frac{1}{2}f_\pi^2(\mu_I^2-m_\pi^2)\left(\frac{\mu_I^2+3m_\pi^3}{\mu_I^2}
+\frac{3\mu_I^2+m_\pi^2}{m_\sigma^2-m_\pi^2}\right)\ .
\end{eqnarray}

\begin{figure}
\centering \includegraphics[width=2.5in]{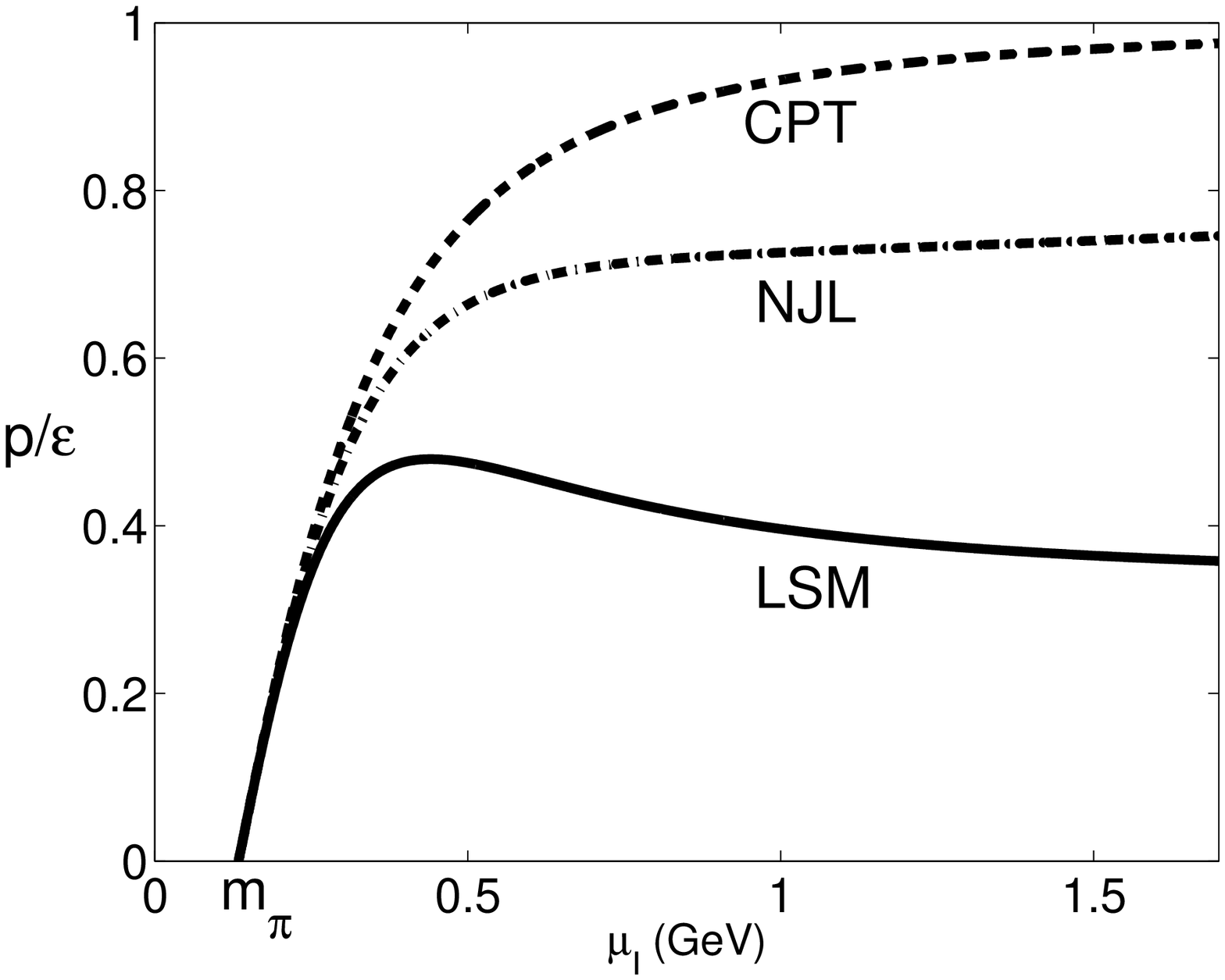}%
\hspace{0.5in}%
\includegraphics[width=2.5in]{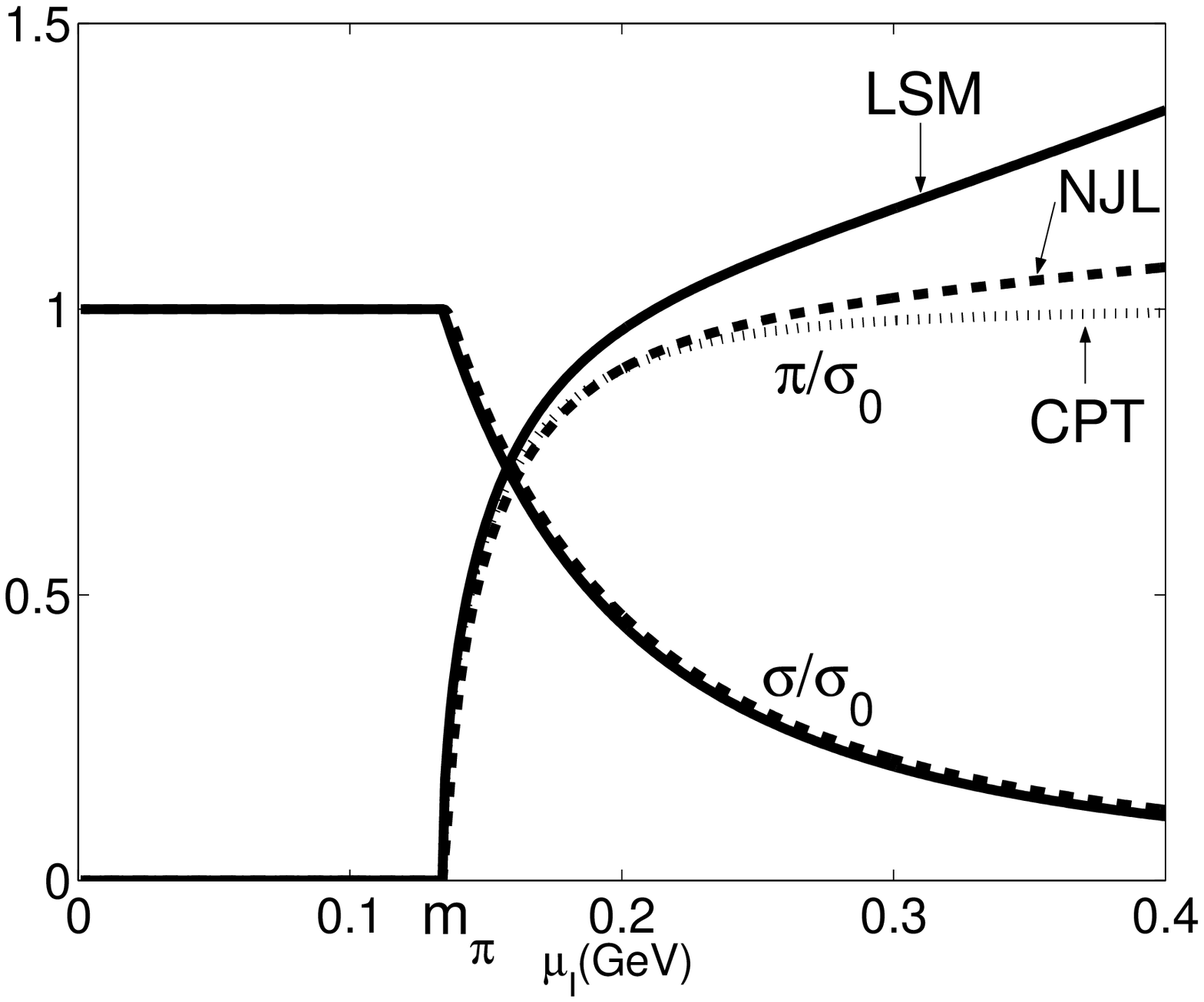}
\caption{The ratio $p/\epsilon$ (upper panel) and the chiral and
pion condensates (lower panel) as functions of isospin chemical
potential $\mu_I$ at $T=\mu_B=0$ in the NJL model (NJL), linear
sigma model (LSM), and chiral perturbation theory (CPT).}
\label{fig16}\end{figure}
If we take the limit $m_\sigma\rightarrow\infty$, we reproduce the
result in the chiral perturbation theory\cite{son,kogut4}.
However, the finite sigma mass leads to a big difference at large
$\mu_I$. At large $\mu_I$, the ratio $p/\epsilon$ approaches $1$
in the chiral perturbation theory, but only about $1/3$ in the
linear sigma model and $0.7$ in the NJL model, as shown in
Fig.(\ref{fig16}). While the three effective models are very
different in high density region, they behavior almost the same in
the region close to the critical point, see also the chiral and
pion condensates shown in the lower panel of (\ref{fig16}).

\begin{figure}
\centering \includegraphics[width=2.7in]{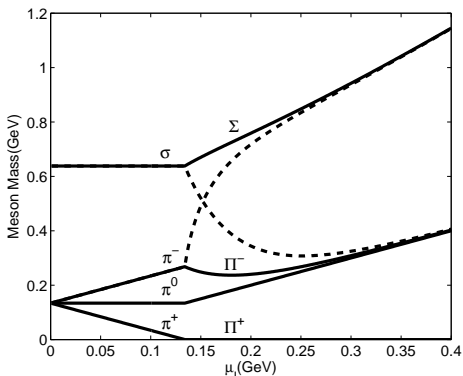} \caption{The
meson masses as functions of isospin chemical potential $\mu_I$ at
$T=0$ in the linear sigma model. The solid lines are the full
calculation and the dashed lines from the approximation neglecting
$\sigma-\pi$ mixing.} \label{fig17}
\end{figure}

The meson mass spectrum is easily obtained by diagonalizing the
quadratic part of the effective Lagrangian (\ref{sigma4}). It is
clear that $\pi_0$ itself is an eigen mode of the system, while
$\pi_+$ and $\pi_-$ are not, like the case in the NJL model. A
simple algebra calculation gives the result
\begin{eqnarray}
\label{sigma9}
&& M_\sigma(\mu_I) = m_\sigma\ ,\nonumber\\
&& M_{\pi^-}(\mu_I) = m_\pi+\mu_I\ ,\nonumber\\
&& M_{\pi^+}(\mu_I) = m_\pi-\mu_I\ ,\nonumber\\
&& M_{\pi_0}(\mu_I) = m_\pi\ ,
\end{eqnarray}
for $\mu_I<m_\pi$ and
\begin{eqnarray}
\label{sigma10}
&& M_{\Sigma}(\mu_I) =\nonumber\\
&& \sqrt{2g_\pi^2f_\pi^2+\frac{7\mu_I^2}{2}-
m_\pi^2+2g_\pi^2\sqrt{\left(\xi^2-\rho^2-\frac{3\mu_I^2}{4g_\pi^2}\right)^2+4\xi^2\rho^2}}\ ,\nonumber\\
&& M_{\Pi_-}(\mu_I)=\nonumber\\
&& \sqrt{2g_\pi^2f_\pi^2+\frac{7\mu_I^2}{2}
-m_\pi^2-2g_\pi^2\sqrt{\left(\xi^2-\rho^2-\frac{3\mu_I^2}{4g_\pi^2}\right)^2+4\xi^2\rho^2}}\ ,\nonumber\\
&& M_{\Pi_+}(\mu_I) = 0\ ,\nonumber\\
&& M_{\pi_0}(\mu_I) = \mu_I\ ,
\end{eqnarray}
for $\mu_I>m_\pi$. The numerical result is shown in
Fig(\ref{fig17}). The $\sigma-\pi$ mixing here plays also an
important role for the meson mass spectra. The calculation without
considering $\sigma-\pi$ mixing is indicated by dashed lines. We
find that the mass spectra are qualitatively consistent with the
NJL model and the chiral perturbation theory.

\subsection {Finite Temperature}
There are two approaches to treat thermal excitation in the linear
sigma model, the Hartree-Fock approximation and the large $N$
expansion. It is well-known that a disadvantage of the
Hartree-Fock approach in the discussion of chiral symmetry
restoration is the lack of the Goldstone mode in the symmetry
breaking phase\cite{petro}. Here we calculate the mass spectra and
phase diagram in the two approaches at finite temperature and
isospin density and compare the results with the NJL calculation.

\begin{figure}
\centering \includegraphics[width=2.5in]{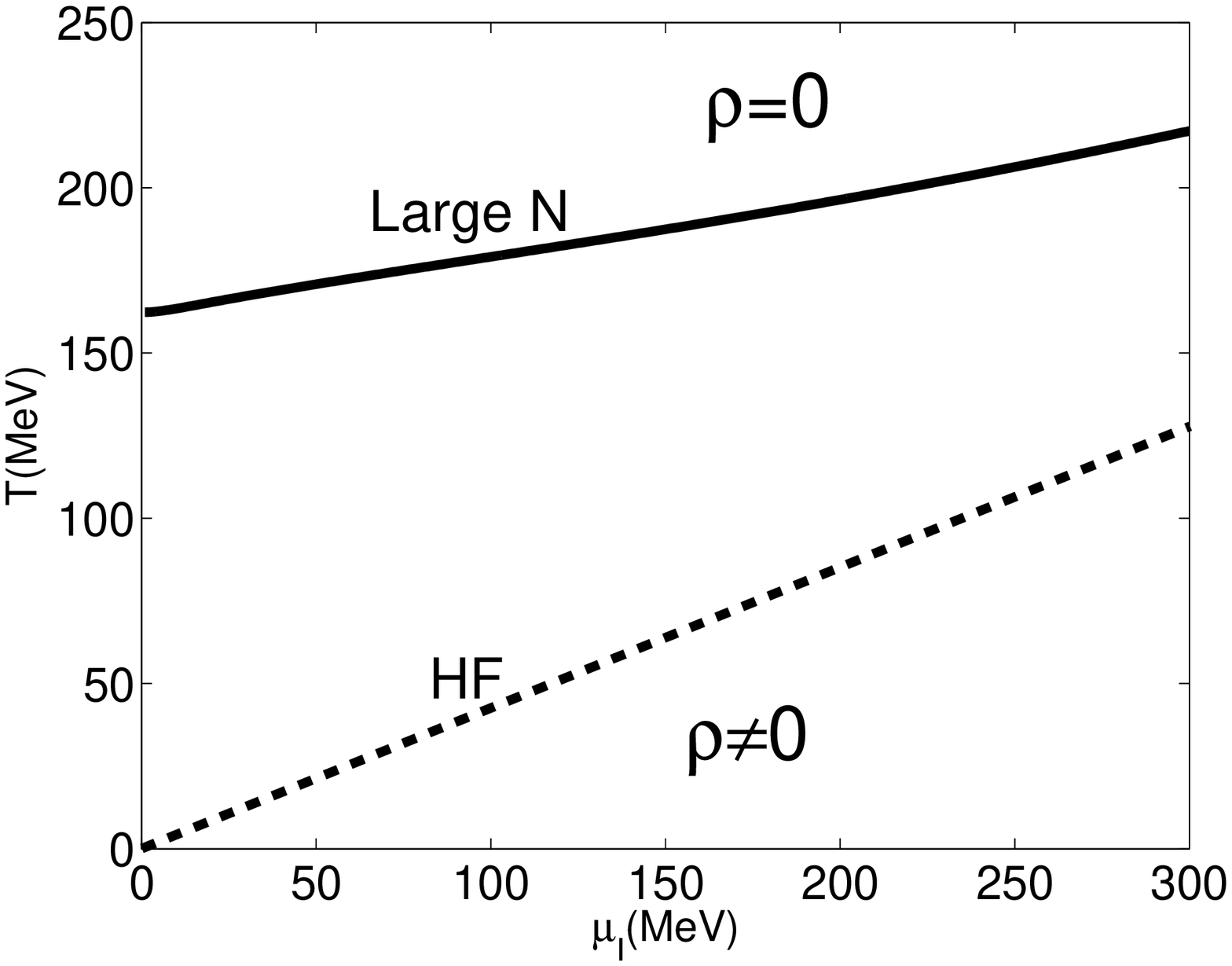}%
\hspace{0.5in}%
\includegraphics[width=2.5in]{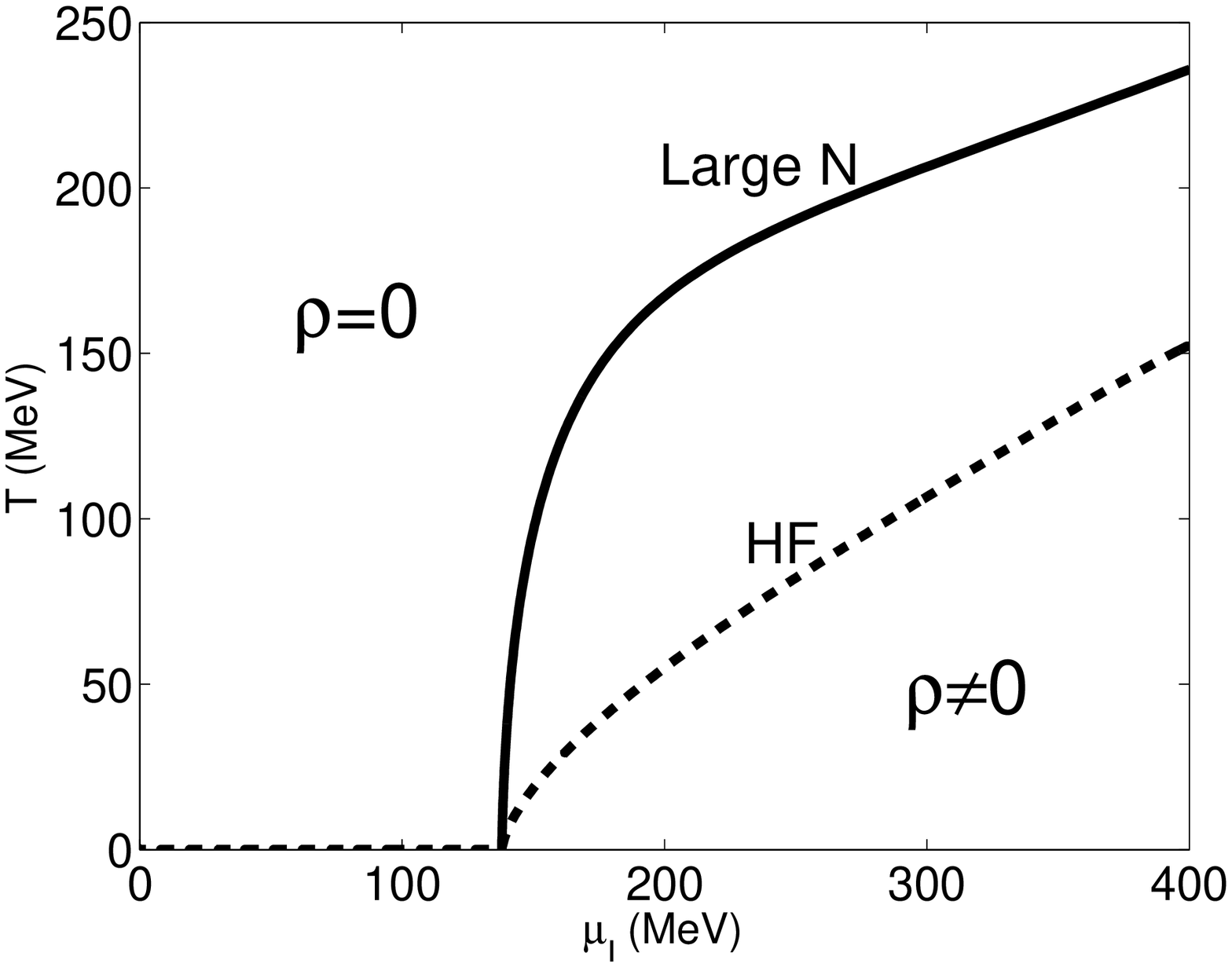}
\caption{The phase diagram of pion superfluidity in the $T-\mu_I$
plane in the chiral limit (upper panel) and the real world (lower
panel) in the linear sigma model. The solid and dashed lines are,
respectively, the results of the large $N$ expansion and the
Hartree-Fock approximation.} \label{fig18}
\end{figure}

Since the sigma model (\ref{sigma1}) can be regarded as a $O(4)$
model, we adopt the large $N$ expansion method in the $O(N)$ sigma
model with boson chemical potential. With the same method and
technics given by Harber and Weldon\cite{haber}, the mass
parameter $M$ and condensates $\xi$ and $\rho$ are determined by
the following gap equations
\begin{eqnarray}
\label{sigma11}
&& M^2 =
2g_\pi^2\left[\xi^2+\rho^2-f_\pi^2+2J'(T,\mu_I,M)\right]+m_\pi^2\
,\nonumber\\
&& \rho(M^2-\mu_I^2)=0\ ,\nonumber\\
&& \xi M^2 - f_\pi m_\pi^2=0\ ,
\end{eqnarray}
the thermal excitation is introduced by the function
\begin{eqnarray}
\label{sigma12}
&& J'(T,\mu_I,M) = \frac{1}{2}\int{d^3{\bf k}\over
(2\pi)^3}\frac{1}{\sqrt{k^2+M^2}}\nonumber\\
&&\times\left[2f_b(T,0,M)+f_b(T,\mu_I,M)+f_b(T,-\mu_I,M)\right]\ ,
\end{eqnarray}
with the Bose-Einstein distribution function
\begin{equation}
\label{sigma13}
f_b(T,\mu_I,M)={1\over e^{\left(\sqrt{{\bf
k}^2+M^2}-\mu_I\right)/T}-1}\ .
\end{equation}
It is easy to derive the solution
\begin{eqnarray}
\label{sigma14}
&& M(T,\mu_I)=\mu_I\ ,\nonumber\\
&& \xi(T,\mu_I)=f_\pi\frac{m_\pi^2}{\mu_I^2}\ ,\nonumber\\
&& \rho(T,\mu_I)=\sqrt{\rho^2(0,\mu_I)-2J'(T,\mu_I,\mu_I)}\ ,
\end{eqnarray}
in the pion superfluidity phase, and the coupled equations
\begin{eqnarray}
\label{sigma15}
&&
M^2=2g_\pi^2\left[\xi^2-f_\pi^2+2J'(T,\mu_I,M)\right]+m_\pi^2\
,\nonumber\\
&& \xi=f_\pi\frac{m_\pi^2}{M^2}\ ,
\end{eqnarray}
for $M$ and $\xi$ in the normal phase with $\rho =0$. From the
above two groups of solutions we can obtain the phase diagram in
the $T-\mu_I$ plane, shown in Fig.(\ref{fig18}).

The Hartree-Fock approach in the superfluidity phase is rather
complicated, and there is no Goldstone mode at finite temperature
in this approach. Here we will not present the treatment in the
superfluidity phase with $\rho\neq0$. However, to calculate the
phase transition line of pion superfluidity in the $T-\mu_I$ plane
is much easier. We need only the formulas in the normal phase with
$\rho =0$. From the detailed derivation given in Appendix
\ref{a3}, the effective meson masses and the chiral condensate in
this phase are governed by the coupled equations,
\begin{eqnarray}
\label{sigma16}
&& M_\sigma^2 =
2g_\pi^2\left(3\xi^2-f_\pi^2+3\langle\sigma\sigma\rangle+2\langle
\pi\pi\rangle+\langle
\pi_0\pi_0\rangle\right)+m_\pi^2\ ,\nonumber\\
&& M_{\pi_0}^2 =
2g_\pi^2\left(\xi^2-f_\pi^2+\langle\sigma\sigma\rangle+2\langle
\pi\pi\rangle+3\langle\pi_0\pi_0\rangle \right)+m_\pi^2\ ,\nonumber\\
&& M_{\pi_\pm}^2 =
2g_\pi^2\left(\xi^2-f_\pi^2+\langle\sigma\sigma\rangle+4\langle\pi\pi\rangle
 +\langle
\pi_0\pi_0\rangle\right)+m_\pi^2\ ,\nonumber\\
&&
2g_\pi^2\xi\left(\xi^2-f_\pi^2+3\langle\sigma\sigma\rangle+2\langle
\pi\pi\rangle+\langle\pi_0 \pi_0\rangle\right)+\xi m_\pi^2= f_\pi
m_\pi^2\ .\nonumber\\
\end{eqnarray}
To illustrate how to determine the critical temperature from the
quantities obtained in the normal phase, we consider the isospin
density derived from the thermodynamic potential shown in Appendix
\ref{a3},
\begin{equation}
\label{sigma17}
n_I=-\frac{\partial\Omega}{\partial
\mu_I}=\int\frac{d^3\bf
k}{(2\pi^3)}\left(f_b(T,\mu_I,M_\pi)+f_b(T,-\mu_I,M_\pi)\right)\ .
\end{equation}
Note that this is only the thermal contribution to the isospin
density and true only in the normal phase. In the superfluidity
phase we should include the contribution from the condensation of
charged pions. From the definition of Bose-Einstein condensation,
the phase transition point where the occupation number density is
divergent at zero momentum is determined by
\begin{equation}
\label{sigma18}
M_\pi(T_c,\mu_I)=\mu_I\ .
\end{equation}
We have checked that this equation can also be derived from the
the superfluidity phase in the limit of $\rho = 0$.

The phase diagram in the $T-\mu_I$ plane calculated with the large
$N$ expansion method and Hartree-Fock approximation is shown in
Fig.(\ref{fig18}) in the chiral limit (upper panel) and the real
world (lower panel). It is clear that the result of the large $N$
expansion is qualitatively consistent with the NJL model
calculation. Especially, in the chiral limit, when
$\mu_I\rightarrow 0$ the critical temperature in the large $N$
approach is very close to the one obtained in the NJL model, while
in the Hartree-Fock approximation the critical temperature
approaches zero, which is certainly wrong.

\section {Finite Baryon Density and Diquark Condensation}
\label{s7}
We have investigated the NJL model at finite isospin chemical
potential $\mu_I$ and temperature $T$ and made comparison with the
other effective models. We now turn to the discussion at finite
baryon chemical potential $\mu_B$. It is well-known
that\cite{karsch,sandy,fodor} the baryon density effect on
deconfinement and chiral restoration is qualitatively different
from the temperature effect: the phase transition is of first
order at high baryon density but of second order or even a smooth
crossover at high temperature. The physics in high baryon density
region is also very different from that in high isospin density
region: the spontaneously broken symmetry is of color symmetry at
sufficiently high baryon chemical potential but isospin symmetry
at sufficiently high isospin chemical potential, the former is
described by the diquark condensate $\langle\psi\psi\rangle$ and
the latter by the pion condensate $\pi$.  We will first consider
the baryon density effect on the pion condensation without
considering the color superconductivity, and then discuss the
competition between the pion superfluidity and color
superconductivity in the $\mu_B-\mu_I$ plane.

\subsection {Finite Baryon Density Effect without Diquark Condensate}
\begin{figure}
\centering \includegraphics[width=2.5in]{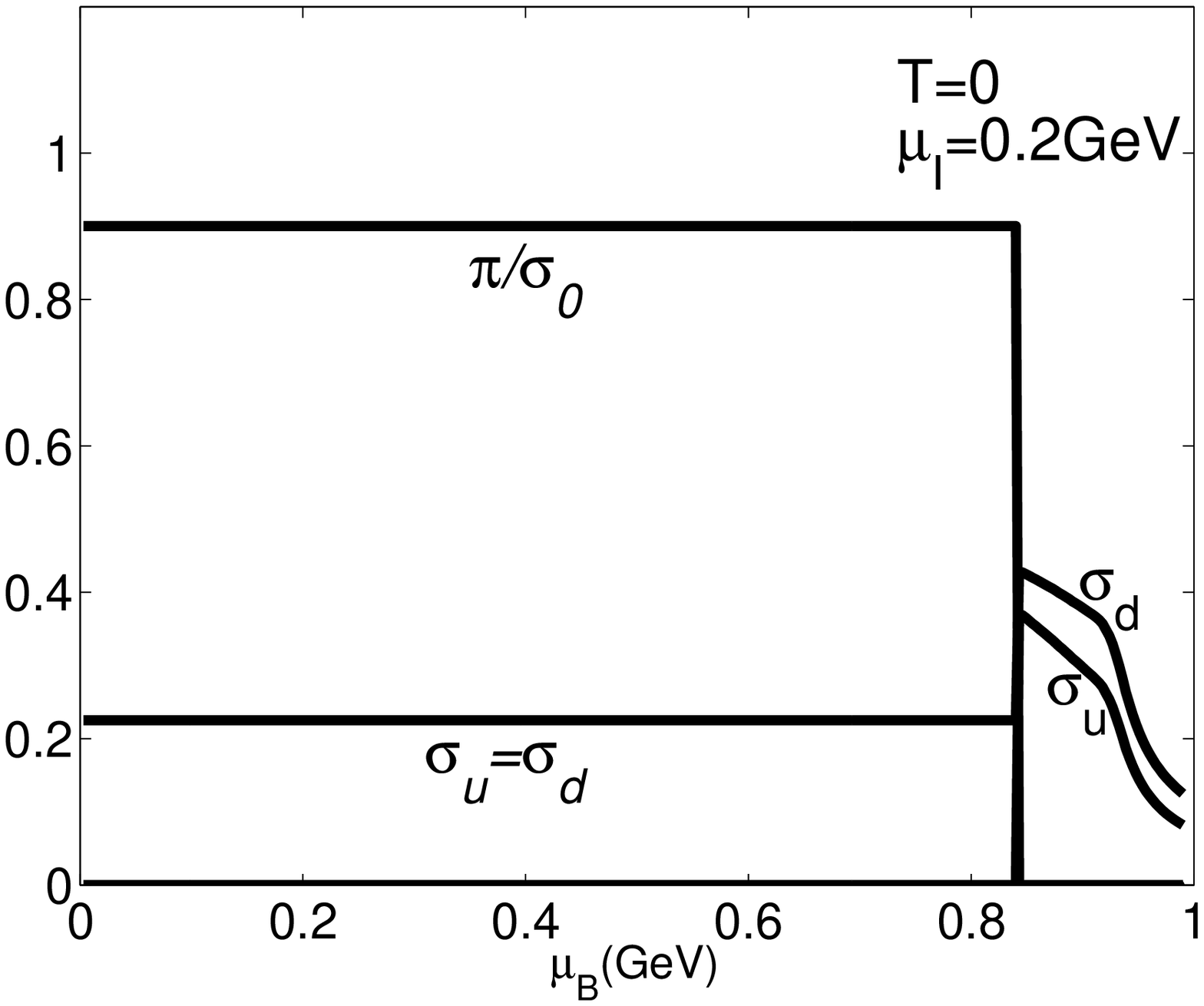}%
\hspace{0.5in}%
\includegraphics[width=2.5in]{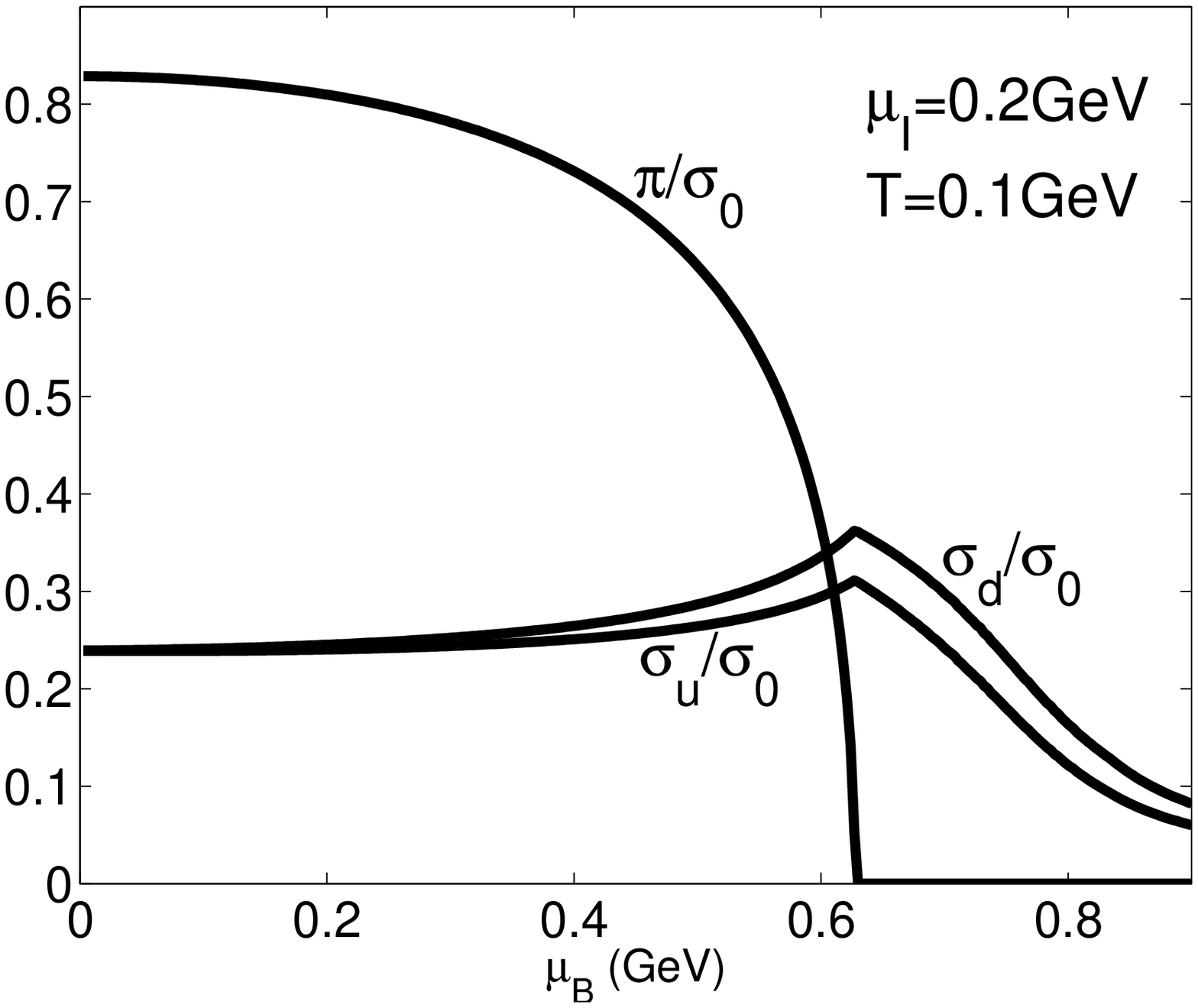}
\caption{The $u$ and $d$ quark condensates and pion condensate,
scaled by the chiral condensate $\sigma_0$ in the vacuum, as
functions of $\mu_B$ at $\mu_I=0.2$ GeV and $T=0$ (upper panel)
and $T=0.1$ GeV (lower panel).} \label{fig19}
\end{figure}

As we indicated in the Section \ref{a2}, when the two chemical
potentials $\mu_B$ and $\mu_I$ are both nonzero, the $u$ and $d$
quark condensates are different from each other, see (\ref{mf19}).
It is clear that the isospin chemical potential $\mu_I$ should be
large enough to guarantee the phase transition of pion
superfluidity at finite baryon chemical potential $\mu_B$ and
temperature $T$, at least it should be larger than the critical
value $m_\pi$ at $T=\mu_B=0$. In our numerical calculation we took
it as $0.2$ GeV. The $\mu_B$ dependence of the chiral and pion
condensates, again scaled by the chiral condensate $\sigma_0$ in
the vacuum, at fixed $T$ and $\mu_I$ is shown in
Fig.(\ref{fig19}). At zero temperature, the $u$ and $d$ quark
condensates $\sigma_u$ and $\sigma_d$ are almost the same, and all
the three condensates, $\sigma_u, \sigma_d$ and $\pi$ keep their
vacuum values in the pion superfluidity phase. At a critical value
$\mu_B \sim 0.85$ GeV, a first order phase transition happens, the
pion condensate jumps down from its vacuum value to zero, and the
$u$ and $d$ quark condensates jump up from their vacuum value to
some larger values. In the normal phase with $\pi=0$, the two
chiral condensates decrease with increasing baryon chemical
potential, and the difference between them becomes remarkable. The
temperature effect changes the phase transition from first order
to second order, see the lower panel for $T=0.1$ GeV, and reduces
the critical baryon chemical potential from about $0.85$ GeV to
about $0.65$ GeV. Also the temperature effect leads to remarkable
difference between the two chiral condensates not only in the
normal phase but also in the pion superfluidity phase.

\begin{figure}
\centering \includegraphics[width=2.5in]{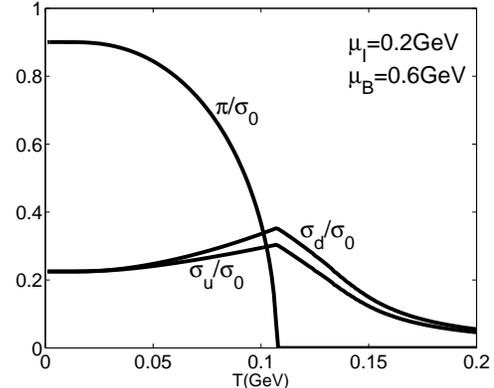} \caption{The
$u$ and $d$ quark condensates and pion condensate, scaled by the
chiral condensate $\sigma_0$ in the vacuum, as functions of $T$ at
$\mu_I=0.2$ GeV and $\mu_B=0.6$ GeV.} \label{fig20}
\end{figure}

\begin{figure}
\centering \includegraphics[width=2.5in]{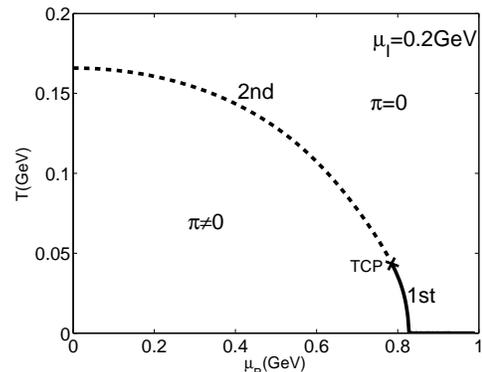} \caption{The
phase diagram of pion superfluidity in the $T-\mu_B$ plane at
$\mu_I=0.2$ GeV. The solid part of the line indicates a first
order phase transition while the dashed part indicates one of
second order. The symbol $\times$ is the position of the
tricritical point. } \label{fig21}
\end{figure}

The temperature dependence of the three condensates is indicated
in Fig.(\ref{fig20}) at fixed chemical potentials. Similar to the
case\cite{hufner,zhuang1} of investigating chiral symmetry
restoration without pion condensation, the temperature effect here
results in a second order phase transition of pion superfluidity.
The phase diagram in the $T-\mu_B$ plane at fixed $\mu_I$ is shown
in Fig.(\ref{fig21}), which is very similar to the chiral phase
transition line at $\mu_I=0$\cite{zhuang1}. The phase transition
is of second order in high temperature region and of first order
in high baryon chemical potential region, and the tricritical
point which connects the first and second order phase transitions
is located at $T=0.045$ GeV and $\mu_B=0.78$ GeV.

\subsection {Competition between Pion and Diquark Condensates at $N_c=2$ }
To study both pion superfluidity and color superconductivity at
finite baryon and isospin chemical potentials, we first consider
the two color NJL model with both quark-antiquark and diquark
channels,
\begin{eqnarray}
\label{di1} {\cal L} &=&
\bar{\psi}\left(i\gamma^{\mu}\partial_{\mu}-m_0\right)\psi
+G_S\left(\left(\bar{\psi}\psi\right)^2+\left(\bar{\psi}i\gamma_5\vec{\tau}\psi\right)^2
\right)\nonumber\\
&+&G_D\left(\bar{\psi}^c
i\gamma_5\tau_2^c\tau_2^f\psi\right)\left(\bar{\psi}i\gamma_5\tau_2^c\tau_2^f\psi^c\right)
\ ,
\end{eqnarray}
where $\tau_2^c$ and $\tau_2^f$ are the second Pauli matrix in
color and flavor spaces, $G_S$ and $G_D$ are the coupling
constants in the quark-antiquark channel and diquark channel, and
from the Fierz transformation we have $G_D = G_S$ for $N_c =2$.

\begin{figure}
\centering \includegraphics[width=2.3in]{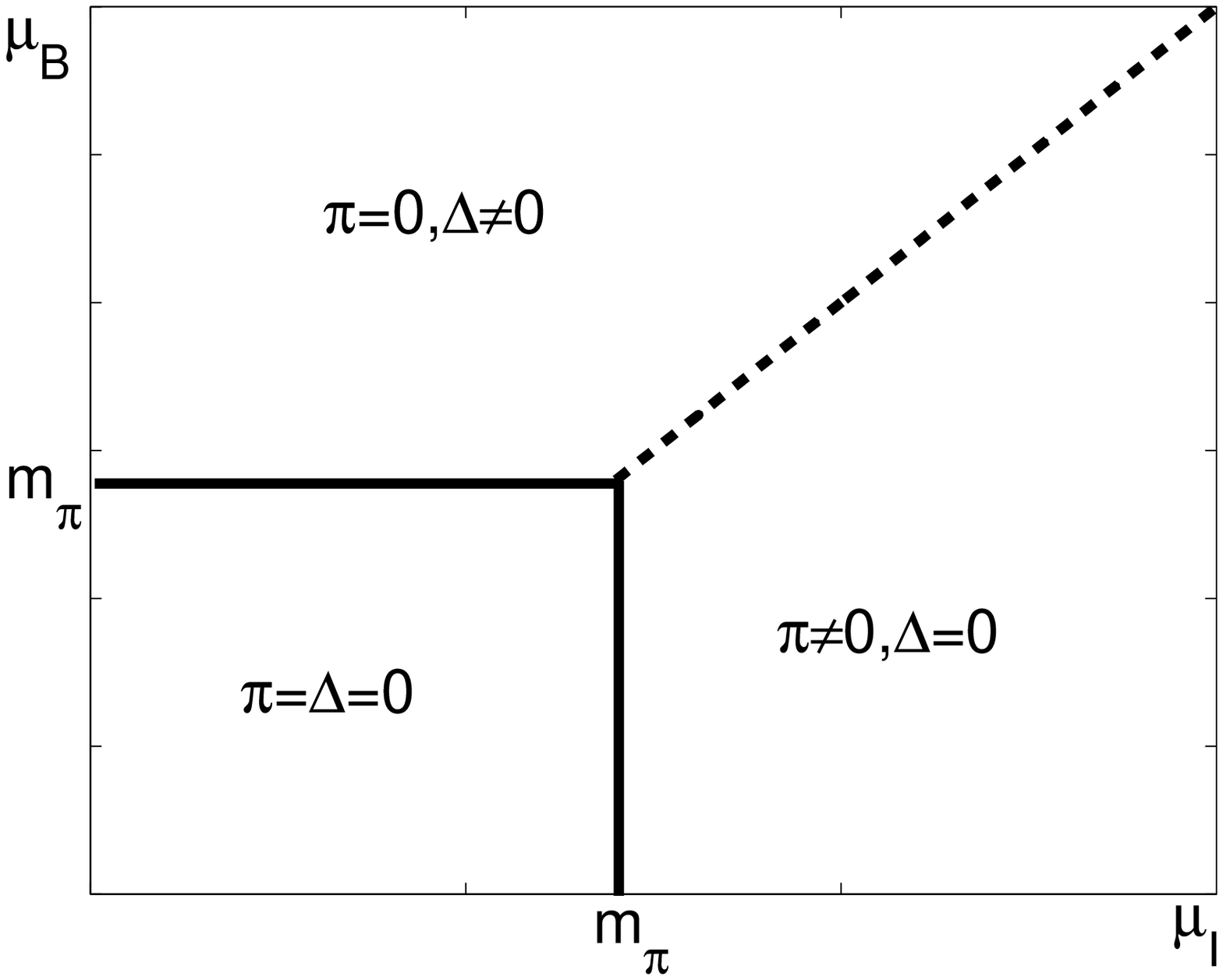}%
\hspace{0.5in}%
\includegraphics[width=2.5in]{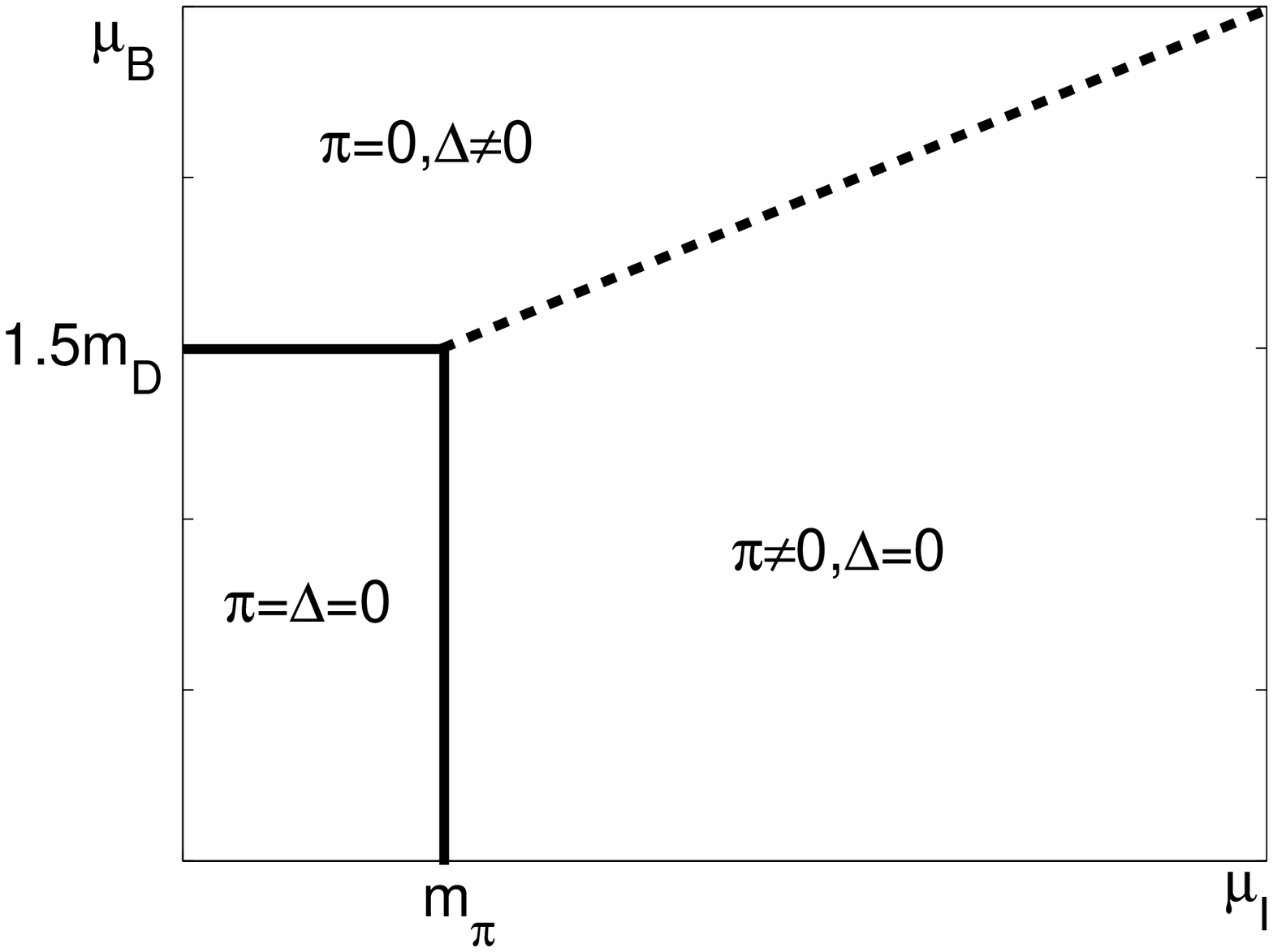}
\caption{The phase diagram of pion superfluidity and color
superconductivity in the $\mu_B-\mu_I$ plane at $T=0$ in the frame
of $N_c=2$ (upper panel) and $N_c=3$ (lower panel) NJL models. }
\label{fig22}
\end{figure}

At sufficiently small $\mu_I$ and large $\mu_B$, there should be
only chiral condensate $\sigma$ and diquark condensate $\Delta$
defined as
\begin{equation}
\label{di2}
\Delta=\langle\bar{\psi}^ci\gamma_5\tau_1^c\tau_1^f\psi\rangle\ .
\end{equation}
Taking the similar way used in Section \ref{s2}, it is
straightforward to derive the quark propagator matrix in
Nambu-Gorkov space and then obtain the gap equations for the two
condensates\cite{huang1,huang2}. At zero temperature they are
\begin{eqnarray}
\label{di3} && \sigma +2N_c M_q\int{d^3{\bf k}\over
(2\pi)^3}{1\over E_k}\Big[{E_k-\mu_B/N_c\over
E_\Delta^-}\Theta(E_\Delta^--|\mu_I|/2)\nonumber\\
&&\ \ \ +{E_k+\mu_B/N_c\over E_\Delta^+}
\Theta(E_\Delta^+-|\mu_I|/2)\Big] = 0\ ,\nonumber\\
&& \Delta\Big[1 - 4N_c G_D\int{d^3{\bf k}\over
(2\pi)^3}\Big({1\over
E_\Delta^-}\Theta(E_\Delta^--|\mu_I|/2)\nonumber\\
&&\ \ \ +{1\over E_\Delta^+}\Theta(E_\Delta^-+|\mu_I|/2)\Big)\Big]
= 0\ ,
\end{eqnarray}
with the energy functions $E_\Delta^\pm$ and effective quark mass
$M_q$ defined as
\begin{eqnarray}
\label{di4} && E_\Delta^\pm = \sqrt{\left(E_k\pm
{\mu_B/N_c}\right)^2+4G_D^2\Delta^2}\ ,\nonumber\\
&& M_q=m_0-2G_S\sigma\ .
\end{eqnarray}
These gap equations for $\sigma$ and $\Delta$ are exactly the same
as the gap equations for $\sigma$ and $\pi$ shown in Section
\ref{s2} at $T=0$, if we do the replacement
$\Delta\rightarrow\pi,\mu_B\rightarrow\mu_I$. Making comparison
with the pion mass equation (\ref{iso4}) in the vacuum and
considering the symmetry between $\Delta$ and $\pi$, we obtain the
critical baryon chemical potential $\mu_B^c$ of color
superconductivity and the critical isospin chemical potential
$\mu_I^c$ of pion superfluidity,
\begin{eqnarray}
\label{di5}
&& \mu_B^c=m_\pi\ ,\ \ \ \ (\mu_I < m_\pi)\ ,\nonumber\\
&& \mu_I^c=m_\pi\ ,\ \ \ \ (\mu_B < m_\pi)\ .
\end{eqnarray}
Therefore, the normal phase without pion and diquark condensations
is in the square box $\mu_I<m_\pi$ and $\mu_B<m_\pi$, shown in
Fig.(\ref{fig22}). The system is in the phase of color
superconductivity with $\pi =0$ and $\Delta \neq 0$ for
$\mu_I<m_\pi$ and $\mu_B>m_\pi$ and in the phase of pion
superfluidity with $\pi \neq 0$ and $\Delta =0$ for $\mu_B <
m_\pi$ and $\mu_I > m_\pi$. In the region of high $\mu_I$ and
$\mu_B$, due to the symmetry between the pion and diquark
condensates proposed by their gap equations, we can conclude that
the pion and diquark condensates are equivalent in the phase
diagram in the $\mu_I-\mu_B$ plane. This can be confirmed by
taking the bosonized version\cite{rapp} of the two color NJL model
with chiral, pion and diquark condensates. The effective potential
in the bosonized version can be derived as
\begin{eqnarray}
\label{di6}
V(\sigma,\pi,\Delta)&=&-a(\sigma^2+\pi^2+\Delta^2)+b(\sigma^2+\pi^2+\Delta^2)^2\nonumber\\
&-&c\sigma-\frac{1}{2}\mu_I^2\pi^2-\frac{1}{2}\mu_B^2\Delta^2\ ,
\end{eqnarray}
where $a, b$ and $c$ are constants. We can easily obtain a first
order phase transition line, shown by the dashed line in
Fig.(\ref{fig22}), by minimizing the effective potential. The
phase diagram calculated here in the frame of NJL model is exactly
the same as the one obtained with effective chiral Lagrangian for
$N_c = 2$\cite{split}.

\subsection {Competition between Pion and Diquark Condensations at $N_c=3$ }
In the real world with color degrees of freedom $N_c =3$, the NJL
model including the diquark sector is defined by the Lagrangian
density
\begin{eqnarray}
\label{di7} {\cal L} &=&
\bar{\psi}\left(i\gamma^{\mu}\partial_{\mu}-m_0\right)\psi
+G_S\left(\left(\bar{\psi}\psi\right)^2+\left(\bar{\psi}i\gamma_5\vec{\tau}\psi\right)^2
\right)\nonumber\\
&+&G_D\left(\bar{\psi}^c_{i\alpha}i\gamma_5\epsilon^{\alpha\beta\gamma}_c
\epsilon^{ij}_f\psi_{j\beta}\right)
\left(\bar{\psi}_{i\alpha}i\gamma_5\epsilon^{\alpha\beta\gamma}_c\epsilon^{ij}_f
\psi^c_{j\beta}\right) \ ,
\end{eqnarray}
where $\epsilon^{ij}_f$ and $\epsilon^{\alpha\beta\gamma}_c$ are
totally antisymmetric tensors in flavor and color spaces.

At sufficiently small $\mu_I$ and large $\mu_B$, there should be
only chiral and diquark condensates. Taking the standard way used
in Section \ref{s2} we can derive the quark propagator matrix in
the Nambu-Gorkov space and then obtain the gap equations
determining the chiral condensate $\sigma$ and diquark condensate
$\Delta$ defined as
\begin{equation}
\label{di8}
\Delta = \langle
\bar\psi^c_{i\alpha}\epsilon_f^{ij}\epsilon_c^{\alpha\beta
3}i\gamma_5\psi_{j\beta}\rangle\ ,
\end{equation}
where it has been regarded that only the first two colors
participate in the condensate, while the third one does not. At
zero temperature, the $\mu_B$ and $\mu_I$ dependence of the two
condensates are governed by the equations
\begin{eqnarray}
\label{di9}
&& \sigma +4M_q\int{d^3{\bf k}\over (2\pi)^3}{1\over
E_k}\Big[{E_k-\mu_B/3\over
E_\Delta^-}\Theta(E_\Delta^--|\mu_I|/2)\nonumber\\
&&\ \ \ \ +{E_k+\mu_B/3\over
E_\Delta^+}\Theta(E_\Delta^+-|\mu_I|/2)\nonumber\\
&&\ \ \ \ +\frac{1}{2}\Theta(E_0^--|\mu_I|/2)+\frac{1}{2}\Theta(E_0^+-|\mu_I|/2)\Big] = 0\ ,\nonumber\\
&& \Delta\Big[1 - 8G_D\int{d^3{\bf k}\over (2\pi)^3}\Big({1\over
E_\Delta^-}\Theta(E_\Delta^--|\mu_I|/2)\nonumber\\
&&\ \ \ \ +{1\over
E_\Delta^+}\Theta(E_\Delta^+-|\mu_I|/2)\Big)\Big]= 0\ .
\end{eqnarray}
Similar to the meson mass equations derived in
RPA\cite{vogl,sandy,zhuang1}, we can obtain the diquark mass in
the vacuum by the pole equation
\begin{equation}
\label{di10}
1 - 8 G_D\int{d^3{\bf k}\over (2\pi)^3}\left({1\over
E_k+m_D/2}+{1\over E_k-m_D/2}\right)= 0\ .
\end{equation}

At sufficiently  small $\mu_B$ and large $\mu_I$, the system
should be in the pion superfluidity phase. The chiral and pion
condensates $\sigma$ and $\pi$ are controlled by the gap equations
shown in Section \ref{s2}. From the comparison of them at zero
temperature with the pion mass equation in the vacuum, and the
comparison of the gap equations (\ref{di9}) with the diquark mass
equation (\ref{di10}), we get the conclusion that in the $\mu_B
-\mu_I$ plane the normal phase without pion and diquark
condensates is in the rectangle defined by
\begin{eqnarray}
\label{di11}
&& \mu_I < m_\pi\ ,\nonumber\\
&& \mu_B < {3\over 2}m_D\ ,
\end{eqnarray}
shown in Fig.(\ref{fig22}). Outside this rectangle, we know that
the system is in color superconductivity phase with $\Delta\neq 0$
and $\pi=0$ for $\mu_I<m_\pi$ and $\mu_B
> {3\over 2} m_D$, and in pion superfluidity phase with $\pi\neq
0$ and $\Delta =0$ for $\mu_B <{3\over 2}m_D$ and $\mu_I >m_\pi$.
As for the phase structure in the region with high baryon and
isospin chemical potentials, it needs to consider the gap
equations for the three condensates $\sigma$, $\pi$ and $\Delta$
simultaneously. The dashed line in the lower panel of
Fig.(\ref{fig22}) which separates the phase with $\pi=0$ and $
\Delta\neq 0$ from the phase $\pi\neq 0$ and $\Delta=0$ is just an
estimation of us.

\section {Extension to $SU(3)$ NJL Model}
\label{s8}
We examine now the possible pion and kaon superfluidity at finite
isospin and strangeness chemical potentials in the frame of flavor
$SU(3)$ NJL model. The Lagrangian density is defined as
\cite{vogl,sandy,volkov,hatsuda}
\begin{eqnarray}
\label{su31} {\cal
L}&=&\bar{\psi}(i\gamma^{\mu}\partial_{\mu}-m_{0})\psi+G\sum_{a=0}^{8}[(\bar{\psi}\lambda_a\psi)^{2}
+(\bar{\psi}i\gamma_{5}\lambda_a\psi)^{2}]\nonumber\\
&-&K[\det\bar{\psi}(1+\gamma_{5})\psi+\det\bar{\psi}
(1-\gamma_{5})\psi]\ ,
\end{eqnarray}
where $m_0=diag(m_{0u},m_{0d},m_{0s})$ is the mass matrix of
current quarks, $G$ and $K$ are coupling constants, and the
t'Hooft's determinant includes six-fermion interaction. The three
flavor NJL Lagrangian can be brought into an effective form
similar to (\ref{mf5}) of the two flavor NJL model by writing the
six-fermion interaction in an effective four-body
form\cite{vogl,sandy,volkov,hatsuda},
\begin{eqnarray}
\label{su32} {\cal
L}_{eff}&=&\bar{\psi}(i\gamma^{\mu}\partial_{\mu}-m_{0})\psi\\
&+&\sum_{i=0}^8 \left[G_i^-
(\bar{\psi}\lambda^i\psi)^2+G_i^+(\bar{\psi}
i\gamma_5\lambda^i\psi)^2\right]\nonumber\\
&+&\left[G_{03}^-(\bar{\psi}\lambda^0\psi)(\bar{\psi}\lambda^3\psi)
+G_{03}^+(\bar{\psi}i\gamma_5\lambda^0\psi)(\bar{\psi}i\gamma_5\lambda^3\psi)\right]
\nonumber\\
&+&\left[G_{30}^-(\bar{\psi}\lambda^3\psi)(\bar{\psi}\lambda^0\psi)
+G_{30}^+(\bar{\psi}i\gamma_5\lambda^3\psi)(\bar{\psi}i\gamma_5\lambda^0\psi)\right]
\nonumber\\
&+&\left[G_{08}^-(\bar{\psi}\lambda^0\psi)(\bar{\psi}\lambda^8\psi)
+G_{08}^+(\bar{\psi}i\gamma_5\lambda^0\psi)(\bar{\psi}i\gamma_5\lambda^8\psi)\right]
\nonumber\\
&+&\left[G_{80}^-(\bar{\psi}\lambda^8\psi)(\bar{\psi}\lambda^0\psi)
+G_{80}^+(\bar{\psi}i\gamma_5\lambda^8\psi)(\bar{\psi}i\gamma_5\lambda^0\psi)\right]
\nonumber\\
&+&\left[G_{38}^-(\bar{\psi}\lambda^3\psi)(\bar{\psi}\lambda^8\psi)
+G_{38}^+(\bar{\psi}i\gamma_5\lambda^3\psi)(\bar{\psi}i\gamma_5\lambda^8\psi)\right]
\nonumber\\
&+&\left[G_{83}^-(\bar{\psi}\lambda^8\psi)(\bar{\psi}\lambda^3\psi)
+G_{83}^+(\bar{\psi}i\gamma_5\lambda^8\psi)(\bar{\psi}i\gamma_5\lambda^3\psi)\right]\
,\nonumber
\end{eqnarray}
with the effective couplings
\begin{eqnarray}
\label{su33}
&&G_0^\pm=G\mp\frac{1}{3}N_cK(\sigma_u+\sigma_d+\sigma_s)\ ,\nonumber\\
&&G_1^\pm=G_2^\pm=G_3^\pm=G\pm\frac{1}{2}N_cK\sigma_s\
,\nonumber\\
&&G_4^\pm=G_5^\pm=G\pm\frac{1}{2}N_cK\sigma_d\ ,\nonumber\\
&&G_6^\pm=G_7^
\pm=G\pm\frac{1}{2}N_cK\sigma_u\ ,\nonumber\\
&&G_8^\pm=G\pm\frac{1}{6}N_cK(2\sigma_u+2\sigma_d-\sigma_s)\ ,\nonumber\\
&&G_{03}^\pm=G_{30}^\pm=\mp\frac{1}{2\sqrt{6}}N_cK(\sigma_u-\sigma_d)\
,\nonumber\\
&&G_{08}^\pm=G_{80}^\pm=\pm\frac{\sqrt{2}}{12}N_cK(\sigma_u+\sigma_d-2\sigma_s)\
,\nonumber\\
&&G_{38}^\pm=G_{83}^\pm=\pm\frac{1}{2\sqrt{3}}N_cK(\sigma_u-\sigma_d)\
,
\end{eqnarray}
where $\sigma_s = \langle\bar s s\rangle$ is the $s$ quark
condensate. The key thermodynamic quantity, the partition function
$Z$ in three flavor case, is defined by
\begin{eqnarray}
\label{su34} Z(T,\mu_I,\mu_B,\mu_S,V)= \text{Tr}
e^{-\beta\left(H-\mu_B B-\mu_I I_3-\mu_S S\right)}\ ,
\end{eqnarray}
with baryon number, isospin number and strangeness number
\begin{eqnarray}
\label{su35} && B=\frac{1}{3}\int d^3{\bf x}\bar{\psi}\gamma^0\psi\
,\nonumber\\
&& I_3=\frac{1}{2}\int d^3{\bf x}\bar{\psi}\gamma^0\lambda_3\psi\
,\nonumber\\
&& S=-\int d^3{\bf x}\bar{s}\gamma^0 s\ ,
\end{eqnarray}
as conserved charges, and baryon, isospin and strangeness chemical
potentials $\mu_B, \mu_I$ and $\mu_S$. In the frame of imaginary
time formulism of finite temperature field theory, the partition
function can be represented as
\begin{eqnarray}
\label{su36} &&
Z(T,\mu_I,\mu_B,\mu_S,V)=\nonumber\\
&&\ \ \ \ \ \int[d\bar{\psi}][d\psi]e^{\int_{0}^{\beta}d\tau\int
d^{3}{\bf x}\left({\cal
L}_{eff}+\bar{\psi}\mu\gamma_{0}\psi\right)}\ ,
\end{eqnarray}
where $\mu=diag(\mu_u,\mu_d,\mu_s)$ is the chemical potential
matrix in flavor space with the chemical potential for the
strangeness quark
\begin{equation}
\label{su37} \mu_{s}=\frac{\mu_B}{3}-\mu_S\ .
\end{equation}

\subsection {Chiral Properties at Low $\mu_I$ and $\mu_S$ }
\begin{figure}
\centering \includegraphics[width=2.5in]{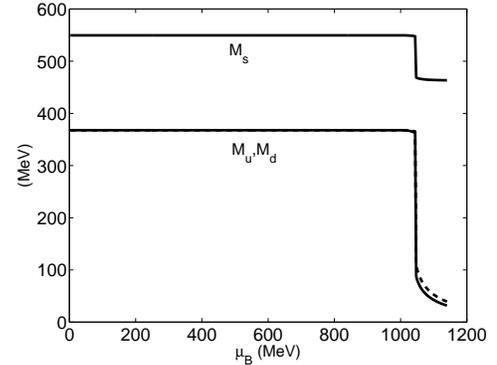} \caption{The
three effective quark masses $M_u, M_d$ and $M_s$ as functions of
baryon chemical potential $\mu_B$ at $T=\mu_S=0$ and $\mu_I=0.06$
GeV in the flavor SU(3) NJL model. } \label{fig23}
\end{figure}

We discuss first the three chiral condensates $\sigma_u, \sigma_d$
and $\sigma_s$ under the condition of low isospin and low
strangeness chemical potentials, $\mu_I < \mu_I^c$ and $\mu_S <
\mu_S^c$, where $\mu_I^c$ is the critical isospin chemical
potential for pion condensation and $\mu_S^c$ the critical
strangeness chemical potential for kaon condensation. They will be
determined in the following subsection. Performing the standard
mean field approach and keeping only the linear terms in the meson
fluctuations, we obtain the Lagrangian in the mean field
approximation
\begin{eqnarray}
\label{su38} {\cal
L}_{mf}&=&\bar{\psi}(i\gamma^{\mu}\partial_{\mu}+\mu\gamma_0-M)\psi\nonumber\\
&-&2G(\sigma_u^2+\sigma_d^2+\sigma_s^2)-4K\sigma_u\sigma_d\sigma_s\
,
\end{eqnarray}
where $M=(M_u, M_d, M_s)$ is the mass matrix in flavor space with
the effective quark masses
\begin{eqnarray}
\label{su39} &&M_i=m_{0i}-4G\sigma_i+2K\sigma_j\sigma_k\
,\nonumber\\
&&\ \ \ \ \ (i=u,d,s,\ i\neq j \neq k,\ j<k)\ .
\end{eqnarray}
Because of the lack of pion and kaon condensates, the mean field
quark propagator is diagonal in flavor space,
\begin{eqnarray}
\label{su310} {\cal S}_{mf}(k)=diag\Big({\cal S}_{u}(k),{\cal
S}_{d}(k),{\cal S}_{s}(k)\Big)\ ,
\end{eqnarray}
with the matrix elements
\begin{eqnarray}
\label{su311} {\cal
S}_{i}(k)=\frac{\Lambda_+^i\gamma_0}{k_0-E_i^-({\bf k})}
+\frac{\Lambda_-^i\gamma_0}{k_0+E_i^+({\bf k})}\ ,
\end{eqnarray}
where $E_i^\pm$ are the effective quark energies
\begin{eqnarray}
\label{su312} E_i^\pm({\bf k})=\sqrt{{\bf k}^2+M_i^2}\pm\mu_i\ ,
\end{eqnarray}
and $\Lambda_\pm^i$  the energy projectors
\begin{equation}
\label{su313} \Lambda_{\pm}^{i} = {1\over
2}\left(1\pm{\gamma_0\left({\bf \gamma\cdot k}+M_{i}\right)\over
\sqrt{{\bf k}^2+M_i^2}}\right)\ .
\end{equation}
In self-consistent Hatree-Fock approximation  the gap equations
which determine the value of the chiral condensates
$\sigma_u,\sigma_d,\sigma_s$ are expressed in terms of the quark
propagators,
\begin{eqnarray}
\label{su314} \sigma_i = -iN_c\int\frac{d^4p}{(2\pi)^4}
\text{Tr}_D{\cal S}_{i}(p)\ .
\end{eqnarray}
After performing the Matsubara frequency summation, we have
\begin{equation}
\label{su315} \sigma_i = -6\int\frac{d^3{\bf
k}}{(2\pi)^3}\frac{M_i}{\sqrt{k^2+M_i^2}}
\left(1-f(E_i^-)-f(E_i^+)\right)\ .
\end{equation}
The three effective quark masses are shown in Fig.(\ref{fig23}) as
functions of baryon chemical potential $\mu_B$ at $T=\mu_S=0$ and
$\mu_I = 0.06$ GeV $< m_\pi$. We use the model parameters fixed in
\cite{rehberg}. The three masses keep their vacuum values till the
common critical point where they suddenly jump down. The mass
difference between $u$ and $d$ quarks arisen from the finite
isospin chemical potential can only be seen after the chiral
transition.

In principle the meson mass spectra can be evaluated by solving
the corresponding pole equation like (\ref{meson6}) for two flavor
NJL model. The meson polarization functions in flavor space are
evaluated in Appendix \ref{a4}. In general case with nonzero pion
and kaon condensates, the off diagonal elements in the
polarization function matrix in flavor space makes it complicated
to solve the pole equation. However, in the region without pion
and kaon condensates we are interested in here, the off diagonal
elements disappear and the pole equation is reduced to
\begin{equation}
\label{su316} 1-2G_i^\pm\Pi_{MM}(k_0=M_M,{\bf k=0}) =0\ ,
\end{equation}
with $G_1^-$ for $M=a^0,a^+,a^-$, $G_4^-$ for
$M=\kappa^+,\kappa^-$, $G_6^-$ for $\kappa^0,\bar{\kappa}^0$,
$G_1^+$ for $M=\pi^0,\pi^+,\pi^-$, $G_4^+$ for $M=K^+,K^-$ and
$G_6^+$ for $K^0,\bar{K}^0$. At zero temperature, we can
analytically obtain the meson masses as functions of $\mu_I$ and
$\mu_S$ by comparing the above mass equations at finite $\mu_I$
and $\mu_S$ with the same equations but in the vacuum,
\begin{eqnarray}
\label{su317} M_M(\mu_I,\mu_S)=m_{M}-Q_I\mu_I-Q_S\mu_S\ ,
\end{eqnarray}
where $Q_I$ and $Q_S$ are the isospin number and strangeness
number of the meson $M$, respectively.

\subsection {Phase Diagram in $\mu_S-\mu_I$ plane}
In this subsection, we analytically determine the critical isospin
chemical potential $\mu_I^c$ for pion condensation and the
critical strangeness chemical potential $\mu_S^c$ for kaon
condensation in the three flavor NJL model, and briefly discuss
the phase structure in $\mu_S-\mu_I$ plane. For simplicity, we
always set $T=\mu_B=0$ in this subsection.

We first consider the case with only pion condensation in the
region of $|\mu_S|<\mu_S^C$ and determine the critical value
$\mu_I^c$. Taking the standard mean field approximation, we obtain
the inverse of the quark propagator
\begin{eqnarray}
\label{su318}
{\cal S}_{mf}^{-1}(k)&=&\left(\begin{array}{ccc}
G_u^{-1}&2iG_1^
+\gamma_5\pi&0\\
2iG_1^+\gamma_5\pi&G_d^{-1}&0\\0&0&G_s^{-1}\end{array}\right)\ ,
\end{eqnarray}
with $G_i^{-1}=\gamma^\mu k_\mu+\mu_i \gamma_0-M_i$ for $i=u, d,
s$. Since in the current case the pion condensate is decoupled
from the $s$ quark propagator, we can take the same procedure as
in the two flavor case and obtain the same critical isospin
chemical potential
\begin{equation}
\label{su319} \mu_I^c = m_\pi\ \ \ \ (\mu_S < \mu_S^c)\ .
\end{equation}

We then study the case with only kaon condensate in the region of
$\mu_I<m_\pi$. We define the $K^+$ and $K^-$ condensates
\begin{eqnarray}
\label{su320}
K_{us} =2\langle\bar{u} i\gamma_5 s\rangle=
2\langle\bar{s} i\gamma_5 u\rangle\ ,
\end{eqnarray}
and $K^0,\bar{K}^0$ condensates
\begin{eqnarray}
\label{su321}
K_{ds} = 2\langle\bar{d} i\gamma_5 s\rangle=
2\langle\bar{s} i\gamma_5 d\rangle\ .
\end{eqnarray}
Which kind of kaon condensation happens depends on the sign of
$\mu_S$ and $\mu_I$. In the physical world with $\mu_I<0$, only
$K^0$ and $K^-$ condensates can be realized. Without losing
generality, we consider here the case $\mu_I>0$ and $\mu_S>0$
only, and then map the result to the other regions in the
$\mu_S-\mu_I$ plane. In this case the possible kaon condensation
is $K^+$.

At mean field level the inverse quark propagator in the flavor
space can be expressed as
\begin{eqnarray}
\label{su322} {\cal S}_{mf}^{-1}(k)&=&\left(\begin{array}{ccc}
G_u^{-1}&2iG_4^
+K_{us}\gamma_5&0\\
2iG_4^+K_{us}\gamma_5&G_s^{-1}&0\\0&0&G_d^{-1}\end{array}\right)\
.
\end{eqnarray}
For convenience, we have changed the basis in the flavor space
from $(u,\ d,\ s)$ in (\ref{su318}) to $(u,\ s,\ d)$ in
(\ref{su322}). It is straightforward to derive the propagator
itself from its inverse,
\begin{eqnarray}
\label{njl323}
{\cal S}_{mf}(k)&=&\left(\begin{array}{ccc} {\cal S}_u(k)&{\cal S}_{us}(k)&0\\
{\cal S}_{su}(k)&{\cal S}_s(k)&0\\0&0&{\cal
S}_d(k)\end{array}\right)\ ,
\end{eqnarray}
with the elements
\begin{eqnarray}
\label{njl324}
{\cal S}_u(k)&=&\Big[(\gamma^\mu k_\mu+\mu_u
\gamma_0-M_u)\\
&+&4(G_4^+)^2K_{us}^2 \gamma_5(\gamma^\mu k_\mu+\mu_s
\gamma_0-M_s)^{-1}\gamma_5\Big]^{-1}\ ,\nonumber\\
{\cal S}_d(k)&=&(\gamma^\mu k_\mu+\mu_d
\gamma_0-M_d)^{-1}\ ,\nonumber\\
{\cal S}_s(k)&=&\Big[(\gamma^\mu k_\mu+\mu_s
\gamma_0-M_s)\nonumber\\
&+&4(G_4^+)^2K_{us}^2\gamma_5 (\gamma^\mu k_\mu+\mu_u
\gamma_0-M_u)^{-1}\gamma_5\Big]^{-1}\ ,\nonumber\\
{\cal S}_{us}(k)&=&-2iG_4^+K_{us}{\cal S}_u(k)\gamma_5(\gamma^\mu
k_\mu+\mu_s \gamma_0-M_s)^{-1}
\ ,\nonumber\\
{\cal S}_{su}(k)&=&-2iG_4^+K_{us}{\cal S}_s(k)\gamma_5(\gamma^\mu
k_\mu+\mu_u \gamma_0-M_u)^{-1}\ .\nonumber
\end{eqnarray}
With the propagator we can express the condensates as
\begin{eqnarray}
\label{njl325} \sigma_i &=& -N_c\int {d^4k\over (2\pi)^4}
\text{Tr}_D \left[i ({\cal S}_{i}(k)\right]\ , \ \ \ \ (i=u,d,s)
\nonumber\\
K_{us} &=& N_c\int {d^4 k\over (2\pi)^4} \text{Tr}_D
\left[\left({\cal S}_{us}(k)+{\cal S}_{su}(k)\right)\gamma_5\right]\
.
\end{eqnarray}
To calculate the critical strangeness chemical potential, we set
$K_{us}\rightarrow0^+$, and the gap equations on the phase
transition line become
\begin{eqnarray}
\label{njl326} && \sigma_i +iN_c\int {d^4k\over (2\pi)^4}
\text{Tr}_D\frac{1}{\gamma^\mu k_\mu+\mu_i \gamma_0-M_i}=0\ ,\\
&&\ \ \ \ \ \ \ \ \ \ \ \ \ \ \ \ \ \ \ \ \ \ \ \ \ \ \ \ \ \ \ \ \ \ \ \ (i=u,d,s)\ ,\nonumber\\
&& 1+ 2N_ciG_4^{+}\int {d^4 k\over (2\pi)^4}\nonumber\\
&& \text{Tr}_D \left[\frac{1}{\gamma^\mu k_\mu+\mu_u
\gamma_0-M_u}\gamma_5\frac{1}{\gamma^\mu k_\mu+\mu_s
\gamma_0-M_s}\gamma_5\right]=0\ .\nonumber
\end{eqnarray}
Doing the trace in Dirac space, performing the Matsubara frequency
summation, and then comparing the gap equations with the mass
equations for kaons in the vacuum, we find
\begin{eqnarray}
\label{njl327}
\mu_S^c=m_K-\mu_I/2\ ,\ \ \ \ (\mu_I < m_\pi)\ .
\end{eqnarray}
This result is certainly independent of the model parameters and
the regularization scheme. The above analytic conclusions
$\mu_I^c=m_\pi$ and $\mu_S^c=m_K-\mu_I/2$ were obtained
numerically in the three flavor NJL model without $U_A(1)$
breaking term\cite{bard3f}.
\begin{figure}
\centering \includegraphics[width=2.5in]{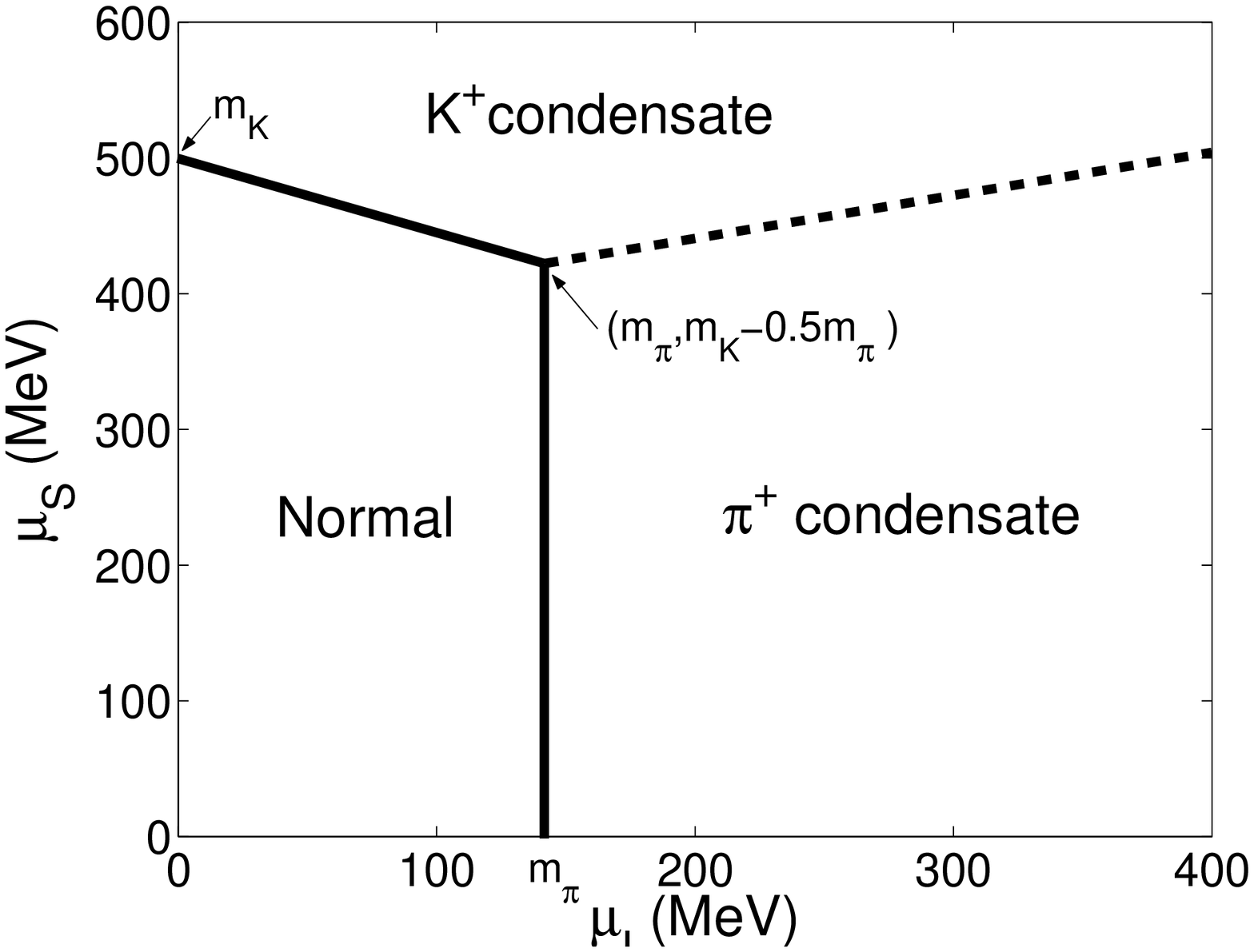}%
\hspace{0.5in}%
\includegraphics[width=2.5in]{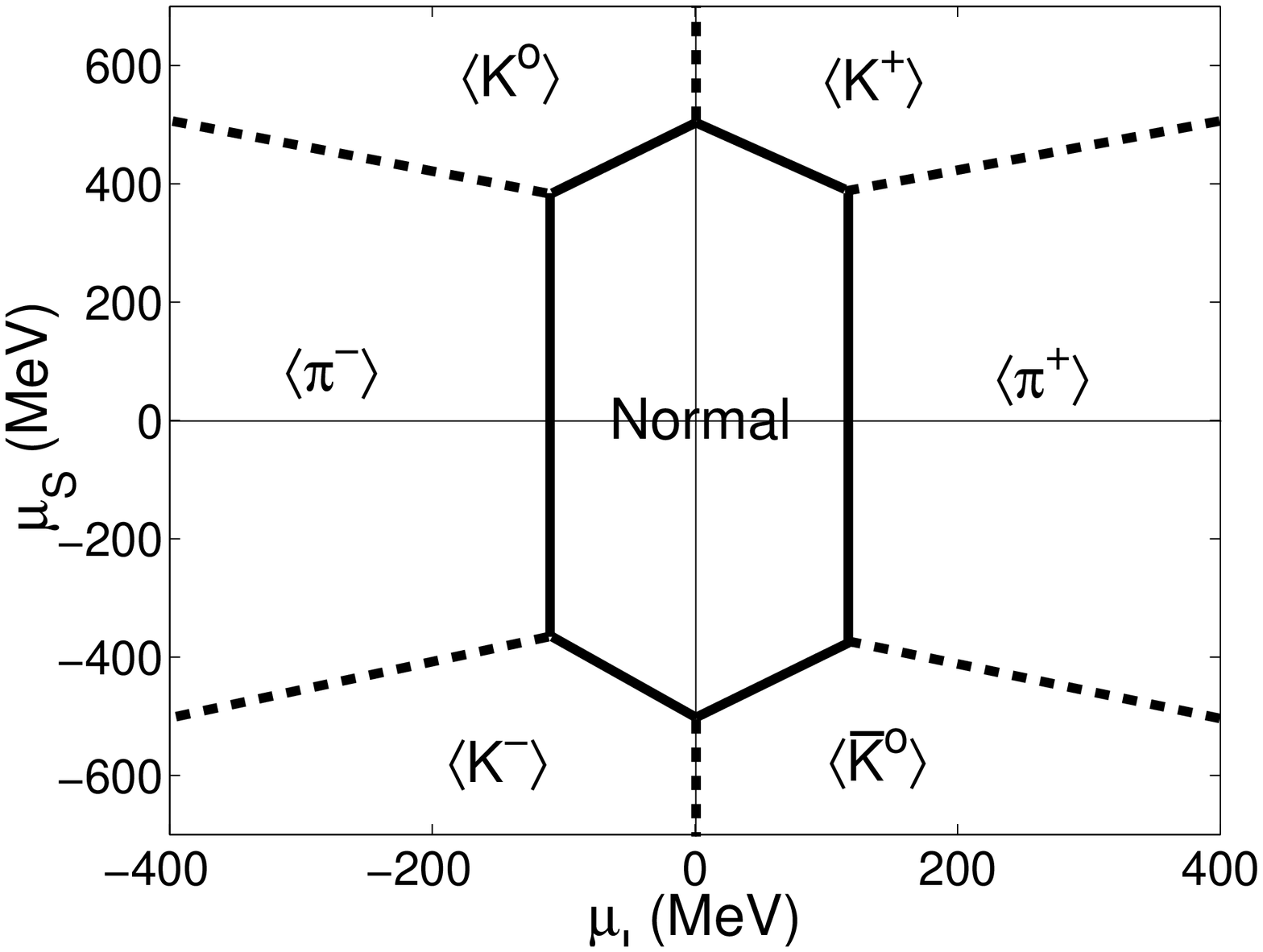}
\caption{The phase structure of pion and kaon condensations in
$\mu_S-\mu_I$ plane at $T=\mu_B=0$ in the flavor $SU(3)$ NJL
model. The upper panel is for $\mu_I >0$ and $\mu_S>0$ and the
lower panel is for all possible $\mu_I$ and $\mu_S$.}
\label{fig24}
\end{figure}

Combining the two critical chemical potentials $\mu_I$ and
$\mu_S$, the system is in the normal phase without pion and kaon
condensations when $\mu_I$ and $\mu_S$ satisfy the constraints
\begin{eqnarray}
\label{njl328}
&& \mu_I < m_\pi\ ,\nonumber\\
&& \mu_S < m_K-\mu_I/2\ .
\end{eqnarray}
In the region of $\mu_I <m_\pi$ and $\mu_S > m_K -\mu_I/2$, there
is only $K^+$ condensation, and in the region of $\mu_S <
m_K-\mu_I/2$ and $\mu_I > m_\pi$ there is only pion condensation.
The phase transition lines in the $\mu_S-\mu_I$ plane are
indicated by solid lines in Fig.(\ref{fig24}). As for the other
region in the $\mu_S - \mu_I$ plane we should consider both pion
and kaon condensations and the calculation will become much
complicated. The dashed line in Fig.(\ref{fig24}) which separates
the region with pion condensate from the region with kaon
condensate is just an estimation. However, with the bosonized
chiral Lagrangian, this dashed line is described by\cite{kogut4}
\begin{equation}
\mu_S^C\left(|\mu_I|>m_\pi\right)=\frac{1}{2\mu_I}\left[\sqrt{(\mu_I^2-m_\pi^2)^2
+4m_K^2\mu_I^2}-m_\pi^2\right]\
.
\end{equation}
Considering the symmetries between $\pi_+$ and $\pi_-$ and among
$K^+, K^-, K^0$ and $\bar K^0$, we display in the lower panel of
Fig.(\ref{fig24}) the phase diagram with different pion and kaon
condensations, which is the same as the one obtained in the frame
of bosonized chiral lagrangian\cite{kogut4}.

\section {Summary}
\label{s9}
It is well-known that physical symmetries of a system dominate its
behavior not only in the vacuum but also in hot and dense medium.
The symmetries in the vacuum are usually changed by the
temperature and density effect. In this paper we have investigated
the changes of isospin symmetry, chiral symmetry and color
symmetry and their reflection in meson spectra and thermodynamic
functions at finite temperature $T$ and isospin, baryon and
strangeness chemical potentials $\mu_I, \mu_B$ and $\mu_S$ in
effective models.

We studied the flavor $SU(2)$ NJL model in detail. With the
standard technics we set up in flavor space the quark propagator
matrix with off diagonal elements arisen from the pion
condensation at finite isospin density. With the propagator we
derived the gap equations determining the temperature and chemical
potential dependence of the pion and chiral condensates, and
obtained the thermodynamic functions in mean field approximation.
By self-consistently solving the gap equations at finite isospin
density and the pion mass equation in the vacuum, we analytically
derived the critical isospin chemical potential for pion
superfluidity, $\mu_I^c=m_\pi$, at quark level. Therefore, in
chiral limit with vanishing pion mass, any small isospin effect
will lead to chiral symmetry restoration and isospin symmetry
spontaneous breaking even in the vacuum. In real world with finite
current quark mass and at finite temperature, there is a strong
competition between the pion and chiral condensates in their
coexistence region. The chiral condensate goes up continuously in
the mixed region and drops down outside, while the pion condensate
decreases monotonously in the whole temperature region. The bulk
properties of the system in the limit of high isospin density is
also quite different from the corresponding ideal gas, the
equation of state is hard and the ration of pressure to energy
density approaches $0.7$. We investigated the effect of pion
superfluidity on the meson properties in the hot and dense
quark-meson plasma, by taking into account the off diagonal
elements in the meson polarization function matrix in isospin
space. We found that the mean field approximation to quarks
together with RPA to mesons can well describe the isospin
spontaneous breaking, there is a massless Goldstone mode which is
a linear combination of the $\pi_+, \pi_-$ and $\sigma$ modes in
the whole superfluidity phase. We proved the Goldstone mode
analytically in general case with finite $T, \mu_I$ and $\mu_B$.
In the superfluidity phase, while the $\pi_0$ mode is still an
eigen mode, the $\sigma, \pi_+$ and $\pi_-$ mixing is very strong,
especially in the region close to the phase transition point. We
discussed also the competition between the pion superfluidity and
color superconductivity at finite baryon density. While we did not
make full calculation with both pion and diquark condensates, we
determined the critical baryon chemical potential $\mu_B^c$ for
color superconductivity. The region of the normal phase without
pion and diquark condensates in $\mu_S-\mu_I$ plane is controlled
by the pion mass $m_\pi$ and diquark mass $m_D$ in the vacuum. In
the case of two colors it is in a square box defined by
$\mu_I<m_\pi$ and $\mu_B<m_\pi$, and in the case of three colors
it is in a rectangle defined by $\mu_I<m_\pi$ and $\mu_B<1.5m_D$.

We briefly extended our discussion on pion superfluidity to the
flavor $SU(3)$ NJL model at finite isospin and strangeness
densities. We found that the conclusion of the critical isospin
chemical potential $\mu_I^c =m_\pi$ for pion superfluidity does
not change, and that the critical strangeness chemical potential
for kaon superfluidity is related not only to the kaon mass $m_K$
but also to the isospin chemical potential, $\mu_S^c=m_K-\mu_I/2$.
We showed the phase diagram of pion and kaon superfluidity in
$\mu_S-\mu_I$ plane with possible charged pion condensates and
kaon condensates.

We have investigated the pion superfluidity and meson properties
at tree level and the thermal excitation at finite isospin density
in the bosonized version of the NJL model, namely, the linear
sigma model, and made comparison of our calculations in the NJL
model and sigma model with the result obtained in chiral
perturbation theory. Many conclusions in the linear sigma model
are qualitatively or even quantitatively consistent with that in
the NJL model. We found that at zero temperature the linear sigma
model can reproduce the result of chiral perturbation theory by
taking the limit $m_\sigma\rightarrow\infty$, but at finite
temperature the Hartree-Fock approximation can not recover the
Goldstone mode in the pion superfluidity phase. The phase diagram
in the $T-\mu_I$ plane in large $N$ expansion approach is more
reasonable than the one in the Hartree-Fock approximation.

It is necessary to discuss the difference among the effective
models. The conclusion $\mu_I^c=m_\pi$ is observed in almost all
model calculations and lattice simulation. While obtaining this
result is easy and looks trivial in the meson effective models
such as chiral perturbation theory and linear sigma model, because
the pion mass is a parameter in these models, it is in the NJL
model a general result of mean field approximation to quarks and
RPA to mesons and diquarks. This is also true for kaon
condensation. In chiral perturbation theory, the chiral and pion
condensates are regarded as a rotation from each other. As a
result, the maximum pion condensate should be the same as the
chiral condensate in the vacuum. In the NJL model, we found that
this rotation point of view is only true around the phase
transition point. Beyond this region the pion condensate can be
larger than the maximal chiral condensate in the vacuum. This
behavior is qualitatively consistent with the lattice
simulation\cite{kogut1,kogut2,kogut3}. As for the linear sigma
model, because the pion condensate increases with $\mu_I$ without
bound, the phase structure at high $\mu_I$ is unphysical. Of
course, due to the lack of confinement mechanism in the NJL model,
we need a momentum cutoff $\Lambda$ in numerical calculations.
Therefore, we can not take the numerical results in high
temperature and high density regions seriously.

{\bf Acknowledgments:}\ The work was supported in part by the
grants NSFC10135030, 10425810, 10435080, G2000077407 and
SRFDP20040003103.

\appendix
\section{ Trace in Dirac Space}
\label{a1}
In obtaining the gap equations for the condensates and the meson
and diquark polarization functions, we need to take the trace of
the quark propagators including energy projectors in Dirac space.
Here we list the results.
\begin{eqnarray}
\text{Tr}\left[\Lambda_+({\bf p})\gamma_0\Lambda_+({\bf
p})\gamma_0\right]&=& \text{Tr}\left[\Lambda_-({\bf
p})\gamma_0\Lambda_-({\bf
p})\gamma_0\right]\nonumber\\
&=&2M^2/E_p^2\ ,\nonumber\\
\text{Tr}\left[\Lambda_+({\bf p})\gamma_0\Lambda_-({\bf
p})\gamma_0\right]&=& \text{Tr}\left[\Lambda_-({\bf
p})\gamma_0\Lambda_+({\bf
p})\gamma_0\right]\nonumber\\
&=&2-2M^2/E_p^2\ ,\nonumber\\
\text{Tr}\left[\Lambda_+({\bf p})\gamma_5\Lambda_+({\bf
p})\gamma_5\right]&=& \text{Tr}\left[\Lambda_-({\bf
p})\gamma_5\Lambda_-({\bf
p})\gamma_5\right]\nonumber\\
&=&2-2M^2/E_p^2\ ,\nonumber\\
\text{Tr}\left[\Lambda_+({\bf p})\gamma_5\Lambda_-({\bf
p})\gamma_5\right]&=& \text{Tr}\left[\Lambda_-({\bf
p})\gamma_5\Lambda_+({\bf
p})\gamma_5\right]\nonumber\\
&=&2M^2/E_p^2\ ,\nonumber\\
\text{Tr}\left[\Lambda_e({\bf p})\gamma_0\Lambda_{e^\prime}({\bf
p})\gamma_5\right]&=& \text{Tr}\left[\Lambda_e({\bf
p})\gamma_5\Lambda_{e^\prime}({\bf
p})\gamma_0\right]\nonumber\\
&=&0\ ,\nonumber\\
\text{Tr}\left[\gamma_5\Lambda_+({\bf
p})\gamma_0\gamma_5\Lambda_+({\bf p})\gamma_0\right]&=&
\text{Tr}\left[\gamma_5\Lambda_-({\bf
p})\gamma_0\gamma_5\Lambda_-({\bf
p})\gamma_0\right]\nonumber\\
&=&0\ ,\nonumber\\
\text{Tr}\left[\gamma_5\Lambda_+({\bf
p})\gamma_0\gamma_5\Lambda_-({\bf p})\gamma_0\right]&=&
\text{Tr}\left[\gamma_5\Lambda_-({\bf
p})\gamma_0\gamma_5\Lambda_+({\bf
p})\gamma_0\right]\nonumber\\
&=&-2\ ,\nonumber\\
\text{Tr}\left[\gamma_5\Lambda_+({\bf
p})\gamma_5\gamma_5\Lambda_+({\bf p})\gamma_5\right]&=&
\text{Tr}\left[\gamma_5\Lambda_-({\bf
p})\gamma_5\gamma_5\Lambda_-({\bf
p})\gamma_5\right]\nonumber\\
&=&2\ ,\nonumber\\
\text{Tr}\left[\gamma_5\Lambda_+({\bf
p})\gamma_5\gamma_5\Lambda_-({\bf p})\gamma_5\right]&=&
\text{Tr}\left[\gamma_5\Lambda_-({\bf
p})\gamma_5\gamma_5\Lambda_+({\bf
p})\gamma_5\right]\nonumber\\
&=&0\ ,\nonumber\\
\text{Tr}\left[\gamma_5\Lambda_e({\bf
p})\gamma_0\gamma_5\Lambda_{e^\prime}({\bf p})\gamma_5\right]&=&
\text{Tr}\left[\gamma_5\Lambda_e({\bf
p})\gamma_5\gamma_5\Lambda_{e^\prime}({\bf p})\gamma_0\right]\nonumber\\
&=&0\ ,\nonumber\\
\text{Tr}\left[\Lambda_e({\bf
p})\gamma_0\gamma_5\Lambda_{e^\prime}({\bf p})\gamma_0\right]&=&
\text{Tr}\left[\gamma_5\Lambda_e({\bf
p})\gamma_0\Lambda_{e^\prime}({\bf p})\gamma_0\right]\nonumber\\
&=&0\ ,\nonumber\\
\text{Tr}\left[\Lambda_e({\bf
p})\gamma_5\gamma_5\Lambda_{e^\prime}({\bf p})\gamma_5\right]&=&
\text{Tr}\left[\gamma_5\Lambda_e({\bf
p})\gamma_5\Lambda_{e^\prime}({\bf p})\gamma_5\right]\nonumber\\
&=&0\ ,\nonumber\\
\text{Tr}\left[\Lambda_+({\bf p})\gamma_0\gamma_5\Lambda_+({\bf
p})\gamma_5\right]&=& \text{Tr}\left[\Lambda_-({\bf
p})\gamma_0\gamma_5\Lambda_-({\bf
p})\gamma_5\right]\nonumber\\
&=&0\ ,\nonumber\\
\text{Tr}\left[\Lambda_+({\bf p})\gamma_0\gamma_5\Lambda_-({\bf
p})\gamma_5\right]&=&2M/E_p\ , \nonumber\\
\text{Tr}\left[\Lambda_-({\bf p})\gamma_0\gamma_5\Lambda_+({\bf
p})\gamma_5\right]&=&-2M/E_p\ ,\nonumber\\
\text{Tr}\left[\Lambda_+({\bf p})\gamma_5\gamma_5\Lambda_-({\bf
p})\gamma_0\right]&=& \text{Tr}\left[\Lambda_-({\bf
p})\gamma_5\gamma_5\Lambda_+({\bf
p})\gamma_0\right]\nonumber\\
&=&0\ ,\nonumber\\
\text{Tr}\left[\Lambda_+({\bf p})\gamma_5\gamma_5\Lambda_+({\bf
p})\gamma_0\right]&=&2M/E_p\ , \nonumber\\
\text{Tr}\left[\Lambda_-({\bf p})\gamma_5\gamma_5\Lambda_-({\bf
p})\gamma_0\right]&=&-2M/E_p\ ,\nonumber\\
\text{Tr}\left[\gamma_5\Lambda_+({\bf p})\gamma_5\Lambda_+({\bf
p})\gamma_0\right]&=& \text{Tr}\left[\gamma_5\Lambda_-({\bf
p})\gamma_5\Lambda_-({\bf p})\gamma_0\right]\nonumber\\
&=&0\ ,\nonumber\\
\text{Tr}\left[\gamma_5\Lambda_-({\bf
p})\gamma_5\Lambda_+({\bf p})\gamma_0\right]&=&2M/E_p\ , \nonumber\\
\text{Tr}\left[\gamma_5\Lambda_+({\bf
p})\gamma_5\Lambda_-({\bf p})\gamma_0\right]&=&-2M/E_p\ ,\nonumber\\
\text{Tr}\left[\gamma_5\Lambda_-({\bf p})\gamma_0\Lambda_+({\bf
p})\gamma_5\right]&=& \text{Tr}\left[\gamma_5\Lambda_+({\bf
p})\gamma_0\Lambda_-({\bf p})\gamma_5\right]\nonumber\\
&=&0\ ,\nonumber\\
\text{Tr}\left[\gamma_5\Lambda_+({\bf
p})\gamma_0\Lambda_+({\bf p})\gamma_5\right]&=&2M/E_p\ , \nonumber\\
\text{Tr}\left[\gamma_5\Lambda_-({\bf p})\gamma_0\Lambda_-({\bf
p})\gamma_5\right]&=&-2M/E_p\ .
\end{eqnarray}

\section {Polarization Functions in SU(2) NJL Model }
\label{a2}
We list the meson polarization functions $\Pi_{ij}(k)$ in the
SU(2) NJL model for $i,j=\sigma, \pi_1, \pi_2, \pi_3$ and for
$i,j=\sigma, \pi_+, \pi_-, \pi_0$. With respect to the basis
$(\sigma, \pi_1, \pi_2, \pi_3)$, they are
\begin{eqnarray}
\Pi_{00}(k) &=& iN_c\int {d^4p\over (2\pi)^4}\text{Tr}_D \Big({\cal
S}_{uu}(q){\cal S}_{uu}(p)+{\cal S}_{dd}(q)
{\cal S}_{dd}(p)\nonumber\\
&+&{\cal S}_{ud}(q){\cal S}_{du}(p)+{\cal S}_{du}(q){\cal S}_{ud}(p)\Big)\ ,\nonumber\\
\Pi_{11}(k) &=& -iN_c\int {d^4p\over (2\pi)^4}\text{Tr}_D
\Big(\gamma_5{\cal S}_{uu}(q)\gamma_5{\cal S}_{dd}(p)\nonumber\\
&+& \gamma_5{\cal S}_{dd}(q)\gamma_5{\cal S}_{uu}(p)
+\gamma_5{\cal S}_{ud}(q)\gamma_5{\cal S}_{ud}(p)\nonumber\\
&+&\gamma_5{\cal S}_{du}(q)\gamma_5{\cal S}_{du}(p)\Big)\ ,\nonumber\\
\Pi_{22}(k) &=& -iN_c\int {d^4p\over (2\pi)^4} \text{Tr}_D
\Big(\gamma_5{\cal S}_{uu}(q)\gamma_5{\cal S}_{dd}(p)\nonumber\\
&+&\gamma_5{\cal S}_{dd}(q)\gamma_5{\cal S}_{uu}(p)-\gamma_5{\cal
S}_{ud}(q)\gamma_5{\cal S}_{ud}(p)\nonumber\\
&-&\gamma_5{\cal S}_{du}(q)\gamma_5{\cal S}_{du}(p)\Big)\ ,\nonumber\\
\Pi_{33}(k) &=& -iN_c\int {d^4p\over (2\pi)^4} \text{Tr}_D
\Big(\gamma_5{\cal S}_{uu}(q)\gamma_5{\cal S}_{uu}(p)\nonumber\\
&+&\gamma_5{\cal S}_{dd}(q)\gamma_5{\cal S}_{dd}(p)-\gamma_5{\cal
S}_{ud}(q)\gamma_5{\cal S}_{du}(p)\nonumber\\
&-&\gamma_5{\cal S}_{du}(q)\gamma_5{\cal S}_{ud}(p)\Big)\ ,\nonumber\\
\Pi_{01}(k) &=& -N_c\int {d^4p\over (2\pi)^4} \text{Tr}_D \Big({\cal
S}_{uu}(q)\gamma_5{\cal S}_{du}(p)\nonumber\\
&+&{\cal S}_{dd}(q)\gamma_5{\cal S}_{ud}(p)+{\cal
S}_{ud}(q)\gamma_5{\cal S}_{uu}(p)\nonumber\\
&+&{\cal S}_{du}(q)
\gamma_5{\cal S}_{dd}(p)\Big)\ ,\nonumber\\
\Pi_{10}(k) &=& -N_c\int {d^4p\over (2\pi)^4}\text{Tr}_D
\Big(\gamma_5{\cal
S}_{uu}(q){\cal S}_{ud}(p)\nonumber\\
&+&\gamma_5{\cal S}_{dd}(q){\cal S}_{du}(p)+\gamma_5{\cal
S}_{ud}(q){\cal S}_{dd}(p)\nonumber\\
&+&\gamma_5{\cal S}_{du}(q){\cal S}_{uu}(p)\Big)\ ,\nonumber\\
\Pi_{02}(k) &=& iN_c\int {d^4p\over (2\pi)^4} \text{Tr}_D \Big({\cal
S}_{uu}(q)\gamma_5{\cal S}_{du}(p)\nonumber\\
&-&{\cal S}_{dd}(q)\gamma_5{\cal S}_{ud}(p)+{\cal
S}_{du}(q)\gamma_5{\cal S}_{dd}(p)\nonumber\\
&-&{\cal S}_{ud}(q)\gamma_5{\cal S}_{uu}(p)\Big)\ ,\nonumber\\
\Pi_{20}(k) &=& iN_c\int {d^4p\over (2\pi)^4} \text{Tr}_D
\Big(\gamma_5{\cal S}_{dd}(q){\cal S}_{du}(p)\nonumber\\
&-&\gamma_5{\cal S}_{uu}(q){\cal S}_{ud}(p)+\gamma_5{\cal
S}_{du}(q){\cal S}_{uu}(p)\nonumber\\
&-&\gamma_5{\cal S}_{ud}(q){\cal S}_{dd}(p)\Big)\ ,\nonumber\\
\Pi_{03}(k) &=& N_c\int {d^4p\over (2\pi)^4} \text{Tr}_D \Big({\cal
S}_{dd}(q)\gamma_5{\cal S}_{dd}(p)\nonumber\\
&-&{\cal S}_{uu}(q)\gamma_5{\cal S}_{uu}(p)+{\cal
S}_{ud}(q)\gamma_5{\cal S}_{du}(p)\nonumber\\
&-&{\cal S}_{du}(q)\gamma_5{\cal S}_{ud}(p)\Big)\ ,\nonumber\\
\Pi_{30}(k) &=& N_c\int {d^4p\over (2\pi)^4} \text{Tr}_D
\Big(\gamma_5{\cal
S}_{dd}(q){\cal S}_{dd}(p)\nonumber\\
&-&\gamma_5{\cal S}_{uu}(q){\cal S}_{uu}(p)+\gamma_5{\cal
S}_{du}(q){\cal S}_{ud}(p)\nonumber\\
&-&\gamma_5{\cal S}_{ud}(q){\cal S}_{du}(p)\Big)\ ,\nonumber\\
\Pi_{12}(k) &=& -N_c\int {d^4p\over (2\pi)^4} \text{Tr}_D
\Big(\gamma_5{\cal S}_{uu}(q)\gamma_5{\cal S}_{dd}(p)\nonumber\\
&-&\gamma_5{\cal S}_{dd}(q)\gamma_5{\cal S}_{uu}(p)+\gamma_5{\cal
S}_{du}(q)\gamma_5{\cal S}_{du}(p)\nonumber\\
&-&\gamma_5{\cal S}_{ud}(q)\gamma_5{\cal S}_{ud}(p)\Big)\ ,\nonumber\\
\Pi_{21}(k) &=& -N_c\int {d^4p\over (2\pi)^4}\text{Tr}_D
\Big(\gamma_5{\cal
S}_{dd}(q)\gamma_5{\cal S}_{uu}(p)\nonumber\\
&-&\gamma_5{\cal S}_{uu}(q)\gamma_5{\cal S}_{dd}(p)+\gamma_5{\cal
S}_{du}(q)\gamma_5{\cal S}_{du}(p)\nonumber\\
&-&\gamma_5{\cal S}_{ud}(q)\gamma_5{\cal S}_{ud}(p)\Big)\
,\nonumber\\
\Pi_{13}(k) &=& iN_c\int {d^4p\over (2\pi)^4} \text{Tr}_D
\Big(\gamma_5{\cal S}_{uu}(q)\gamma_5{\cal S}_{ud}(p)\nonumber\\
&-&\gamma_5{\cal S}_{dd}(q)\gamma_5{\cal S}_{du}(p)+\gamma_5{\cal
S}_{du}(q)\gamma_5{\cal S}_{uu}(p)\nonumber\\
&-&\gamma_5{\cal S}_{ud}(q)\gamma_5{\cal S}_{dd}(p)\Big)\ ,\nonumber\\
\Pi_{31}(k) &=& iN_c\int {d^4p\over (2\pi)^4} \text{Tr}_D
\Big(\gamma_5{\cal
S}_{uu}(q)\gamma_5{\cal S}_{du}(p)\nonumber\\
&-&\gamma_5{\cal S}_{dd}(q)\gamma_5{\cal S}_{ud}(p)+\gamma_5{\cal
S}_{ud}(q)\gamma_5{\cal S}_{uu}(p)\nonumber\\
&-&\gamma_5{\cal S}_{du}(q)\gamma_5{\cal S}_{dd}(p)\Big)\
,\nonumber\\
\Pi_{23}(k) &=& N_c\int {d^4p\over (2\pi)^4} \text{Tr}_D
\Big(\gamma_5{\cal
S}_{du}(q)\gamma_5{\cal S}_{uu}(p)\nonumber\\
&+&\gamma_5{\cal S}_{ud}(q)\gamma_5{\cal S}_{dd}(p)-\gamma_5{\cal
S}_{uu}(q)\gamma_5{\cal S}_{ud}(p)\nonumber\\
&-&\gamma_5{\cal S}_{dd}(q)\gamma_5{\cal S}_{du}(p)\Big)\ ,\nonumber\\
\Pi_{32}(k) &=& N_c\int {d^4p\over (2\pi)^4} \text{Tr}_D
\Big(\gamma_5{\cal
S}_{uu}(q)\gamma_5{\cal S}_{du}(p)\nonumber\\
&+&\gamma_5{\cal S}_{dd}(q)\gamma_5{\cal S}_{ud}(p)-\gamma_5{\cal
S}_{ud}(q)\gamma_5{\cal S}_{uu}(p)\nonumber\\
&-&\gamma_5{\cal S}_{du}(q)\gamma_5{\cal S}_{dd}(p)\Big)\ ,
\end{eqnarray}
where the momentum $q$ is defined as $q=p+k$.

For determining the meson masses in RPA, we need only the
polarization functions at ${\bf k} =0$. With the calculated trace
shown in Appendix 1, we have
\begin{eqnarray}
\Pi_{03}=\Pi_{30}=\Pi_{13}=\Pi_{31}=\Pi_{23}=\Pi_{32}=0.
\end{eqnarray}
If we further set $k_0=0$ which is used to determine the
temperature and chemical potentials corresponding to the zero
meson mass, we obtain
\begin{eqnarray}
\Pi_{02}=\Pi_{20}=\Pi_{12}=\Pi_{21}=0.
\end{eqnarray}
After performing the Matsubara frequency summation we derive the
expressions as explicit functions of temperature and chemical
potentials,
\begin{eqnarray}
&&\Pi_{22}(k_0=0) = -2N_c\int{d^3{\bf p}\over
(2\pi)^3}\Big[\frac{1}{E_p^-}\left(f(E^-_-)-f(-E^-_+)\right)\nonumber\\
&&\ \ \ \ +\frac{1}{E_p^+}\left(f(E^+_-)-f(-E^+_+)\right)\Big]\
,\\
&&\Pi_{33}(k_0) = 2N_c\int{d^3{\bf p}\over
(2\pi)^3}\nonumber\\
&&\ \ \ \ \Bigg[{E_p^-
E_p^+-\left(E_p^2-\mu_I^2/4\right)-4G^2\pi^2\over k_0^2 -
\left(E_p^- - E_p^+\right)^2}\left(\frac{1}{E_p^+} -\frac{1}{
E_p^-}\right)\nonumber\\
&&\ \ \ \ \left(f(E^-_-)+f(-E^+_+)-f(-E^-_+)-f(E^+_-)\right)\nonumber\\
&&+{E_p^- E_p^++\left(E_p^2-\mu_I^2/4\right)+4G^2\pi^2\over k_0^2
- \left(E_p^- + E_p^+\right)^2}\left(\frac{1}{E_p^+} +\frac{1}{
E_p^-}\right)\nonumber\\
&&\ \ \ \ \left(f(E^-_-)+f(E^+_-)-f(-E^-_+)-f(-E^+_+)\right)\Bigg]\
.\nonumber
\end{eqnarray}

The meson polarization functions in the basis $(\sigma, \pi_+,
\pi_-, \pi_0)$ are defined as
\begin{eqnarray}
\Pi_{\sigma\sigma}(k) &=& iN_c\int {d^4p\over (2\pi)^4} \text{Tr}_D
\Big({\cal S}_{uu}(q){\cal S}_{uu}(p)+{\cal S}_{dd}(q){\cal
S}_{dd}(p)\nonumber\\
&+&{\cal S}_{ud}(q){\cal S}_{du}(p)+{\cal S}_{du}(q){\cal S}_{ud}(p)\Big)\ ,\nonumber\\
\Pi_{\pi_+\pi_+}(k) &=& -2iN_c\int {d^4p\over (2\pi)^4} \text{Tr}_D
\Big(\gamma_5{\cal S}_{uu}(p+k)\gamma_5{\cal S}_{dd}(p)\Big)\ ,\nonumber\\
\Pi_{\pi_-\pi_-}(k) &=& -2iN_c\int {d^4p\over (2\pi)^4} \text{Tr}_D
\Big(\gamma_5{\cal S}_{dd}(p+k)\gamma_5{\cal S}_{uu}(p)\Big)\ ,\nonumber\\
\Pi_{\pi_+\pi_-}(k) &=& -2iN_c\int {d^4p\over (2\pi)^4} \text{Tr}_D
\Big(\gamma_5{\cal S}_{ud}(p+k)\gamma_5{\cal S}_{ud}(p)\Big)\ ,\nonumber\\
\Pi_{\pi_-\pi_+}(k) &=& -2iN_c\int {d^4p\over (2\pi)^4} \text{Tr}_D
\Big(\gamma_5{\cal S}_{du}(p+k)\gamma_5{\cal S}_{du}(p)\Big)\ ,\nonumber\\
\Pi_{\sigma\pi_+}(k) &=& -\sqrt{2}N_c\int {d^4p\over (2\pi)^4}
\text{Tr}_D
\Big({\cal S}_{uu}(p+k)\gamma_5{\cal S}_{du}(p)\nonumber\\
&+&{\cal
S}_{du}(p+k)\gamma_5{\cal S}_{dd}(p)\Big)\ ,\nonumber\\
\Pi_{\sigma\pi_-}(k) &=& -\sqrt{2}N_c\int {d^4p\over (2\pi)^4}
\text{Tr}_D
\Big({\cal S}_{ud}(p+k)\gamma_5{\cal S}_{uu}(p)\nonumber\\
&+&{\cal
S}_{dd}(p+k)\gamma_5{\cal S}_{ud}(p)\Big)\ ,\nonumber\\
\Pi_{\pi_+\sigma}(k) &=& -\sqrt{2}N_c\int {d^4p\over (2\pi)^4}
\text{Tr}_D \Big(\gamma_5{\cal S}_{uu}(p+k){\cal
S}_{ud}(p)\nonumber\\
&+&\gamma_5{\cal
S}_{ud}(p+k){\cal S}_{dd}(p)\Big)\ ,\nonumber\\
\Pi_{\pi_-\sigma}(k) &=& -\sqrt{2}N_c\int {d^4p\over (2\pi)^4}
\text{Tr}_D \Big(\gamma_5{\cal S}_{du}(p+k){\cal
S}_{uu}(p)\nonumber\\
&+&\gamma_5{\cal S}_{dd}(p+k){\cal S}_{du}(p)\Big)\ .
\end{eqnarray}
The explicit expressions as functions of temperature and chemical
potentials at ${\bf k}=0$ can be written as
\begin{eqnarray}
&&\Pi_{\sigma\sigma}(k_0) =  2N_c\int{d^3{\bf p}\over
(2\pi)^3}\frac{{\bf p}^2}{E_p^2}\nonumber\\
&&\ \ \ \ \Bigg[{E_p^-
E_p^+-\left(E_p^2-\mu_I^2/4\right)-4G^2\pi^2\over k_0^2 -
\left(E_p^- - E_p^+\right)^2}\left(\frac{1}{E_p^+} -\frac{1}{
E_p^-}\right)\nonumber\\
&&\ \ \ \ \left(f(E^-_-)+f(-E^+_+)-f(-E^-_+)-f(E^+_-)\right)\nonumber\\
&&\ \ \ \ +{E_p^-
E_p^++\left(E_p^2-\mu_I^2/4\right)+4G^2\pi^2\over k_0^2 -
\left(E_p^- + E_p^+\right)^2}\left(\frac{1}{E_p^+} +\frac{1}{
E_p^-}\right)\nonumber\\
&&\ \ \ \ \left(f(E^-_-)+f(E^+_-)-f(-E^-_+)-f_(-E^+_+)\right)\Bigg]\ ,\nonumber\\
&&\ \ \ \ +4N_c\int{d^3{\bf p}\over
(2\pi)^3}\frac{M_q^2}{E_p^2}\nonumber\\
&&\ \ \ \ \Bigg[\frac{4G^2\pi^2}{k_0^2
-4(E_p^-)^2}\frac{1}{E_p^-}(f(E^-_-)-f(-E^-_+))\nonumber\\
&&\ \ \ \ +\frac{4G^2\pi^2}{k_0^2-4(E_p^+)^2}\frac{1}{E_p^+}
(f(E^+_-)-f(-E^+_+))\Bigg]\ ,\nonumber\\
&&\Pi_{\pi_+\pi_+}(k_0)= 4N_c\int{d^3{\bf p}\over
(2\pi)^3}\nonumber\\
&&\ \ \ \ \Bigg[{\left(E_p^-\right)^2
+\left(E_p-\mu_I/2\right)^2+k_0\left(E_p-\mu_I/2\right)\over k_0^2
-4\left(E_p^-\right)^2}\nonumber\\
&&\ \ \ \ {1\over E_p^-}
\left(f(E^-_-)-f(-E^-_+)\right)\nonumber\\
&&\ \ \ \ +{\left(E_p^+\right)^2
+\left(E_p+\mu_I/2\right)^2-k_0\left(E_p+\mu_I/2\right)\over k_0^2
- 4\left(E_p^+\right)^2}\nonumber\\
&&\ \ \ \ {1\over E_p^+}
\left(f(E^+_-)-f(-E^+_+)\right)\Bigg]\ ,\nonumber\\
&&\Pi_{\pi_-\pi_-}(k_0)=  4N_c\int{d^3{\bf p}\over
(2\pi)^3}\nonumber\\
&&\ \ \ \ \Bigg[{\left(E_p^-\right)^2
+\left(E_p-\mu_I/2\right)^2-k_0\left(E_p-\mu_I/2\right)\over k_0^2
- 4\left(E_p^-\right)^2}\nonumber\\
&&\ \ \ \ {1\over E_p^-}
\left(f(E^-_-)-f(-E^-_+)\right)\nonumber\\
&&\ \ \ \ +{\left(E_p^+\right)^2
+\left(E_p+\mu_I/2\right)^2+k_0\left(E_p+\mu_I/2\right)\over k_0^2
- 4\left(E_p^+\right)^2}\nonumber\\
&&\ \ \ \ {1\over E_p^+}
\left(f(E^-_+)-f(-E^+_+)\right)\Bigg]\ ,\nonumber\\
&&\Pi_{\pi_+\pi_-}(k_0)=\Pi_{\pi_-\pi_+}(k_0)=-4N_c\int{d^3{\bf
p}\over
(2\pi)^3}\nonumber\\
&&\ \ \ \
\Big[\frac{4G^2\pi^2}{k_0^2-4(E_p^-)^2}\frac{1}{E_p^-}(f(E^-_-)-f(-E^-_+))\nonumber\\
&&\ \ \ \ +\frac{4G^2\pi^2}{k_0^2
-4(E_p^+)^2}\frac{1}{E_p^+}(f(E^+_-)-f(-E^+_+))\Big]\ ,\\
&&\Pi_{\sigma\pi_+}(k_0)=\Pi_{\pi_+\sigma}(k_0) =
4\sqrt{2}N_cG\pi\int{d^3{\bf p}\over
(2\pi)^3}\frac{M_q}{E_p}\nonumber\\
&&\ \ \ \
\Bigg[\frac{2E_p-\mu_I+k_0}{k_0^2-4(E_p^-)^2}\frac{1}{E_p^-}(f(E^-_-)-f(-E^-_+))\nonumber\\
&&\ \ \ \ +\frac{2E_p+\mu_I-k_0}{k_0^2-4(E_p^+)^2}\frac{1}{E_p^+}(f(E^+_-)-f(-E^+_+))\Bigg]\ ,\nonumber\\
&&\Pi_{\sigma\pi_-}(k_0)=\Pi_{\pi_-\sigma}(k_0) =
4\sqrt{2}N_cG\pi\int{d^3{\bf p}\over
(2\pi)^3}\frac{M_q}{E_p}\nonumber\\
&&\ \ \ \
\Bigg[\frac{2E_p-\mu_I-k_0}{k_0^2-4(E_p^-)^2}\frac{1}{E_p^-}(f(E^-_-)-f(-E^-_+))\nonumber\\
&&\ \ \ \
+\frac{2E_p+\mu_I+k_0}{k_0^2-4(E_p^+)^2}\frac{1}{E_p^+}(f(E^+_-)-f(-E^+_+))\Bigg]\
.\nonumber
\end{eqnarray}

\section {Hatree-Fock Approximation in Linear Sigma Model }
\label{a3}
In Hatree-Fock approximation, the interaction terms among $\sigma$
and $\pi_i$ in the effective Lagrangian (\ref{sigma4}) of the
linear sigma model are fully absorbed into the $\sigma$ and
$\pi_i$ effective masses, and the model behaviors like a
quasi-particle system. At sufficiently high temperature, the pion
condensation vanishes and the system contains only chiral
condensate. This is the case we considered in deriving the
critical isospin chemical potential $\mu_I^c$ for pion
superfluidity in Section \ref{s6}. In this case the Lagrangian
density is simplified as
\begin{eqnarray}
\label{a31}
 {\cal L}_{HF} &=&
\frac{1}{2}\Big((\partial\sigma)^2 +(\partial {\bf
\pi}_3)^2+\left(\partial_t\pi_1-\mu_I\pi_2\right)^2\\
&+&\left(\partial_t\pi_2+\mu_I\pi_1\right)^2-\left({\bf
\nabla}\pi_1\right)^2+\left({\bf
\nabla}\pi_2\right)^2\Big)\nonumber\\
&-&\frac{1}{2}\left[M_\sigma^2 \sigma^2+M_0^2 \pi_3^2+M_\pi^2
(\pi_1^2+\pi_2^2)\right]-\bar{U}(\xi)\ .\nonumber
\end{eqnarray}
with the effective meson masses
\begin{eqnarray}
\label{a32}
&& M_\sigma^2 =
2g_\pi^2\left(3\xi^2-f_\pi^2+3\langle\sigma\sigma\rangle+2\langle
\pi\pi\rangle+\langle
\pi_0\pi_0\rangle\right)+m_\pi^2\ ,\nonumber\\
&& M_0^2 =
2g_\pi^2\left(\xi^2-f_\pi^2+\langle\sigma\sigma\rangle+2\langle
\pi\pi\rangle+3\langle\pi_0\pi_0\rangle \right)+m_\pi^2\ ,\nonumber\\
&& M_\pi^2 =
2g_\pi^2\left(\xi^2-f_\pi^2+\langle\sigma\sigma\rangle+4\langle\pi\pi\rangle
 +\langle
\pi_0\pi_0\rangle\right)+m_\pi^2\ ,\nonumber\\
\end{eqnarray}
and the effective potential
\begin{eqnarray}
\label{a33} \bar{U}(\xi)
&=&\frac{1}{2}(2g_\pi^2f_\pi^2-m_\pi^2)\xi^2+{g_\pi^2\over
2}\xi^4-f_\pi m_\pi^2\xi\nonumber\\
&+&{g_\pi^2\over
2}\Big[\Big(\langle\sigma\sigma\rangle+2\langle\pi\pi\rangle+\langle\pi_0\pi_0\rangle\Big)^2\nonumber\\
&+&2\Big(\langle\sigma\sigma\rangle^2+2\langle\pi\pi\rangle^2+\langle\pi_0\pi_0\rangle^2\Big)\Big]\
 ,
\end{eqnarray}
where the thermal excitation functions
$\langle\sigma\sigma\rangle, \langle\pi\pi\rangle$ and
$\langle\pi_0\pi_0\rangle$ are calculated from the mean field
propagators,
\begin{eqnarray}
\label{a34} \langle\sigma\sigma\rangle &=& \int{d^3{\bf k}\over
(2\pi)^3}\frac{1}{\sqrt{k^2+M_\sigma^2}}\frac{1}{e^{\sqrt{\bf
k^2+M_\sigma^2}/T}-1}\ ,\\
\langle\pi_0\pi_0\rangle &=& \int{d^3{\bf k}\over
(2\pi)^3}\frac{1}{\sqrt{k^2+M_0^2}}\frac{1}{e^{\sqrt{\bf
k^2+M_0^2}/T}-1}\ ,\nonumber\\
\langle\pi\pi\rangle &=&\int{d^3{\bf k}\over
(2\pi)^3}\frac{1}{2\sqrt{k^2+M_\pi^2}}\nonumber\\
&&\left[\frac{1}{e^{(\sqrt{\bf
k^2+M_\pi^2}-\mu_I)/T}-1}+\frac{1}{e^{(\sqrt{\bf
k^2+M_\pi^2}+\mu_I)/T}-1}\right]\ .\nonumber
\end{eqnarray}
The thermodynamic potential of the system in Hartree-Fock
approximation can be expressed as the summation of the
quasi-particle contributions plus the effective potential,
\begin{eqnarray}
\label{a35} \Omega&=&\bar{U}(\xi)+T\int\frac{d^3\bf
k}{(2\pi^3)}\ln\left[1-e^{-\sqrt{\bf
k^2+M_\sigma^2}/T}\right]\nonumber\\
&+&T\int\frac{d^3\bf k}{(2\pi^3)}\ln\left[1-e^{-\sqrt{\bf
k^2+M_0^2}/T}\right]\nonumber\\
&+&T\int\frac{d^3\bf k}{(2\pi^3)}\ln\left[1-e^{-(\sqrt{\bf
k^2+M_\pi^2}-\mu_I)/T}\right]\nonumber\\
&+&T\int\frac{d^3\bf k}{(2\pi^3)}\ln\left[1-e^{-(\sqrt{\bf
k^2+M_\pi^2}+\mu_I)/T}\right]\ .
\end{eqnarray}
The gap equation for the chiral condensate $\xi$ is derived by
minimizing the thermodynamic potential,
\begin{equation}
\label{a36}
2g_\pi^2\xi\left(\xi^2-f_\pi^2+3\langle\sigma\sigma\rangle+2\langle
\pi\pi\rangle+\langle\pi_0 \pi_0\rangle\right)+\xi m_\pi^2= f_\pi
m_\pi^2\ .
\end{equation}
The mass equations (\ref{a32}) with the thermal excitation
functions (\ref{a34}) and the gap equation (\ref{a36}) form a
group of coupled equations and determine self-consistently the
temperature and chemical potential behavior of the chiral
condensate and the effective meson masses.

\section{ Polarization Functions in $SU(3)$ NJL Model}
\label{a4}
We define the following matrices
\begin{eqnarray}
\lambda_{ab}^\pm=\frac{1}{\sqrt{2}}(\lambda_a \pm i\lambda_b)\ .
\end{eqnarray}
The meson polarization functions are defined as
\begin{equation}
\Pi_{MM}(k) = i\int{d^4p\over (2\pi)^4} \text{Tr}\left(\Gamma_M^*
{\cal S}_{mf}(p+k)\Gamma_M {\cal S}_{mf}(p)\right)\ ,
\end{equation}
with the couplings
\begin{eqnarray}
\Gamma_M &=& \left\{\begin{array}{ll}
\lambda_{12}^+ & M=a^+ \\
\lambda_{12}^- & M=a^- \\
\lambda_3 & M=a^0\\
\lambda_{45}^+ & M=\kappa^+ \\
\lambda_{45}^- & M=\kappa^- \\
\lambda_{67}^+ & M=\kappa^0 \\
\lambda_{67}^- & M=\bar{\kappa}^0 \\
i\gamma_5\lambda_{12}^+ & M=\pi^+ \\
i\gamma_5\lambda_{12}^- & M=\pi^- \\
i\gamma_5\lambda_3 & M=\pi^0\\
i\gamma_5\lambda_{45}^+ & M=K^+ \\
i\gamma_5\lambda_{45}^- & M=K^- \\
i\gamma_5\lambda_{67}^+ & M=K^0 \\
i\gamma_5\lambda_{67}^- & M=\bar{K}^0
\end{array}\right.\ ,\nonumber\\
\Gamma_M^* &=& \left\{\begin{array}{ll}
\lambda_{12}^- & M=a^+ \\
\lambda_{12}^+ & M=a^- \\
\lambda_3 & M=a^0\\
\lambda_{45}^- & M=\kappa^+ \\
\lambda_{45}^+ & M=\kappa^- \\
\lambda_{67}^- & M=\kappa^0 \\
\lambda_{67}^+ & M=\bar{\kappa}^0 \\
i\gamma_5\lambda_{12}^- & M=\pi^+ \\
i\gamma_5\lambda_{12}^+ & M=\pi^- \\
i\gamma_5\lambda_3 & M=\pi^0\\
i\gamma_5\lambda_{45}^- & M=K^+ \\
i\gamma_5\lambda_{45}^+ & M=K^- \\
i\gamma_5\lambda_{67}^- & M=K^0 \\
i\gamma_5\lambda_{67}^+ & M=\bar{K}^0 \\
\end{array}\right.\ .
\end{eqnarray}
Doing the trace in color and flavor space first, we obtain
\begin{eqnarray}
\Pi_{a_0a_0}(k) &=& iN_c\int {d^4p\over (2\pi)^4} \text{Tr}_D
\Big({\cal
S}_{u}(p+k){\cal S}_{u}(p)\nonumber\\
&+&{\cal
S}_{d}(p+k){\cal S}_{d}(p)\Big)\ ,\nonumber\\
\Pi_{a_+a_+}(k) &=& 2iN_c\int {d^4p\over (2\pi)^4} \text{Tr}_D
\Big({\cal S}_{u}(p+k){\cal S}_{d}(p)\Big)\ ,\nonumber\\
\Pi_{a_-a_-}(k) &=& 6i\int {d^4p\over (2\pi)^4} \text{Tr}_D
\Big({\cal
S}_{d}(p+k){\cal S}_{u}(p)\Big)\ ,\nonumber\\
\Pi_{k_+k_+}(k) &=& 2N_ci\int {d^4p\over (2\pi)^4} \text{Tr}_D
\Big({\cal S}_{u}(p+k){\cal S}_{s}(p)\Big)\ ,\nonumber\\
\Pi_{k_-k_-}(k) &=& 2N_ci\int {d^4p\over (2\pi)^4} \text{Tr}_D
\Big({\cal
S}_{s}(p+k){\cal S}_{u}(p)\Big)\ ,\nonumber\\
\Pi_{k^0k^0}(k) &=& 2N_ci\int {d^4p\over (2\pi)^4} \text{Tr}_D
\Big({\cal S}_{d}(p+k){\cal S}_{s}(p)\Big)\ ,\nonumber\\
\Pi_{\bar{k}^0\bar k^0}(k) &=& 2N_ci\int {d^4p\over (2\pi)^4}\text{Tr}_D \Big({\cal S}_{s}(p+k){\cal S}_{d}(p)\Big)\ ,\nonumber\\
\Pi_{\pi_0\pi_0}(k) &=& -N_ci\int {d^4p\over (2\pi)^4} \text{Tr}_D
\Big(\gamma_5{\cal S}_{u}(p+k)\gamma_5{\cal
S}_{u}(p)\nonumber\\
&+&\gamma_5{\cal
S}_{d}(p+k)\gamma_5{\cal S}_{d}(p)\Big)\ ,\nonumber\\
\Pi_{\pi_+\pi_+}(k) &=& -2N_ci\int {d^4p\over (2\pi)^4} \text{Tr}_D
\Big(\gamma_5{\cal S}_{u}(p+k)\gamma_5{\cal S}_{d}(p)\Big)\ ,\nonumber\\
\Pi_{\pi_-\pi_-}(k) &=&- 2N_ci\int {d^4p\over (2\pi)^4} \text{Tr}_D
\Big(\gamma_5{\cal
S}_{d}(p+k)\gamma_5{\cal S}_{u}(p)\Big)\ ,\nonumber\\
\Pi_{K_+K_+}(k) &=& -2N_ci\int {d^4p\over (2\pi)^4} \text{Tr}_D
\Big(\gamma_5{\cal S}_{u}(p+k)\gamma_5{\cal S}_{s}(p)\Big)\ ,\nonumber\\
\Pi_{K_-K_-}(k) &=& -2N_ci\int {d^4p\over (2\pi)^4} \text{Tr}_D
\Big(\gamma_5{\cal
S}_{s}(p+k)\gamma_5{\cal S}_{u}(p)\Big)\ ,\nonumber\\
\Pi_{K^0K^0}(k) &=& -2N_ci\int {d^4p\over (2\pi)^4} \text{Tr}_D
\Big(\gamma_5{\cal S}_{d}(p+k)\gamma_5{\cal S}_{s}(p)\Big)\ ,\nonumber\\
\Pi_{\bar{K}^0\bar K^0}(k) &=& -2N_ci\int {d^4p\over (2\pi)^4}
\text{Tr}_D \Big(\gamma_5{\cal S}_{s}(p+k)\gamma_5{\cal
S}_{d}(p)\Big)\
.\nonumber\\
\end{eqnarray}

\end{document}